\documentstyle[aaspp4]{article}
\def \sun {$_{\scriptscriptstyle \odot}$}
\def\gtaprx {\lower .1ex\hbox{\rlap{\raise .6ex\hbox{\hskip .3ex
	{\ifmmode{\scriptscriptstyle >}\else
		{$\scriptscriptstyle >$}\fi}}}
	\kern -.4ex{\ifmmode{\scriptscriptstyle \sim}\else
		{$\scriptscriptstyle\sim$}\fi}}}
\def\ltaprx {\lower .1ex\hbox{\rlap{\raise .6ex\hbox{\hskip .3ex
	{\ifmmode{\scriptscriptstyle <}\else
		{$\scriptscriptstyle <$}\fi}}}
	\kern -.4ex{\ifmmode{\scriptscriptstyle \sim}\else
		{$\scriptscriptstyle\sim$}\fi}}}
\lefthead{FRYER, WOOSLEY, \& HARTMANN}
\righthead{BHAD GRBs}
\begin{document}
\begin{center} To be submitted to {\em The Astrophysical Journal}
\end{center}
\vspace{1.cm}
\title{Formation Rates of Black Hole Accretion Disk Gamma-Ray Bursts}
\author{Chris L. Fryer, S. E. Woosley}
\affil{Astronomy Department, University of California
\\  Santa Cruz, CA 95064 \\ cfryer@ucolick.org}
\authoremail{cfryer@ucolick.org, woosley@ucolick.org}
\author{Dieter H. Hartmann}
\affil{Clemson University \\ Clemson, SC, 29634}
\authoremail{hartmann@grb.phys.clemson.edu}

\begin{abstract}
The cosmological origin of at least an appreciable fraction of
classical gamma-ray bursts (GRBs) is now supported by redshift
measurements for a half-dozen faint host galaxies. Still, the nature
of the central engine (or engines) that provide the burst energy
remains unclear. While many models have been proposed, those currently
favored are all based upon the formation of and/or rapid accretion
into stellar mass black holes. Here we discuss a variety of such
scenarios and estimate the probability of each. Population synthesis
calculations are carried out using a Monte Carlo approach in which the
many uncertain parameters intrinsic to such calculations are varied.
We estimate the event rate for each class of model as well as the
propagation distance for those having significant delay between
formation and burst production, i.e., double neutron star (DNS)
mergers and black hole -- neutron star (BH/NS) mergers. One conclusion
is a one to two order of magnitude decrease in the rate of DNS and
BH/NS mergers compared to that previously calculated using invalid
assumptions about common envelope evolution.  Other major
uncertainties in the event rates and propagation distances include the
history of star formation in the universe, the masses of the galaxies
where merging compact objects are born, and the radii of the hydrogen
stripped cores of massive stars. For reasonable assumptions regarding
each, we calculate a daily event rate in the universe for i) merging
neutron stars: $\sim$100/day; ii) neutron-star black hole mergers:
$\sim$450/day; iii) collapsars: $\sim$10$^4$/day; iv) helium star
black hole mergers: $\sim$1000/day; and v) white dwarf black hole
mergers: $\sim$20/day.  The range of uncertainty in these numbers
however, is very large, typically two to three orders of
magnitude. These rates must additionally be multiplied by any relevant
beaming factor ($f_{\Omega} < 1$) and sampling fraction (if the entire
universal set of models is not being observed). Depending upon the
mass of the host galaxy, half of the DNS mergers will happen within
60\,kpc (for a galaxy with a mass comparable to that of the Milky Way)
to 5\,Mpc (for a galaxy with negligible mass) from the galactic center. 
The same numbers characterize BH/NS mergers.  Because of the delay time, 
neutron star and black hole mergers will happen at a redshift 0.5 to 0.8 
times that of the other classes of models. Information is still lacking 
regarding the hosts of short hard bursts, but we suggest that they are due to
DNS and BH/NS mergers and thus will ultimately be determined to lie
outside of galaxies and at a closer mean distance than long complex
bursts (which we attribute to collapsars). In the absence of a
galactic site, the distance to these bursts may be difficult to
determine.

\end{abstract}


\section{Introduction}

The mechanisms responsible for gamma-ray bursts (GRBs) are not yet
known with any certainty, but as evidence for a cosmological origin
mounts (e.g., Metzger et al. 1997; Frail et al. 1997; Bloom, et
al. 1998; Bloom et al. 1999a), models based upon black hole accretion
disks (BHAD models) have gained prominence. Such models have several
advantages, especially their large inherent energy and potentially
short time scales. In the neutrino-powered GRB paradigm, BHAD models
can overcome the strong angle dependence of neutrino annihilation and
extract some of the gravitational energy of the disk, possibly without
ejecting too much baryonic matter (Paczynski 1991; Woosley 1993;
Ruffert et al. 1997; Popham, Woosley, \& Fryer 1998; MacFadyen \&
Woosley 1999).  BHAD models also provide favorable conditions for
magnetic fields to extract energy, either from the gravitational
potential energy of the accretion torus itself, or from the rotational
energy of spinning black holes (Blandford \& Znajek 1977; MacDonald et
al. 1986; Paczynski 1991, 1998; Narayan et al. 1992; Hartmann \&
Woosley 1995; Meszaros \& Rees 1997a; Katz 1997; Livio, Ogilvie, \&
Pringle 1999).  The merger of double neutron star (DNS) systems is
probably the best studied representative of this class, but a variety
of progenitor systems lead to similar conditions: mergers of other
compact systems such as binaries consisting of a neutron star and a
black hole (BH/NS) or of a white dwarf and black hole (WD/BH), the
inspiral of a compact object into its companion's helium core during
common envelope evolution (helium-mergers), and the accretion of
material into a black hole in a ``failed supernova'' (collapsar).

Estimates of the rates for these various occurrences are plagued by a
variety of uncertainties in the supernova explosion mechanism, stellar
evolution, and binary evolution.  Although the current literature
abounds with population synthesis calculations, especially of DNS
systems, and the closely-related BH/NS binaries (Clark, van den
Heuvel, \& Sutantyo 1979; Lipunov, Postnov, \& Prokhorov 1987; Hills,
Bender \& Webbink 1991; Tutukov \& Yungelson 1993; Lipunov et
al. 1995; Portegies-Zwart \& Spreeuw 1996; Fryer, Burrows, \& Benz
1998; Portegies-Zwart \& Yungelson 1998, Bethe \& Brown 1998;
Belczy\'nski \& Bulik 1998; Bloom, Sigurdsson, \& Pols 1999b), little
attention has been directed towards either exploring the uncertainties
in the models or in examining the other kinds of models.

In this paper, we shall study the rate of formation of all varieties
of BHAD progenitors, including our best estimates of the (admittedly
uncertain) physics. We will show that the discrepancy between rates
calculated from population studies and extrapolations from the data of
DNS systems are reconcilable if one includes both the uncertainties in
the population synthesis calculations and the error in the measured
rates. BHAD models other than DNS will also be considered within
the same parameterized framework. Comparisons with observed compact
systems can be used to constrain some of the parameters, e.g., the
pulsar velocity distribution and the observed properties of X-ray
binaries and DNS systems. However, even with these constraints, we
shall find that most rates for GRB progenitor formation are uncertain
by over two orders of magnitude.

From these studies, we shall also determine the velocity distributions
of the GRB progenitor systems and their corresponding merger time
scales. By evolving these systems in representative gravitational
potentials for both high and low mass galaxies, we can determine the
distances each GRB progenitor travels before producing a GRB. A
fraction of the compact binary systems receive large enough systemic
velocities to be ejected from their host galaxies, but many remain
bound even in such low mass galaxies as the host of GRB970508 (Bloom
et al. 1998).

In $\S$2, we describe formation scenarios for each of the black hole
accretion disk GRB progenitor systems. Section 3 describes our
population synthesis code and the parameters of population synthesis,
noting the effects of the uncertainty in each.  Section 4 gives our
predictions, including the effects of varying systemic velocities and
merger time scales to obtain both the spatial and temporal distribution
of GRBs from each progenitor. We conclude with a summary of the merits
of the different models.

\section{Black Hole Accretion Disk progenitors}

Modeling the progenitors of BHAD derived GRBs involves all the
uncertainties inherent in any study of compact object formation.
Moreover, since the systems must be in short-period binaries to merge,
the problem is compounded by uncertainties in the parameters of
binaries (mass ratios, separations) and their evolution.  Before
delving farther into these uncertainties, it is first helpful to have
in mind some general scenarios (Table 1, Fig. 1).

\subsection{Double Neutron Star Binaries}\label{sec:dns}

When two neutron stars merge, they quickly form an object too large to
be supported by nuclear and degeneracy pressure (for most choices of
equation of state). A black hole forms on a dynamic time scale, but a
significant amount of mass, typically 0.03 -- 0.3 M$_\odot$, has too
much angular momentum to enter the hole promptly. An accretion disk
forms. What happens next depends upon the viscosity and entropy of
this disk, the evolution of its magnetic field, and the interaction of
the disk with the rotating black hole. All three may be quite
complicated. If the viscosity is low and the temperature inadequate to
drive rapid energy loss by neutrinos, the disk may persist for a long
time, perhaps over 10 s in the models envisioned by Meszaros \& Rees
(1997a). Any relativistic outflow that develops must then be a
consequence of MHD processes in the disk or of the magnetic extraction
of rotational energy from the hole. These same MHD processes can also
function in a viscous disk (indeed some appreciable magnetic field
must be present to provide the viscosity), but the time scale is much
shorter and the temperature higher. Most models of DNS mergers that
have been carried out on the computer are of this latter variety
(Davies et al. 1994;  Rasio \& Shapiro 1994; Ruffert, Janka, Sch\"afer 1995; 
Janka \& Ruffert 1996; Wilson, Mathews, \& Marronetti 1996; Mathews \& Wilson 
1997;  Ruffert et al. 1997, Ruffert \& Janka 1998; Rossweg at al. 1999). In
the ``hot model'' The entire accretion process then lasts for
10-100\,ms and most of the binding energy of the disk is radiated as
neutrinos. Indeed the neutrino luminosity can be so great that a
significant fraction, on the order of 1\%, is converted, via neutrino
annihilation, to thermal energy along the rotational axis. This is
enough to power energetic jets that produce a GRB either by internal
or external shocks. The hot DNS model might therefore be more
appropriate for the short, hard subset of GRBs (Fishman \& Meegan
1995) which are very difficult to make other ways (except in the
closely allied BH/NS model). Observationally, these short hard bursts
are not yet constrained, either in energy or host, by the observations
of Beppo-Sax, which hitherto has employed a 5 s trigger time.

The progenitor systems here are double neutron stars in binaries so
close that merger occurs in a Hubble time as a consequence of
gravitational radiation.  DNS systems have a long observable lifetime
because the primary's neutron star accretes matter and becomes a
recycled radio pulsar.  Recycled pulsars are thought to be produced in
binaries where accretion from a companion spins up the neutron star
whose magnetic field has partially decayed (see reviews by Verbunt
1993; Phinney \& Kulkarni 1994). Due to the low magnetic field
strengths, spin down of these pulsars occurs slowly, and consequently
recycled radio pulsars have observable lifetimes $\gtrsim 10-100$
times longer than normal pulsars. It is not surprising then that
essentially all pulsars observed in double neutron star systems are
recycled pulsars.

The standard scenario for forming close orbit DNS systems (Scenario I,
Fig. 2) begins with two massive stars. The more massive star (primary)
evolves off the main sequence, overfills its Roche-lobe, and transfers
mass to its companion (secondary).  The primary then evolves to the
end of its life, forming a neutron star in a supernova explosion.  The
binary system remains bound, unless large kicks\footnote{The pulsar
velocity distribution requires that neutron stars are somehow
accelerated to mean velocities of roughly $\sim 450 {\rm \, km \,
s^{-1}}$, most likely at or near the time of birth.  We discuss this
in more detail in \S \ref{sec:sn}.} are imparted to the neutron star
during formation, and a binary consisting of a neutron star and
massive companion is formed.  This system passes through an X-ray
binary phase which then evolves through a common envelope as the
secondary star expands.  During this common envelope phase in the
standard model, the neutron star spirals into the massive secondary
and the orbital energy released ejects the secondary's hydrogen
mantle, forming a neutron star/helium star binary. Mass accretion onto
the neutron star primary is alleged to add angular momentum to the
neutron star, ``recycling'' it as a pulsar. After the explosion of the
helium star as a supernova, a close DNS binary remains which, in time,
merges via gravitational wave emission.

However, recent calculations of neutron star accretion (Chevalier
1993, 1996; Houck \& Chevalier 1992; Brown 1995; Fryer, Benz, \&
Herant 1996) reveal that, during the common envelope phase, the
neutron star accretes at the Bondi-Hoyle rate, not the Eddington rate.
During its spiral in, the neutron star can accrete over a solar mass
(Bethe \& Brown 1998) and collapse to form a black hole. Thus, the
``standard'' scenario for DNS systems may in fact form BH/NS binaries.

How, then, are close DNS systems formed? Brown (1995) suggested an
alternate scenario (Scenario II, Fig. 3) in which the initial binary
system is comprised of two massive stars of nearly equal mass. The
secondary (whose mass is assumed to be within 5\% of the primary)
evolves off the main sequence before the explosion of the primary as a
supernova.  The two stars then enter a common envelope phase with two
helium cores orbiting within one combined hydrogen envelope. After the
hydrogen envelope is ejected, first one, then the other, helium stars
explode.  If the kicks imparted to the neutron stars during the
supernova explosion are sufficiently low, a double neutron star system
can form. To generate the observed systems, after the first supernova,
the newly formed neutron star must accrete material, possibly from its
companion's helium star wind, to become the observed ``recycled''
radio pulsar. If accretion is insufficient, a ``silent'' double
neutron star binary is formed in which neither neutron star is
recycled. Such systems, though only briefly detectable as radio 
pulsras, might still evolve to produce GRBs.

A third scenario (Scenario III, Fig. 4) avoids any common envelope
phase, but requires that the neutron star formed in the explosion of
the secondary receive a kick that places it into an orbit that will
allow the two neutron stars to merge within a Hubble time.  However,
in general, to avoid a common envelope phase, the pre-supernova
orbital separation must be extremely wide (greater than 1 AU). The
range of kick magnitudes and directions which will produce such orbits
is so small that, as we shall show in this paper, this formation scenario is
probably not important.

From the number of known Galactic DNS binaries (PSR 1913+16, PSR
1534+12), a formation rate can be estimated. However, correcting for
time evolution of pulsar luminosities, beaming fractions, and distance
estimations proves tricky for a sample of only two merging double
neutron star systems.  From the number of observed DNS systems
($N_{\rm obs}$), and including an estimate of all the selection
effects: beaming fraction, fraction of galaxy currently observed,
etc., the formation rate is calculated as
\begin{equation}\label{eq:rate}
{\rm Rate}_{\rm DNS}=f_{\rm b} \frac{N_{\rm obs}}
{T_{\rm PSR}},
\end{equation}
where $T_{\rm PSR}$ is the observable lifetime of the pulsar and all
other uncertainties have been represented by a single factor, $f_{\rm
b}$, whose value may be uncertain by 1-2 orders of
magnitude.  With best guesses for $f_{\rm b}$,
the current rate inferred for our Galaxy from the
observations fall in the range: $\sim 10^{-6}-10^{-5} \, {\rm
yr^{-1}}$ (Phinney 1991; Narayan, Piran, \& Shemi 1991; Curran \& Lorimer 1995;
van den Heuvel \& Lorimer 1996). Bailes (1996) estimated an upper
limit for the formation rate of these systems using the fact that
all of the observed pulsars in DNS systems are recycled and none of
the companion neutron stars are observed as normal pulsars. Although
normal pulsars have much shorter lifetimes, they dominate the observed
pulsar population because they do not require binary accretion to
``recycle'' them. The formation rate of single pulsars is roughly
$\sim0.008$\,yr$^{-1}$ and, at the time of the calculation by Bailes
(1996), roughly 650 normal radio pulsars were observed.  Using
eq. (\ref{eq:rate}), we can then estimate the value of $f_{\rm
b}/T_{\rm PSR}$.  If we assume that the collapse and
pulsar formation of the secondary is not too dissimilar from that of
single stars (since the core of a massive star is not affected greatly
by its binary companion, this assumption may not be too bad), then we
can use the value of $f_{\rm b}/T_{\rm PSR}$ derived from
the pulsar population to place an upper limit on the DNS formation
rate. Since the number of normal secondary pulsars is 0 ($<1$), the
current DNS formation rate in our Galaxy is then (Bailes 1996)
\begin{equation}
{\rm Rate}_{\rm DNS}< \left( \frac{f_{\rm b}}{T_{\rm
PSR}} \right)_{\rm PSR}=10^{-5} \, {\rm yr^{-1}}.
\end{equation}
Since this calculation, the number of observed pulsars, the number 
of DNS systems has increased, and the pulsar lifetime has changed, 
and this upper limit has risen to $10^{-4} \, {\rm yr^{-1}}$
(Arzoumanian, Cordes, \& Wasserman 1999).

\subsection{Black Hole + Neutron Star Binaries}\label{sec:bhns}

As BH/NS binaries merge, the neutron star is tidally disrupted. Some
of the material accretes directly, producing little emission, but the
remainder forms an accretion disk of a few tenths of solar mass which
can power a GRB (Lee \& Kluzniak 1995; Eberl 1998; Eberl et al. 1999).
Like the DNS mergers, if the disk is hot and viscous, the relativistic
outflows created by BH/NS mergers are probably of short duration, but
somewhat more energetic than their DNS counterparts because of the
larger mass reservoir.

Merging BH/NS binaries are formed by pathways that are similar to DNS
binaries.  In general, the collapse of a massive star can form black
holes in two distinct ways (Woosley \& Weaver 1995; Fryer 1999). The
star can explode, but with such a weak shock that much of the stellar
mantle of heavy elements falls back onto the neutron star, causing it
to further collapse into a black hole. Alternatively, especially for
the more massive stars, the ram pressure of the infalling stellar
material may be too large to allow the shock to propagate at all and
the star may collapse directly into a black hole.  The two limits for
black hole formation from fallback and direct collapse are,
respectively, $M_{\rm BH} \approx 25$M\sun\,, $M_{\rm Coll} \approx
40$M\sun\,(Fryer 1999).  The ``standard'' formation scenario for BH/NS
systems (Scenario IV, Fig. 5) begins with two massive stars, the
primary having a mass greater than $M_{\rm BH}$.  As with the standard
DNS formation scenario, the primary evolves off the main sequence,
overfills its Roche-lobe, and transfers mass to its companion. The
primary continues to evolve to the end of its life, forming a black
hole and, possibly, a supernova explosion.  With the exception of
extremely large kicks, the system remains bound, creating a binary
consisting of a black hole and a massive star. This system passes
through an X-ray binary phase (e.g., Cyg X-1, LMC X-1) and then
evolves through a common envelope as the secondary star
expands. During this common envelope phase (in the standard model) the
black hole spirals into the massive secondary and ejects the
secondary's hydrogen envelope.  After the secondary's supernova
explosion, a BH/NS binary forms which then merges via gravitational
wave emission.

BH/NS binaries can also form from neutron star binaries in which 
the neutron star of the primary gains too much matter during common 
envelope phase and collapses into a black hole (\S \ref{sec:dns},
Scenario V, Fig. 6). This formation scenario produces binaries 
consisting of a low-mass black hole ($\sim 3 M_\odot$) and a neutron 
star which could generate the most energetic BH/NS GRBs (Eberl 1998, 
Eberl et al. 1999).   

Finally, just as with the DNS systems, a BH/NS binary can form under a
third mechanism avoiding any common envelope evolution for appropriate
kick magnitudes and directions (Scenario VI, Fig. 7). Like the DNS
kick formation scenario, the kick scenario is probably not important 
for BH/NS formation.

Although BH/NS binaries have not yet been observed, their formation
rate is probably comparable or even larger than that of DNS
systems. As with the double neutron stars, we can determine an
observational upper limit to the BH/NS rate in the Galaxy (assuming the
observational biases are similar to those for single pulsars) of $\sim
10^{-4}$\,yr$^{-1}$.

\subsection{Black Hole -- White Dwarf Binaries}

As WD/BH binaries merge, the white dwarf is tidally disrupted and most
of its matter is converted into an accretion disk around the black
hole. The accretion of this disk material onto the black hole can
drive a gamma-ray burst (Fryer et al. 1999). Although the accretion
disk is more massive than that formed in DNS mergers, its radius is
much larger so the accretion time scale is longer and neutrino energy
transport less efficient. However, the black hole is rapidly rotating
and up to 10$^{53}$ erg of either rotational energy or disk binding
energy is potentially available on a viscous time scale for the
disk. MHD powered jets might provide long-duration GRBs.

Not surprisingly, there is also more than one way to produce a binary
containing a white dwarf and a black hole.  Our first scenario begins
with a binary consisting of a primary with mass greater than $M_{\rm BH}$
and a low-mass ($< 10$M\sun) companion (Scenario VII, Fig. 8).  
After the primary collapses to form
a black hole, one has a binary consisting of a black hole and
a main sequence star. If the mass of the main sequence companion is
less than $\sim 2-3 M_\odot$, magnetic braking will drive the binary
together, forming a low-mass X-ray binary (LMXB). More massive
companions will not become X-ray binaries until they evolve off the
main sequence (e.g. LMC X-3, 2023+338 Nova Cyg, J0422+32 Nova
Per). For masses of $5-10 {\rm M_\odot}$, a common envelope phase is
likely to occur which decreases the orbital separation. These stars
eventually evolve into high mass white dwarfs, a fraction of which
will merge with the black hole primary in a Hubble time.

Alternatively, these systems may form from neutron star binaries with
low-mass ($< 10 M_\odot$) companions (Scenario VIII, Fig. 9). If the
neutron star enters a common envelope phase when the low-mass
companion expands into a giant, it can accrete too much material and
collapse to a black hole. BH/WD binaries may also form avoiding the
first common envelope phase (where the low-mass companion spirals into
the massive primary) entirely with a ``well-placed'' kick (Scenario
IX, Fig. 10). Just as with Scenario VII, the system evolves briefly
through an X-ray binary phase and possible common envelope phase which
tightens the orbit.

Just as with BH/NS binaries, there are no observed closely interacting
BH/WD binaries and our estimates must be very uncertain. The lack of a
detectable system so far implies only
\begin{equation}
{\rm Rate}_{\rm WD/BH} \lesssim \frac{V_{\rm Galaxy}}
{V_{\rm WD \, Obs}} T_{\rm WD}^{-1}
\approx \frac{(10 {\rm \, kpc})^2 (1 {\rm \, kpc})}{(10 \, {\rm pc})^3}
\times (10^{10} \, {\rm yr})^{-1} 
\approx 0.01 \, {\rm yr^{-1}}.
\end{equation}
In this paper, we will show in fact that the predicted rate is $\sim
5$ orders of magnitude less than this upper limit.

\subsection{Collapsars}

The only BHAD GRB progenitor which does not necessarily require a
binary system (though, as we shall see, being in a binary still helps)
is the collapsar (Woosley 1993).  Collapsars are formed when the
collapse of the iron core of a rotating massive supernova progenitor
proceeds directly to a black hole. As the stellar mantle falls into
the newly formed black hole, angular momentum slows the collapse along
the equator, ultimately forming an accretion disk which powers the GRB
(Woosley 1993, MacFadyen \& Woosley 1999).  Collapsars can produce
long duration bursts ($\gtaprx10$\,s) with a large range of burst
energies by either neutrino or MHD processes. In order that the jet
not dissipate its energy in the hydrogen envelope of a red supergiant
star, it is necessary that the collapsar occur in a hydrogen stripped,
Wolf-Rayet star.

The collapsar model requires that no supernova shock is launched, 
and hence collapsars form only from stars whose masses exceed 
$M_{\rm Coll}$.  One obvious consequence is that GRBs from this model 
would closely trace star formation. There will also be significant amounts 
of circum-burster material from the pre-SN stellar wind and the gas in 
the surrounding star forming medium. The afterglow (and even some bursts) 
would be produced by shock interactions with this wind, not with the ordinary
interstellar medium. Scenario X (Fig. 11) is the simplest possibility for
producing a collapsar. The iron core of a single very massive star
collapses promptly to a black hole.

Current simulations of collapsars (MacFadyen \& Woosley 1999) suggest
that a successful collapsar model requires specific angular momenta of
the massive star cores to be in the range: $j \approx 3 \times 10^{16}
- 2 \times 10^{17}$ and that the massive star has lost most of its
hydrogen envelope, but retains a sufficiently large helium core to
collapse and power a GRB. If the angular momentum is below this range,
an accretion disk does not form and the material falls more or less
directly into the black hole.  If the angular momentum is too large,
the disk does not efficiently dissipate energy through neutrino
emission and a weak outflow instead of an accretion disk results. The
requisite high mass of the helium star may be more easily attained in
a star of low metallicity (MacFadyen \& Woosley 1999).

The biggest uncertainty in estimating the rate of collapsar models,
however, is the angular momentum.  There are diverse views regarding
how much angular momentum is retained in the stellar cores of massive
stars when they die.  Heger (1998) modeled the evolution of rotating
massive stars and found that the cores of most stars naturally have
angular momenta close to the range required by MacFadyen \& Woosley
(1999).  However, Spruit \& Phinney (1998) have argued that magnetic
fields will couple the cores of stars to their envelopes, causing the
star to rigidly rotate.  As the star became a red supergiant (or even
a blue one), the core rotation would essentially halt. No GRB could be
produced if this occurred.  However, the results of Spruit \& Phinney
rest upon uncertain assumptions involving the evolution of magnetic
field instabilities, especially those responsible for producing a
radial component in the region where the core and envelope are
interacting. Here we will assume that there is no such magnetic
coupling and all massive stars are rotating within the range required
by MacFadyen \& Woosley (1999). It will turn out however that, because
of the stellar mass loss prescription we adopt, massive stars other
than those interacting binaries, will produce a minority of collapsar
progenitors. If the stellar envelope is removed very early by a
companion star, perhaps the rotation of the core is not slowed as
much. However, one may want to multiply our final numbers for
collapsar rates by an uncertain factor less than one to account for
inadequate angular momentum.

Binary companions can also remove hydrogen envelopes during a common
envelope phase. After common envelope evolution, which uncovers the
helium core, the primary star (if massive enough) collapses into a
black hole.  If the angular momentum of the primary is appropriate, a
collapsar is formed (Scenario XI, Fig. 12). Note that this system may
remain in a binary and later evolve into a BH/NS or WD/BH binary.

Even if both Scenarios X and XI fail (e.g. because of magnetic braking
of the core's rotation; Spruit \& Phinney 1998), it is still possible
to form a collapsar in a binary in which the two stars are of
comparable mass, and evolve off the main sequence together. If
the two helium cores merge during their common envelope evolution,
they will form a massive helium core with a large amount of angular
momentum even if their individual cores were rotating slowly (Scenario
XII, Fig. 13). Scenario XII by itself then might be thought of as giving
a lower bound to the collapsar rate.

\subsection{Helium Star Mergers}

The helium star merger model (Fryer \& Woosley 1998) is another
outcome possible for the common envelope phase so central to making
DNS and BH/NS binaries.  If the inspiralling compact object does not
have sufficient orbital energy to eject its companion's hydrogen
envelope, it moves on into the helium core (e.g., in Scenarios I, IV,
V; see Figs. 2,5,6).  As the helium core and compact object merge, the
Bondi-Hoyle accretion rate may become extremely high ($1
M_\odot$\,s$^{-1}$) and the compact object quickly accretes enough to
become a black hole if it has not already. The angular momentum of the
helium core/black hole binary is injected into the helium core,
forming a massive disk ($\sim 4 M_\odot$) around a spinning black hole
(Scenario XIV). This BHAD GRB model has not been simulated in detail,
but it is likely to produce bursts of long duration similar to that of
the collapsar model, though probably longer. Unlike the collapsar
model, there is no concern about having enough angular
momentum. Indeed the problem may be too much of it.

Alternatively, the merging process can be initiated by a kick that
places the neutron star inside its companion, ultimately leading to
the merger of the neutron star with its companion's helium core, again
forming an accretion disk around a black hole.  However, the odds of
such a well-placed kick are again slim, and this scenario never occurred in
our simulations and hence we do not list it separately.

\section{The Population Synthesis Calculation}

For our numerical calculations, we used a modified version of the
Monte Carlo population synthesis code of Fryer, Burrows, \& Benz
(1998).  This code has previously been tested against, and, within the
errors, agrees with the low-mass X-ray binary formation calculations
of Kalogera \& Webbink (1998) and the high-mass X-ray binary formation
simulations of Dalton \& Sarazin (1995). Calculations of compact
object formation begin with a set of initial binary conditions, each
of which brings some element of uncertainty to the result: the initial
mass function, the fraction of binary stars, the binary mass ratio
distribution, the distribution of orbital separations, and initial
eccentricities. Monte-Carlo calculations then evolve the binary stars
and algorithms for both the single-star evolution (e.g., He core mass,
radii during giant phase, winds) and the binary-star evolution (e.g.,
mass transfer, common envelope evolution) must also be
assumed. Finally, one must make some approximations to the supernova
explosion mechanism and its results - especially neutron star (and
black hole) kicks.  We allow all these parameters to range over a
reasonable range of uncertainties, applying constraints where we can,
to derive a range in resultant GRB rate.  The effects of the free
parameters are demonstrated in Tables 2-7. An estimated range of rates,
including all the uncertainties, is given Table 8.

With such a large number of free parameters, we cannot survey the
entire parameter space, so we instead chose a ``standard'' model from
which to examine the effects of each individually.  Our standard model
uses the following assumptions (to be described in greater detail in
subsequent subsections): $\alpha_{\rm IMF}=2.7$, circular initial orbits, a
logarithmic distribution of initial separations [$P(A) \propto 1/A$], a flat
probability distribution for the mass ratio, a binary fraction of
40\%, a supernova rate of $0.02\,$yr$^{-1}$, isotropic supernova kicks
given to both neutron stars and black holes, a lower mass limit for
black hole formation of $25 M_\odot$, stellar radii for mass-transfer
and common-envelope evolution set to 1/4 that given by Kalogera \&
Webbink (1998), helium star masses as given by Kalogera \& Webbink
(1998) but with wind mass-loss rates set to 10$\%$ of those used by
Woosley, Langer, \& Weaver (1995), mass transfer parameters
($\alpha_{\rm MT}$, $\beta_{\rm MT}$) of 1.,0.8 respectively, and 
common envelope efficiency, ($\alpha_{\rm CE}$), equal to 0.5.  For other
parameters not discussed in this section, we used the standard values
given in Fryer, Burrows, \& Benz (1998).  Unless otherwise stated,
each Monte-Carlo simulation models 20 million massive binaries and the
plots and tables include only those binaries which will merge within a
Hubble time.  With 20 million binaries, a supernova rate of
$0.02\,$yr$^{-1}$ and a binary fraction of 40\%, a binary merger rate
of 1 Myr$^{-1}$ corresponds to 5000 systems and statistical errors of
roughly 2\%.

How calculations normalized to the supernova rate in our Galaxy are
translated into universal rates based upon a history of star formation
is discussed in $\S$4.2.

\subsection{Initial Conditions}

The initial conditions in our population synthesis calculations
include distributions of the zero-age main-sequence masses,
of orbital separations, and of orbital eccentricities for the 
binaries. The primary mass is determined using an initial mass 
function (IMF):
\begin{equation}
f(M_{p,0}) = f(1) M_{p,0}^{-\alpha_{\rm IMF}} 
{\rm stars \, yr^{-1} \, M^{-1}_\odot},
\end{equation}
where $M_{p,0}$ is the initial mass of the primary, and 
$\alpha_{\rm IMF}$ is typically set to 2.7 (Scalo 1986).
Rather than fix $f(1)$, we fix the supernova rate (\S \ref{sec:sn}). 
Since we are only interested in compact binaries, we require 
that the primaries have a mass of at least $10 M_\odot$.  
We allow massive stars up to 150 $M_\odot$ to be formed,
but our choice of the initial mass function severely 
restricts the role these massive stars play in our 
population synthesis. Repeating the calculation 
with $\alpha_{\rm IMF}=2.35$ yields similar (within 
a factor of 3) merger rates (Table 2).

Secondary masses are much more difficult to determine.  
The standard technique (see Hogeveen for a review 1990) prescribes 
a mass ratio ($q=\frac{M_{s}}{M_{p}}$) distribution 
$P(q)$ by 
\begin{equation}
P(q) \propto q^{-\alpha_{\rm MR}}.
\end{equation}
Because it is extremely difficult to detect low-mass companions of
massive stars, the value of $\alpha_{\rm MR}$ ranges from -1 or 0
(Garmany, Conti, \& Massey 1980) to 2.7 for $q>0.35$ and 0. for
$q<0.35$ (Hogeveen 1990).  The discrepancy in these mass ratios arises
not from different data, but a different interpretation of the data.
Of the 67 massive stars measured by Garmany, Conti, \& Massey (1980),
30\% contained close binaries with mass ratios exceeding 0.4, and, according
to their analysis, 40\% of massive stars were in binaries. Hogeveen
(1990), on the other hand, argues that Garmany, Conti, \& Massey
(1980) underestimated the number of binaries with low-mass companions
and instead estimates a much larger fraction of binary massive stars
($\sim $70\%) peaked toward extreme mass ratios ($\alpha_{\rm MR}=2.7$). For
the merger rates in Table 2, we always assume a binary fraction of
40\%.  However it is likely that the binary fraction, and hence the
merger rates, for the $\alpha_{\rm MR}=2.7$ calculations are a factor
of 2 higher. Even with this increase of 2 of the binary fraction,
these extreme mass ratios predict over an order of magnitude fewer DNS
binaries (which, dominated by Scenario II, require nearly equal mass
binaries) and nearly a factor of 3 more WD/BH binaries (Table 2).  In
this case, WD/BH mergers are more common than DNS mergers!

For initial orbital separations, we use the 
distribution introduced by Kraicheva et al. (1979):
\begin{equation}
P(A_0) \propto 1/A_0
\end{equation}
with constraints on the innermost initial separation
(equal to 3 times the sum of the stellar radii) and 
on the outermost initial separation (equal to $10^4 R_\odot$).
Increasing this inner separation by a factor of 2, or increasing 
the outer separation by 2 orders of magnitude, raises (or lowers 
for the outer limit) the merger rates by less than 20-30\%.  
Likewise, assuming the initial eccentricity is evenly distributed 
between 0 and 0.9 (versus assuming initially circular orbits) 
also has less than a 40\% effect (Table 2).

With the exception of the mass ratio distribution, uncertainties in 
the initial conditions have little effect on the merger rates 
of neutron star binaries. For our standard set of parameters, 
we use, as do most of the previous calculations of DNS binaries, 
a flat mass ratio distribution ($\alpha_{\rm MR}=0$) with 
a binary mass fraction of 40\%.  
However, the standard formation scenario for low-mass X-ray binaries 
requires that the mass ratio distribution is peaked 
toward extreme mass ratios (Kalogera \& Webbink 1998).   
Unless low-mass X-ray binaries are formed via some alternative 
mechanism, $\alpha_{\rm MR}$ must be higher ($\sim 2.7$), 
and we are overestimating the DNS merger rate by over an 
order of magnitude and underestimating the WD/BH merger 
rate by a factor of 3. For the rest of our standard 
set of parameters, we use the Scalo (1986) initial mass function, 
initially circular orbits, and the distribution of separations and 
inner and outer limits discussed in the previous paragraph.  

\subsection{Uncertainties in the Supernova Model}\label{sec:sn}

To form the binaries containing compact objects that might eventually
produce BHAD GRBs, we must not only follow the evolution of
binary stars, but also take into account the many uncertainties
associated with the supernova process.  One key quantity, to which all
our formation scenarios are proportional, is the supernova rate and
its history in the universe.  We assume a current Galactic supernova
rate of 0.02\,yr$^{-1}$, consistent with the results of Cappellaro et
al. (1997), but even that is uncertain by at least a factor of 2. As
one moves to the distant past, the rate is presumably larger for our
Galaxy, but then galaxies too were evolving in size and luminosity. We
discuss in \S 4 the scaling of our supernova rate to distant
redshifts.

Another source of great uncertainty in estimating the formation
rate of DNS and BH/NS systems is an inadequate understanding of the
kicks imparted to neutron stars (or black holes) at birth.  Growing
evidence suggests that neutron stars receive a momentum boost, or
``kick'', ($v_{\rm kick}^{\rm mean} \sim 300-500$\,km\,s$^{-1}$)
during their formation, probably due to some asymmetry in the
explosion itself.  Evidence includes the pulsar velocity distribution
(Dewey \& Cordes 1987; Lyne \& Lorimer 1994; Fryer, Burrows, Benz
1998; Cordes \& Chernoff 1998), associations of neutron stars with
supernova remnants (Caraveo 1993; Frail, Goss \& Whiteoak 1994
although see Gaensler \& Johnston 1995), observations of bow shocks
produced by neutron stars as they plow through the interstellar medium
(Cordes, Romani, \& Lundgren 1993), and the formation scenarios for
DNS systems and many other close binary systems(DNS systems: Flannery
\& van den Heuvel 1975; Burrows \& Woosley 1986; Yamaoka, Shigeyama,
\& Nomoto 1993; Fryer \& Kalogera 1997; spin/orbit angles in binaries:
Kaspi {\it et al.} 1996; Wasserman,Cordes, \& Chernoff 1996; eccentric
binaries: van den Heuvel \& Rappaport 1986).  Although small kicks
($\sim 200$\,km\,s$^{-1}$) are required to match the observed DNS
systems, larger kicks tend to disrupt the binaries. Recent analyses of
the pulsar velocity distribution suggests a bimodal kick distribution
with roughly half of the neutron stars receiving kicks smaller than
$200\,$km\,s$^{-1}$ and the other half receiving kicks greater than
$500-600$\,km\,s$^{-1}$ (Fryer, Burrows, \& Benz 1998). The binary
systems whose neutron stars receive very large kicks become
unbound. The remaining systems may form DNS and BH/NS binaries. The
choice of kick magnitude has a strong effect on the formation rate of
these two compact binary progenitors (Fig. 14).

During the supernova explosion, the binary only remains bound only if
$M_{\rm ejecta} < 0.5 M_{\rm sys}r/a_0$ where $M_{\rm sys}$, $r$, and
$a_0$ are the total mass, orbital separation, and semi-major axis of
the binary (Hills 1983). With the circular orbits important for
close-binary formation (common envelope evolution circularizes the
orbits), this equation becomes $M_{\rm ejecta} < 0.5 M_{\rm sys}$.
When a kick is imparted onto the neutron star during the supernova
explosion, the criterion for bound systems is:
\begin{equation}\label{eq:kick}
M_{\rm ejecta} < 0.5 M_{\rm sys} \left[ 1-2 (v_{\rm kick}/v_{\rm orb})
cos \theta cos \phi - (v_{\rm kick}/v_{\rm orb})^2 \right],
\end{equation}
where $v_{\rm kick}$ and $v_{\rm orb}$ are the magnitudes of the 
kick and orbital velocities respectively. 
$\theta$ and $\phi$ are the polar 
angle directions of the kick velocity with respect the the 
orbital velocity. Although for some systems, a kick might 
actually prevent the disruption of the binary (this is 
important for low-mass X-ray binary formation:  Kalogera \& 
Webbink 1998; Kalogera 1998, Fryer, Burrows, \& Benz 1998), 
in general, the kick tends to unbind binaries. For an isotropic kick 
distribution, the probability that a system remains 
bound after the supernova explosion is given by (Brandt \& 
Podsiadlowski 1995):
\begin{equation}
P(v_{\rm kick},v_{\rm orb},M_{\rm ejecta},M_{\rm sys})=
0.5+0.5\frac{v_{\rm orb}}{v_{\rm kick}}\left[ 
\frac{M_{\rm sys}-M_{\rm ejecta}}{M_{\rm sys}}-0.5 \left(
\frac{v_{\rm kick}}{v_{\rm orb}} \right)^2 -0.5 \right].
\end{equation}

We performed a series of Monte-Carlo simulations assuming delta
function kicks (Fig. 14). Note that the merger rates of DNS binaries
initially increase as the mean kick velocities increase from $0 \,
{\rm km \, s^{-1}}$. The actual binary {\it formation} rates are much
higher for lower kicks (Fig. 15), but higher kicks are required to
form the close binaries which will merge within a Hubble time. The
presupernova orbital separation must be sufficiently wide to avoid a
common envelope phase which would destroy the binary and a kick is
required to tighten the orbits and make close merging binaries. In DNS
binary formation, the primary's neutron star must avoid common
envelope evolution with its secondary's helium star (Fryer \& Kalogera
1997) or it will accrete too much material and become a BH/NS GRB.
Similarly, low-mass stars must have sufficiently wide orbital
separations so that they do not merge with their massive companions
before the massive star collapses to form a black hole.  Clearly, the
radius of the helium star is important for these calculations. We
will return to this effect in \S \ref{sec:sevo}.

The actual kick distribution is probably not a delta function, but
spans wide range of velocities. In Table 3, we give the merger rates
for both a Maxwellian velocity distribution (with a mean of 450$\,
{\rm km \, s^{-1}}$) and a double peaked distribution similar to that
described by Fryer, Burrows, \& Benz (1998) with two peaks near 100$\,
{\rm km \, s^{-1}}$ and 600$\, {\rm km \, s^{-1}}$ with $\sigma=50,150
{\rm km \, s^{-1}}$ respectively (``FBB kick distribution''). Note
that although the FBB and Maxwellian kick distributions have roughly
the same average magnitude (and both roughly explain the pulsar
velocity distribution), the FBB kick distribution produces many more
binary systems. Because the merger rates depend so sensitively upon
the kick distribution, we will employ several kick velocities in our
studies.

At the present time, no known mechanism provides the large kicks
implied by the pulsar observations. It is therefore doubly difficult
to predict the distribution for black holes since the velocities 
of single black holes have not been measured.  Kick mechanisms
driven by asymmetries in the supernova explosion seem likely to impart
smaller kick velocities to black holes, if only because they are more
massive. If anisotropic neutrino emission is involved, for example,
the black hole may also form before much neutrino momentum has been
emitted and therefore experience less recoil. As one might expect,
setting the black hole kick magnitudes to 10\% that of neutron stars
implies much larger black hole binary merger rates (Table 3). 

Another uncertainty in supernova explosions is the limiting main
sequence mass star that will form a black hole, either by fallback
($M_{\rm BH}$) or prompt collapse ($M_{\rm Coll}$). Current supernova
models (Fryer 1999) suggest that $M_{\rm BH}$ is not
much above $25 {\rm M_\odot}$ which agrees roughly with observations
(Maeder 1992; Kobulnicky \& Skillman 1997; Portegies Zwart,
Verbunt, \& Ergma 1992; Ergma \& van den Heuvel 1998).  In Fig. 16
(and Table 3), we show results from simulations with $M_{\rm BH}$
ranging from $25 M_\odot$ to $85 M_\odot$ and $M_{\rm Coll}=M_{\rm
BH}+15$M\sun\, using the FBB kick distribution.  A higher mass limit
leads to the formation of fewer black holes, and hence, fewer BH/WD
binaries and collapsars. The merger rate of BH/NS binaries does not
depend sensitively on the black hole mass limit.  This is because the
dominant formation scenario (Sc. V) requires the primary to collapse
into a neutron star which then accretes matter in a common envelope
phase and collapses to a black hole (see \S \ref{sec:bev}).  If
hyper-critical accretion does not occur, formation scenario V does not
work and, as shown in Fig. 16, the formation rate of BH/NS binaries
depends more sensitively upon the critical black hole mass limit.  For
our standard set of simulations, we set $M_{\rm BH}=25 $M\sun and
$M_{\rm Coll}=40 $M\sun.

In our simulations, we also assume that massive stars collapse to form
black holes with masses set to 1/3 the primary star's mass at the time
of collapse.  For stars with masses between $M_{\rm BH}$ and $M_{\rm
Coll}$, a supernova ejects some of the mass.  Even beyond $M_{\rm
Coll}$, not all of the star will collapse to form a black hole.  As
the gamma-ray burst forms, it also may drive off part of the helium
core.  The resultant mass distribution of black holes in binaries with
massive companions is double peaked (Fig. 17).  The large peak at
2.4\,M\sun ($> 50$\% of the black holes) is caused by those black
holes formed by hyper-critical accretion onto neutron stars in common
envelope evolution.  These low-mass black holes will produce the most
energetic BH/NS merger GRBs.  Their actual masses depend sensitively
upon common envelope evolution, but will probably not exceed 4\,M\sun.
The second, spread-out distribution, is even more difficult to
quantify, as it depends not only upon binary evolution, but also the
rates of stellar winds and the mass of the ejecta in supernova
explosions.  In fact, assuming that the mass of the black hole is only
1/3 that of the primary star's mass is probably a lower limit to
the black hole mass (Fryer 1999).  Setting the black hole mass equal
to the collapsing star's mass slightly increases the rates of black
hole binaries for low kicks. However, for high kicks, the high
momentum imparted to these more massive systems tends to unbind the
system.

\subsection{Stellar Evolution}\label{sec:sevo}

Uncertainties in stellar models are often neglected but, as we shall
show in this section, they can have a large effect on the rates for
the BHAD models. The most important uncertainty in the stellar models
is the stellar radius, especially as the star moves off the main
sequence. The radius of a star is important because it determines if,
and when, a binary system passes through a common envelope or mass
transfer phase. Accurate stellar radii are difficult to determine
observationally and current theoretical models concentrate on the
cores of massive stars. Many of the theoretically predicted radii
depend upon artificial outer boundary conditions. Indeed, especially
for massive giant stars and for Wolf Rayet stars, the radius is not a very
precisely defined quantity.  Observations tell us only about the 
radius of the photosphere.  This size can be much larger than the 
radius which is important for population synthesis calculations, 
that is, the radius where the stellar density is sufficient to drive 
common envelope evolution (see \S \ref{sec:bev}).

For both giant and helium stars, our standard model assumes that the
common envelope radius is 25\% the radius given in the stellar model
fits by Kalogera \& Webbink (1998). If we raise this to the values
given by Kalogera \& Webbink (1998), both the rates of DNS and BH/NS
binaries decrease over an order of magnitude in some cases (Table 4).
Increasing stellar radii causes more binaries to go into common
envelope phases, but, more importantly, increasing helium-star radii
cause more binaries to merge (producing fewer compact binaries and
more helium merger GRBs).

In addition, many of the formation scenarios (e.g. I, II, IV, V)
produce compact stars (neutron star, black hole) around helium-star
companions. As the helium-star radius increases, the orbital
separations of compact binaries become wider and consequently fewer
{\it merging} compact binaries are produced. Thus, for a given kick
distribution, increasing the helium-star radius causes both the number
of binaries and the fraction of those binaries which will merge in a
Hubble time to decrease (see Fryer \& Kalogera 1997).

Winds affect both helium star masses and radii. For instance, if we
use a mass loss rate like that of Woosley, Langer, \& Weaver
(1993,1995), helium masses are relatively small (e.g., a massive star
forming a $20 M_\odot$ helium core will end its life with a $3.5
M_\odot$ helium core), and their radii are greater. For larger helium
star radii, fewer close binaries form, and the merger rates of both
DNS systems and BH/NS systems decrease (Table 4). In addition, high
winds raise the critical mass for black hole formation (less massive
stars simply do not have massive enough cores at time of collapse to
prevent a supernova explosion) which reduces the rates of all black
hole systems (the collapsar rate decreases by an order of
magnitude). Note that this mass loss is an extreme case and some
helium stars must not have such drastic winds if the formation
scenarios for black hole X-ray binaries such as V404 Cyg, Cyg X-1, and
XN Mus 92, are correct (White \& Paradijs 1996). In our simulations,
we parameterize the mass loss rate as a fraction of the rate used by
Woosley, Langer, \& Weaver (1993,1995): $\alpha_{\rm Wind}=$(Mass
Loss)/(Mass Loss$_ {\rm WLW}$). Current mass loss estimates suggest
that the WLW models overestimate the maximum mass-loss by a factor of
two to three.  The effect of mass-loss is minimal unless $\alpha_{\rm
Wind}$ exceeds 0.6 (Fig. 18).  If the mass loss never exceeds this
value, only the collapsar model is affected at all by mass loss.  As
the mass loss rises from $\alpha_{\rm Wind}=0.$ to $\alpha_{\rm
Wind}=0.6$, the collapsar rate is cut by a factor of 2-3.

We have simplified the uncertainties by assuming the error in the red
supergiant wind is the same as that of Wolf-Rayet winds.  We decrease
the mass loss in the red supergiant wind by the same factor that we
decrease the mass loss for helium stars by lumping the mass loss
uncertainty into one parameter.  With this assumption, we are probably
underestimating the number of collapsars formed via winds (Sc. X),
perhaps grossly so.  Using just one parameter, those stars with winds
strong enough to uncover the helium core also lose too much of their
helium core to collapse directly into a black hole.  Observations
suggest that single stars above ($\sim 40$\,M\sun\,) must lose their
hydrogen envelope (Chiosi \& Maeder 1986), which would require a high
value for the wind parameter: $\alpha_{\rm Wind}>0.8$.  However, to
match the black hole binaries, the wind parameter for helium stars
must be low: $\alpha_{\rm Wind}<0.5$ (Kalogera 1999).  Assuming 
that black hole binaries are formed in systems which do not go 
into common envelope evolution until after central helium exhaustion, 
Wellstein \& Langer (1999) are able to explain the black hole binaries 
but maintain a high mass loss rate.  If their models are correct, 
very few massive stars, single or binary, would have enough mass 
prior to collapse to form black holes directly, and the formation 
rate of collapsars in the present-day universe would drop dramatically.
However, since mass loss is metal dependent, the collapsar rate might be 
larger at higher redshifts than it is today (MacFadyen \& Woosley 1999).  

\subsection{Binary Evolution}\label{sec:bev}

For binaries to merge and form GRBs, the progenitor systems must have
short periods. consequently, most will undergo some sort of active
binary evolution. We have assumed that if the mass of the expanding
star ($M_{\rm exp}$) is less than $\sim$2.5 times the mass of its
companion ($M_{\rm comp}$), the companion is unable to incorporate
more than a small fraction of the overflowing mass (Webbink 1979). A
fraction ($\beta_{\rm MT}$) of the expanding star's envelope is
accreted by the companion, and the remainder is lost from the system,
carrying away a specific angular momentum ($\alpha_{\rm MT}$). The
orbital separation of the system ($A$) in terms of its initial
separation ($A_0$) after mass transfer is given by (Podsiadlowski,
Joss, \& Hsu 1992):
\begin{equation}
\frac{A}{A^0}=\frac{M_{\rm exp}+M_{\rm comp}}{M_{\rm exp}^0+M_{\rm comp}^0}
\left ( \frac{M_{\rm exp}}{M_{\rm exp}^0} \right)^{C_1} 
\left ( \frac{M_{\rm comp}}{M_{\rm comp}^0} \right)^{C_2}
\end{equation}
where the values of the constants are:
\begin{equation}
C_1 \equiv 2 \alpha_{\rm MT} (1-\beta_{\rm MT}) - 2
\end{equation}
\begin{equation}
C_2 \equiv \frac{-2 \alpha_{\rm MT}}{\beta_{\rm MT}} (1-\beta_{\rm MT}) - 2
\end{equation}
and
\begin{equation}
M_{\rm comp}=\beta_{\rm MT} (M_{\rm exp}^0-M_{\rm exp}) + M_{\rm comp}^0,
\end{equation}
and where the superscript 0 denotes pre-mass transfer phase values.
Although the values of $\alpha_{\rm MT}$ and $\beta_{\rm MT}$ are
relatively uncertain, varying $\alpha_{\rm MT}$ from $0.5-2.0$ does
not change the GRB rates by more than 30\%(Table 5).  However,
changing the amount of material lost by the system during mass
transfer has a similar effect to that of changing the mass loss by
winds. Increasing $\beta_{\rm MT}$ from 0.5 to 1.0 decreases the BH/NS
merger, collapsar, and helium merger rates by up to an order of
magnitude (Table 5).

If $M_{\rm exp} \gtrsim 2.5 M_{\rm comp}$ (if the companion is a
compact object, this factor of 2.5 reduces to 1.0), the companion can
not adjust to the accreting mass and instead enters into a common
envelope phase. Once a common envelope is created, the orbital
separation decreases rapidly, probably in $1-1000$\,yr (see Sandquist
et al. 1998 and references therein). This inspiral continues until the
binary is able to shed the expanding star's hydrogen envelope leaving
a bare helium core. One can estimate the final orbital separation
($A_f$) in terms of the initial separation ($A_i$) of the binary after
common envelope evolution (Webbink 1984):
\begin{equation}
\frac{A_{f}}{A_{i}}=\frac{\alpha_{\rm CE}\, r_{L}\, q}{2} \left(\frac{M_{He}}
{M_{s}-M_{He} + \frac{1}{2}\alpha_{\rm CE} r_{L} M_{NS}} \right),
\end{equation} 
where $M_{s}$, $M_{He}$, and $M_{NS}$ are, respectively, the masses of
the secondary, the secondary's helium core, and of the neutron star
and $r_{L}=R_L/A_i$ is the dimensionless Roche lobe radius of the
secondary.  $\alpha_{\rm CE}$ is a parameter designed to represent how
efficiently the orbital energy ejects the hydrogen envelope and
includes assumptions about the binding energy of the expanding star
and the amount of orbital energy injected, per gram, into the hydrogen
envelope. The value of $\alpha_{\rm CE}$ is not well constrained, nor
is it constant.  It depends both upon the mass and evolutionary stage
of the expanding star when a common envelope phase occurs. Varying the
common envelope efficiency over a range (0.25-1.0) varies the GRB
rates by less than a factor of 3 (Table 6), and one might naively
assume that common envelope evolution does not affect the GRB rates.

However, common envelope evolution may be drastically different from
that given by the simple $\alpha_{\rm CE}$ prescription.  Nearly all
past population synthesis calculations ignore the possible merger of
an inspiralling compact object with the core of its companion in a
common envelope phase. In most cases, previous population synthesis
codes simply did not check to see when the merging compact object fell
within the helium core radius. However, such an assumption may not be
entirely incorrect. Terman, Taam, \& Hernquist (1995) found that, if
the star is sufficiently evolved to produce a steep density gradient
between the helium core and the hydrogen envelope, the inspiral of the
compact object might halt just above the helium core, preventing the
merger of the helium core and compact object. Although their findings
were greatly influenced by inadequate resolution ($\sim 3000$
particles) and the subsequent effects of the numerical artificial
viscosity inherent in their smooth particle hydrodynamical technique
(Rasio \& Livio 1996), nature may still produce similar results. At
this time, common envelope evolution is not well enough understood to
derive from it any reliable constraints.  If we ignore the non-zero
helium radius, the rate of DNS and BH/NS mergers increases nearly up
to $\sim 10^{-4} \, {\rm yr^{-1}}$(Table 7).  However, since this
particular orbital evolution is currently only justified by an
incorrect hydrodynamical simulation, this high merger rate should be
considered with suspicion.

Another critical aspect of common envelope evolution is the amount of
accretion experienced by a compact object spiraling into the hydrogen
envelope of a companion star.  In past calculations, it has been
assumed that in this state a neutron star will accrete at the
Eddington rate.  However, studies of the conditions arising near the
base of a neutron star suggest that neutron stars in common envelopes
will actually accrete at the Bondi-Hoyle rate since the accreting
material loses its energy via neutrino emission (Chevalier 1993, 1996;
Houck \& Chevalier 1992; Brown 1995; Fryer, Benz, \& Herant 1996).
Assuming all neutron stars inspiral to the surface of the expanding
star's helium core, Bethe \& Brown (1998) estimated that all neutron
stars in common envelope would accrete roughly $1 M_\odot$ of material
and collapse to form black holes.  This collapse can only be avoided
only if: angular momentum in the accretion hydrogen is sufficient to
lower the accretion rate (Chevalier 1996), neutrino driven explosions
eject the accreting hydrogen allowing only small spurts of accretion
(Fryer, Benz \& Herant 1996), or the star simply doesn't inspiral to
the surface of the helium core where the densities of the hydrogen
envelope lead to high temperatures at the surface of the neutron star.

Although angular momentum can halt accretion along the axis, 
it is unlikely to prevent the rapid accretion onto the neutron 
star simply because accretion is unimpeded along the poles.  
Especially at the high densities near the helium core surface 
(where Bethe \& Brown 1998 predict most of the accretion to occur), 
it is unlikely that angular momentum will limit the accretion 
rate sufficiently to prevent neutrino cooling (Fryer \& Kalogera 1998).
Similarly, although neutrino driven explosions may prevent 
accretion in the outer layers of a hydrogen giant, near the 
helium core surface, this explosion mechanism is ineffective 
(Fryer, Benz, Herant 1996). However, if the neutron star does 
not inspiral all the way down to the helium core surface, it 
may not accrete enough to become a black hole and will remain 
a neutron star. Note that by ignoring hypercritical accretion and 
assuming that compact objects will not spiral into the helium core 
produces a DNS merger rate comparable to the rates of past work, 
most of which made these two, physically unlikely assumptions(Table 7).  
In addition, these binaries have much shorter orbits on average, 
and hence shorter merger times.  In our standard model, we assumed that 
all the common envelope neutron stars form black holes and, hence, 
scenario I (see \S \ref{sec:dns}) is excluded from these simulations.  

\subsection{Summary and Constraints}

Table 8 summarizes the range of results from all the population
synthesis calculations. The basic uncertainties in the rates can be
grouped into three categories: kicks, progenitor evolution, and
orbital separations. Except for the collapsar model, all the BHAD GRB
models depend sensitively (up to three orders of magnitude in the
rate) upon the magnitude and distribution of the kick velocities.
Uncertainty here dominates the errors in these BHAD GRB rates. On the
observational front, additional data on pulsar velocities would be
useful. From the theorist we still seek an understanding of why the
kicks are so large.

Uncertainties in the progenitor evolution include the mass ratio
distribution in binaries, which determines the mass of the secondary
and the compact object into which the secondary will evolve, and the
mass limits for black hole and collapsar formation.  The mass ratio
distribution is difficult to determine from first principles, but it
better observations could certainly constrain this uncertainty. The
mass limit for collapsars on the other hand is something theorists can
work on. Together, these uncertainties affect all the GRB rates, and
can vary the WD/BH merger and collapsar rates by $\sim 2$ orders of
magnitude.

Finally, we have lumped all the uncertainties of binary evolution 
into the seemingly innocuous category: orbital separation distribution.  
This category is dominated by uncertainties in the stellar radii and 
the common envelope evolution. Most of the high DNS rates listed in 
Lipunov, Postnov, \& Prokhorov (1987) are derived by assuming that 
neutron stars do not accrete hypercritically and do not merge with 
their companion's helium core (most previous work used population 
synthesis codes which do not even determine whether the neutron 
star spirals into its companion's helium core during a common 
envelope phase). If we take these assumptions as 
incorrect and instead rely upon the $\alpha_{\rm CE}$ prescription 
for common envelope evolution, the uncertainties in this category 
result in roughly an order of magnitude decrease in the DNS GRB rates.  

Comparison of population synthesis models with observations 
of compact binary systems can place some constraints upon the 
allowed parameter space. For instance, in Fig. 19 we plot 
the simulated distribution of eccentricities and separations 
of double neutron star systems superimposed on the 4 observed 
galactic double neutron star systems\footnote{van Kerkwijk \& 
Kulkarni 1999 have discovered an observational candidate of PSR 2303+46 
($A=30$R\sun\,,$e=0.66$), which would mean that this system consists 
of a neutron star and a white dwarf and is not a DNS system at all.}.  
Although with so few systems and no knowledge of the observational biases, 
we do not dare do any statistics on this sample, we can probably
rule out very large stellar radii and low kick velocities 
($v_{\rm kick}=50$) because such a set of parameters would have
a very small probability of producing the observed systems (see 
Fryer \& Kalogera 1998 for more 
details). The remaining three simulations all 
reproduce the observed systems, but predict a range of 
merger rates: $10^{-7}$\,yr$^{-1}<\,{\rm Merger Rate}_{\rm DNS}\,
<5 \times 10^{-5}$\,yr$^{-1}$. Similarly, the existence of short-period 
black hole binaries requires that not all helium stars have 
the strong winds proposed by Woosley, Langer \& Weaver (1995).  

Additional constraints exist and may well provide more stringent 
limits on the binary formation rate. Until these categories 
improve and better limits are constructed, we must be satisfied 
with the rates in Table 8. However, the rates in table 8 provide 
useful constraints nonetheless.  To get the extremely high 
merger rates of DNS and BH/NS binaries, one must assume that 
compact objects do not merge with their companion's helium 
core during a common envelope phase.  Otherwise, the rate of 
mergers is at least an order of magnitude lower which, although 
this decrease may prove troublesome for LIGO and VIRGO, it is certainly 
sufficient to explain GRBs (see \S 4).  Collapsars and He-mergers also 
have rates large enough to explain the GRB statistics. It is unlikely 
that BH/WD mergers explain a sizable fraction of GRBs, but they may be 
required to explain some peculiar GRBs.

\section{Implications For Observations}  

Table 9 summarizes the event rate, distance traveled before producing
a burst, and mean redshift for the models we have studied. In addition
we give estimates of the burst energy and duration (from other works,
e.g., Popham et al. 1999; Meszaros \& Rees 1997a).  There is
effectively no delay for collapsars and He-mergers which produce
bursts occurring, respectively, within $\sim$10\,Myr and $\sim$
50\,Myr of their formation.  However most DNS and BH/NS systems
experience a considerable delay between their formation and merger
and, during that tie, may travel a considerable distance. How far
depends upon the mass of the host galaxy. Galaxies similar to the
Milky Way have escape velocities of order 400-700 km/s, depending on
the location inside the galaxy.  The angular separation between GRBs
and the centers of their host galaxies is thus an important diagnostic
(e.g., Portegies-Zwart \& Yungelson 1998; Bloom, Sigurdsson, \& Pols
1999b; Bulik, Belczy\'nski, \& Zbijewski 1999).  There is of course
the possibility of a strong selection effect. It might be that GRBs
which occur far from their host galaxy are less likely to produce
observable afterglows because of the low density of the ISM in
galactic halos (Meszaros \& Rees 1997b). This would then mean that we
have to multiply the event rates by a corresponding factor much less
than one.

In this section, we restrict our analysis to 5 simulations that we
feel represent the range of most-likely population synthesis
parameters.  These include the standard set of parameters, a
simulation with winds set to 50\% that of 
Woosley, Langer \& Weaver (1995), and a
simulation with the stellar radii raised to those of Kalogera \&
Webbink (1998) all using the FBB kick distribution.  Two simulations
were run using a Maxwellian kick distribution, one with the standard
set of parameters, and one with the stellar radii equal to Kalogera \&
Webbink (1998).  However the latter of these produced so few binaries
that we do not include it in the figures (although the results are in
the tables).

\subsection{The angular separation between OTs and host candidates}

There now exists a small, but growing sample of optical GRB afterglows
(OTs) associated with host galaxies. In almost every case the angular
offset of the OT from the photometric center of the host is of
order arcseconds. A linear distance of 10 kpc corresponds to an
angular separation of 0.7 h$_{100}$ F(z) arcsec, where F $\sim 3$ for
redshifts in the range 1-3 (e.g., Peebles 1993) and h$_{100}$ is the
Hubble constant in units of 100 km\,s$^{-1}$\,Mpc$^{-1}$.  Host
redshifts are known for GRB970508 (z = 0.838; Metzger {\it et al.}
1997; Bloom {\it et al.} 1998), GRB971214 (z = 3.42; Kulkarni {\it et
al.} 1998), GRB980703 (z = 0.966; Djorgovski {\it et al.} 1998), and
GRB980613 (z = 1.096; Djorgovski {\it et al.} 1999). These
observations clearly argue against any model that invokes massive
black holes located at the centers of active or normal galaxies.
They also have important implications for models discussed in this paper.

We have modeled the spatial distribution of DNS and BH/NS GRBs by
calculating orbits in three representative galactic mass models of
Miyamoto-Nagai type (see Hartmann, Epstein, \& Woosley 1990 for
details of the dynamic model): a massive galaxy resembling the Milky
Way (with V$_{\rm rot}$(R$_\odot$) = 220 km/s), a system with a mass
four times smaller, and a system with 1/100th the mass of the Milky
Way (essentially a negligible mass). The average initial positions of
the GRB progenitors in each class were taken to be an axisymmetric
ring with galactocentric radius R = 5 kpc with a Gaussian spread of 1
kpc around that radius. The orbits were following the formation of 
the primary's compact object.  For the DNS and BH/NS binaries, the 
binaries, this process required two steps, the evolution before 
and after the collapse of the secondary. Systemic velocities
were added as isotropic vectors, while the initial velocity was
assumed to be that for circular motion with V$_{\rm rot}$. The orbits
were followed for a period given from the Monte Carlo population
synthesis model. Several thousand orbits were used for each BHAD
model. At the moment of final merger the distance to the galactic
center was recorded.  Fig. 20 shows the resulting distance
distribution for 4 different sets of population synthesis parameters.
Integrating these curves, it is clear that the models predict a
substantial fraction of DNS and BH/NS GRBs at large distances from
their hosts (see Fig. 21, Table 10).

In Fig. 21, we plot the distribution of DNS and BH/NS binaries
separately, but clearly the two GRB progenitors are very similar
reflecting their similar formation history.  In addition, because many
of the binaries are composed of low-mass black holes (Fig. 17), the
velocity boost received to the system as the secondary forms a neutron
star is very similar to that received in the case of DNS systems.  
Our results agree well with those of Bulik, Belczy\'nski, \& 
Zbijewski (1999) which have a wider spread than that of Bloom et al. 
(1999b).  The population synthesis studies of Bloom et al. (1999b) 
probably did not include a non-zero helium star radius, allowing them to 
form DNS binaries with shorter periods, and hence shorter merger 
times (see \S 3.4).  

Fig. 21 also shows the distribution of separations compared to the the
observations of OTs corresponding to long hard bursts.  It appears
that DNS and BH/NS mergers are ruled out as the likely explanation for
most of these events. The observations instead support the idea that
long duration bursts are either collapsars, helium-mergers, or WD/BH
mergers, all of which tend to occur well within the galaxy (see
Fig. 21).  Indeed, collapsars should all occur at their star formation
site. The spread in Fig. 21 reflects only the assumption regarding
galactocentric radius for star formation (and is thus also the birth
line for the progenitors of DNS and BH/NS systems). Ideally this
radius should be adjusted to the size of the observed GRB host
galaxy, so the agreement with the ``promptly dying'' models is really
better than the figure indicates.

\subsection{GRBs and the Star Formation History of the Universe}

Since only very massive stars are believed to produce black holes, all
of our models are in one way or another dependent upon the star
formation history of the universe. Helium-mergers and collapsars trace
it most directly since the delay time between star birth and the GRB
is, by cosmic standards, instantaneous. The relatively infrequent
WD/BH mergers may be delayed by a longer time ($\sim 100$\,Myr) but
still should trace star formation directly. DNS and BH/NS systems, on
the other hand have a delay and though they reflect the star formation
rate, they do so with a spread of delays, depending especially on the
gravitational radiation time scale.

Fig. 22 shows the distribution of DNS+BH/NS merger times for our four
different sets of population synthesis parameters. Note that all four
have some short-period mergers, but the typical case gives a GRB that
lags behind the star formation rate. Fig. 23 shows the GRB redshift
history for an Einstein-deSitter universe using two different star
formation histories under the functional form given by Watanabe et
al. (1998): flattened - $A=3., \, z_p=1.5, \, B=0.$ and peaked -
$A=1., \, z_p=1., \, B=0.5$.  The flattened and peaked formation
histories represent extremes in the star formation.  Recent studies of
the cosmic star formation rate indicates that the SFR(z) density
function does not turn over past z $\sim$ 1.5 (e.g., Pascarelle,
Lanzetta, $\&$ Fernandez-Soto 1998), but instead remains roughly
constant to redshifts of $\sim$ 5.  Beyond a redshift of 5, very
little is known about the star formation history.  At some redshift,
the star formation rate must drop.  Differences in the population
synthesis models cause considerable scatter in the median redshift of
GRBs (Table 11).  For a flattened star formation rate, the median
redshift varies from 0.85-1.2.  Note that the peaked star formation
rate leads to much lower median burst redshifts (0.54-0.81).
Collapsars and helium-mergers will produce bursts which trace the star
formation more directly and hence their median redshifts are higher:
1.8, 0.96 respectively for the flattened, peaked star formation
histories (Table 10).  The median redshift of collapsar GRBs is likely
to be even higher, as its rate increases with decreasing mass loss.
The low metallicity stars born at high redshift are likely to have
weaker winds and, hence, produce more collapsars.
  
The current redshift estimates of z=1-3 from the observations (e.g.,
Hogg and Fruchter 1999) lie within our two star formation rates for
all of the GRB models.  Unfortunately, the uncertainty in the star
formation rate beyond a redshift of 1.5 makes it impossible to rule
out any of the models.  However, even with the uncertainties in the
dominant BHAD model and the population synthesis models, the two star
formation rates can be easily distinguished.  If an increased sample
of GRB redshifts pushes toward a median GRB redshift of 2, then no
BHAD GRB model can explain the burst distribution with the peaked star
formation history.  In this way, GRBs can be used to place constraints
on the star formation history.

Another constraint on the star formation history is the fact that no
lensed GRBs have been observed (Holz, Miller, \& Quashnock 1999).
Using their analysis of constant redshift GRB distributions, we can
constrain the GRB histories calculated here. For a flat SFR history
combined with our standard model and the FBB kick distribution, only
30\% of the DNS+BH/NS GRBs have redshifts greater than 4, and less
than 10\% have redshift greater than 8.  The flat SFR history again
must be restricted if the he-mergers and collapsars are the dominant
long-duration bursts.  The peaked SFR history predicts few lensed
GRBs, and hence a lensed event would argue against this star formation
history.

Table 11 gives the total number of bursts per day for the three star
formation rates.  To then determine how many bursts we will observe
requires a convolution with the detection efficiencies and the beaming
factor, which lower this rate by a factor of 30-1000.  Within the
uncertainties in the conversion from actual bursts to observed bursts,
the star formation history, and the rates, all of the BHAD GRB
progenitors can explain the observed bursts.

\section{BHAD GRB Models}

The merger rates, durations, energies, median redshifts and locations 
of the BHAD GRB models are summarized in Table 9. Our estimated rates, 
median redshifts and locations, coupled with the previously calculated 
energies and durations (Popham, Woosley, \& Fryer 1998; Ruffert \& Janka 
(1998); Janka, Ruffert, \& Eberl 1998) compare well with the current 
observational limits.  From the data in Table 9, we can 
construct a general picture of GRB progenitors in which 
short-duration bursts are dominated by DNS and BH/NS mergers, 
and long-duration bursts are dominated by collapsars and helium mergers.  
If this picture is correct, we can make several observational predictions 
which will be tested in the next 10 years.  We predict that all 
long-duration bursts should reside in their host galaxy, whereas 
some short-duration bursts should occur beyond their host galaxy.
In addition, the median redshift of short-duration bursts should 
be less than that of their long-duration counterparts.

\subsection{Short-Duration Bursts}

DNS and BH/NS binary mergers are capable of producing short duration
bursts.  Their total rate very likely lies in the range of
0.1-30\,Myr$^{-1}$ in the Milky way, which corresponds to an overall 
rate in the universe of 1-1500\,day$^{-1}$.  To determine an observed 
rate, one must then include the beaming factor and detector efficiencies.  
Note that this rate is much less than has been predicted in previous 
calculations (Lipunov, Postnov, \& Prokhorov 1987).  Although this lower 
rate does not prevent DNS and BH/NS binary mergers from playing a major 
role in explaining GRBs, it has discouraging implications for the predicted 
rate of gravity wave events detectable with LIGO and VIRGO (for a review, 
see Finn 1999).

DNS and BH/NS binaries are the only BHAD GRBs which can occur 
outside of their host galaxy and the only models capable of giving
bursts much shorter than one second. Roughly 3-8\% of these bursts 
occur more than a few arc minutes from a host galaxy of Milky Way mass.
This fraction can increase beyond 20\% for lower massed host galaxies.
In addition, the median redshift of DNS and BH/NS merger GRBs 
is less, in some cases significantly less, than that of the 
median star formation redshift.  The current bursts whose locations 
have been ``pinpointed'' by BeppoSAX are all long-duration bursts and 
hence, no constraints can be placed on the locations and redshift 
of short-duration bursts.  However, since none of the observed 
long-duration bursts have been found outside of their host galaxy, 
the evidence is beginning to suggest that DNS and BH/NS binaries 
are limited to short-duration bursts.  But much more data is 
required to draw any firm conclusions from the observations.

DNS and BH/NS binaries also experience a significant delay (up to a 
Hubble time) between their formation and their merger to form a 
GRB.  This delay causes their median redshift to be lower by 20-50\% than 
that of stars, and hence, the long-duration gamma-ray bursts.  

The major uncertainty in these calculations comes from uncertainties 
in common envelope evolution:  accretion during common envelope 
evolution, the amount of inspiral that occurs and even stellar radii, 
which determine if a binary goes into a common envelope.  
Whereas hypercritical accretion requires a new mechanism for DNS formation 
(Brown 1995), it creates a new formation mechanism for BH/NS systems 
(Bethe \& Brown 1998).  The inspiral in common envelope helps to 
produce short-period binaries, but if the inspiraling compact object merges 
with its companion, it forms a helium-merger GRB, not a DNS or BH/NS 
binary.

\subsection{Long-Duration Bursts}

Collapsars, helium-mergers, and WD/BH binaries all definitely make
long-duration bursts. Because we do not know how many stars will have
the necessary angular momentum and mass to produce collapsars, the
collapsar rate is particularly uncertain. A relatively conservative
estimate comes from employing only Scenario XII (Fig. 13) - the late
time merger of two helium stars. Using standard parameters (\S 3)
this mechanism alone gives a daily rate of 400\,day$^{-1}$ 
(Table 3). A very liberal estimate, valid if magnetic fields do
not slow the rotation of the helium core,  additionally includes
Scenario XI and a comparable number of single star collapsars
(Scenario X with low helium core mass loss rates and high red giant
mass loss rates). This gives a value about two orders of magnitude larger.
 
Likewise, since the merger of black holes with helium stars has not
been studied in detail, its rate is uncertain (10-4000\,day$^{-1}$). 
The WD/BH binary rate is significantly smaller (0.1-50\,day$^{-1}$). 

Collapsars and helium-mergers occur in their host galaxy and trace the
star formation rate.  Hence, any long duration burst occurring outside
of its host galaxy would present a problem for the simple picture
assuming DNS and BH/NS mergers produce only short-duration bursts.
Collapsars and He-mergers also occur nearly simultaneously with the
star formation rate and will have a higher median redshift than
short-duration bursts (Table 9).  These bursts may also provide ideal
diagnostics of the high redshift star formation history.  A much larger
sample of well localized GRBs is essential to test these concepts, and
to lend support or to disprove the BHAD scenario for cosmological
GRBs.

Just like DNS and BH/NS systems, calculations of helium mergers and 
WD/BH binaries depend most sensitively upon the common envelope 
evolution.  The uncerainties in collapsars are due primarily to 
the mass loss and spin rate of their progenitors.  The mass loss 
depends upon stellar winds and the spin rate depends upon the coupling of 
the core to the stellar envelope.

\acknowledgements This research has been supported by NASA (NAG5-2843,
MIT SC A292701, and NAG5-8128), the NSF (AST-97-31569), and the US DOE
ASCI Program (W-7405-ENG-48).  It is a pleasure to thank Alex Heger
for his invaluable advice.  We are grateful for conversations with
Alex Heger, Andrew MacFadyen, Norbert Langer, Bob Popham, and Thomas Janka 
that helped to elucidate some of the uncertain aspects of gamma-ray bursts, 
massive stellar evolution and hyperaccreting black holes.

\begin{deluxetable}{rclll}
\tablewidth{42pc}
\tablecaption{GRB Formation Scenarios}
\tablehead{ \colhead{Scenario} 
& \colhead{Common Envelope\tablenotemark{a}} 
& \colhead{Primary Mass\tablenotemark{b}} 
& \colhead{Secondary Mass} 
& \colhead{Comments\tablenotemark{c}}}

\startdata
DNS I & NS in Sec. & $ M_{\rm SN} < M^0_{\rm p} < M_{\rm BH}$ & 
$ M_{\rm SN} < M^0_{\rm s} < M_{\rm BH}$ & 
Hyp. Acc. $\rightarrow$ V \nl
 & & & & Low $\alpha_{\rm CE} \rightarrow$ XIII \nl
 II & He Cores & $M_{\rm SN}< M^0_{\rm p} < M_{\rm BH}$ & 
$M^0_{\rm s}=M^0_{\rm p}\pm $5\% &
Low $\alpha_{\rm CE} \rightarrow$ Single \nl
 III & none & $ M_{\rm SN} < M^0_{\rm p} < M_{\rm BH}$ & 
$ M_{\rm SN} < M^0_{\rm s} < M_{\rm BH}$ & 
Kick \nl
 & & & & \nl
BH/NS IV & BH in Sec. & $ M^0_{\rm p} > M_{\rm BH}$ & 
$ M_{\rm SN} < M^0_{\rm s} < M_{\rm BH}$ & 
Low $\alpha_{\rm CE} \rightarrow$ XIII \nl
V & NS in Sec. & $ M_{\rm SN} < M^0_{\rm p} < M_{\rm BH}$ & 
$ M_{\rm SN} < M^0_{\rm s} < M_{\rm BH}$ & 
no Hyp. Acc. $\rightarrow$ I \nl
 & & & & Low $\alpha_{\rm CE} \rightarrow$ XIII \nl
 VI & none & $ M^0_{\rm p} > M_{\rm BH}$ & 
$ M_{\rm SN} < M^0_{\rm s} < M_{\rm BH}$ & 
Kick \nl
 & & & & \nl
WD/BH VII & Sec. in Pri. & $  M^0_{\rm p} > M_{\rm BH}$ & 
$ M^0_{\rm s} < M_{\rm SN} $ & 
Low $\alpha_{\rm CE} \rightarrow$ Single \nl
  VIII & NS in Sec. & $ M_{\rm SN} < M^0_{\rm p} < M_{\rm BH}$ & 
$ M^0_{\rm s} < M_{\rm SN} $ & 
Low $\alpha_{\rm CE} \rightarrow$ XIII \nl
 IX & none & $  M^0_{\rm p} > M_{\rm BH}$ & 
$ M^0_{\rm s} < M_{\rm SN} $ & 
Kick \nl
 & & & &  \nl
Collapsar X & Single Star & $M^0 > M_{\rm Coll}$ & N/A & Winds \nl
XI & Sec. in Pri. & $M^0 > M_{\rm Coll}$ & none & Hyd. ejected \nl
XII & Double He Cores & $M^0 \gtrsim M_{\rm Coll}$ & none & Merger \nl
 & & & & \nl
He-Merger XIII & NS,BH in Sec. & $M^0_{\rm p} > M_{\rm SN}$ & none & 
High $\alpha_{\rm CE} \rightarrow$ \nl
 & & & & I,IV,V,VII \nl

\tablenotetext{a}{Many scenarios require a common envelope phase, 
one of the most poorly understood aspects of the population 
synthesis of binary systems.  If a common envelope phase exists,
hypercritical accretion, the common envelope efficiency 
($\alpha_{\rm CE}$), and the radius of the helium star all 
become important.  Sec.$\equiv$Secondary Star, 
Pri.$\equiv$Primary Star.}
\tablenotetext{b}{$M_{\rm SN}\equiv$ critical mass above which the core 
of the massive star will collapse to form a neutron star or black hole
($M_{\rm SN}\approx8-12 M_\odot$).  $M_{\rm BH}\equiv$ critical mass
above which the collapse forms a black hole 
($M_{\rm BH}\approx25-35 M_\odot$).}
\tablenotetext{c}{Sensitivity of the scenario on the population
synthesis parameters and the inter-relation of the scenarios.  
Hyp. Acc.$\equiv$Hypercritical Accretion, $\alpha_{\rm CE} \equiv$
common envelope efficiency, Single$\equiv$Single Star, 
Kick $\equiv$ result extremely sensitive to neutron star kick.  A 
high helium radius has the same effect as a low common envelope 
efficiency.}

\enddata
\end{deluxetable}

\begin{deluxetable}{rlllll}
\tablewidth{36pc}
\tablecaption{\bf Population Synthesis: Initial Conditions}
\tablehead{  
& \multicolumn{5}{c}{\bf Formation Rate (Myr${^{-1}}$ per galaxy)} \\
\colhead{Scenario} & \colhead{Stand.} & \colhead{$\alpha_{\rm IMF}=2.35$} 
& \colhead{$\alpha_{\rm MR}=-1.0$} & \colhead{$\alpha_{\rm MR}=2.7$} 
& \colhead{$P(e_0) \propto 1$}}
\startdata
\multicolumn{6}{c}{${\bf V_{\rm kick}=100 \, {\rm km \, s^{-1}}, 
\sigma=50 {\rm \, km \, s^{-1}}}$} \nl
& & & & & \nl
{\bf DNS} II\tablenotemark{a} & 3.2 & 4.2 & 8.9 & 0.14 & 3.4 \nl
III & 0.0012 & 0.00069 & 0.0019 & 0.00026 & 0.0 \nl
{\bf Total} & {\bf 3.2} & {\bf 4.2} & {\bf 8.9} & {\bf 0.14} & {\bf 3.4} \nl
{\bf BH/NS} IV & 1.7 & 2.6 & 1.6 & 0.59 & 0.93 \nl
V & 10. & 14. & 13. & 3.2 & 7.1 \nl
VI & 0.00029 & 0.00034 & 0. & 0.00026 & 0. \nl
{\bf Total} & {\bf 12.} & {\bf 16.} & {\bf 14.} & {\bf 3.8} & {\bf 8.1} \nl
{\bf WD/BH} VII & 0.18 & 0.22 & 0.10 & 0.24 & 0.22 \nl
VIII & 0.0032 & 0.0021 & 0.00063 & 0.0065 & 0.0023 \nl 
IX & 0.18 & 0.27 & 0.085 & 0.41 & 0.44 \nl
{\bf Total} & {\bf 0.36} & {\bf 0.49} & {\bf 0.19} & {\bf 0.66} & {\bf 0.44} \nl
{\bf Collapsar} X & 5.0 & 7.5 & 17 & 0.18 & 5.6 \nl
XI & 620. & 1100. & 880. & 130. & 590. \nl
XII & 0. & 0. & 5.1 & 0. & 0. \nl
{\bf Total} & {\bf 630.} & {\bf 1100.} & {\bf 900.} & {\bf 130.} 
& {\bf 600.} \nl  
{\bf He-Merg.} XIII & {\bf 39.} & {\bf 44.} & {\bf 34.} & {\bf 29.} & {\bf 34.} \nl
\hline
\multicolumn{6}{c}{${\bf V_{\rm kick}=500 \, {\rm km \, s^{-1}}, 
\sigma=50 {\rm \, km \, s^{-1}}}$} \nl
& & & & & \nl
{\bf DNS} II\tablenotemark{a} & 0.31 & 0.30 & 0.86 & 0.010 & 0.32 \nl
III & 0. & 0. & 0. & 0. & 0. \nl
{\bf Total} & {\bf 0.31} & {\bf 0.30} & {\bf 0.86} & {\bf 0.010} & {\bf 0.32} \nl
{\bf BH/NS} IV & 0. & 0. & 0. & 0. & 0. \nl
V & 0.11 & 0.18 & 0.12 & 0.11 & 0.096 \nl
VI & 0.0020 & 0.0016 & 0.0032 & 0. & 0.0010 \nl
{\bf Total} & {\bf 0.11} & {\bf 0.18} & {\bf 0.12} & {\bf 0.11} & {\bf 0.097} \nl
{\bf WD/BH} VII & 0.0014 & 0.0020 & 0.00040 & 0.0068 & 0.00040 \nl 
VIII & 0.00040 & 0.00040 & 0.00040 & 0.0024 & 0.00060 \nl
IX & 0.0058 & 0.0070 & 0.00020 & 0.019 & 0.0050 \nl
{\bf Total} & {\bf 0.0076} & {\bf 0.0094} & {\bf 0.0010} & {\bf 0.028} & {\bf 0.0060} \nl
{\bf Collapsar} X & 4.7 & 7.3 & 16 & 0.18 & 5.3 \nl
XI & 600. & 1100. & 840. & 130. & 570. \nl
XII & 0. & 0. & 5.4 & 0.12 & 0. \nl
{\bf Total} & {\bf 600.} & {\bf 1100.} & {\bf 860.} & {\bf 130.} 
& {\bf 580.} \nl  
{\bf He-Merg.} XIII & {\bf 2.4} & {\bf 2.8} & {\bf 3.3} & {\bf 1.1} & {\bf 2.8} \nl

\tablenotetext{a}{In these models, hypercritical accretion drives all neutron 
stars which pass through a common envelope phase into black holes.  Thus, 
no DNS binaries are formed via Scenario I.}
\enddata
\end{deluxetable}

\begin{deluxetable}{rllllllll}
\tablewidth{40pc}
\tablecaption{\bf Population Synthesis: Supernovae}
\tablehead{  
& \multicolumn{8}{c}{\bf Formation Rate (Myr${^{-1}}$ per galaxy)} \\
\colhead{Scenario} & \multicolumn{2}{c}{Standard} & \multicolumn{2}{c}
{$V^{\rm BH}_{\rm Kick}=0.1V^{\rm NS}_{\rm Kick}$} & 
\multicolumn{2}{c}{$M_{\rm BH,Coll}^{\rm crit}=
75,90 M_{\odot}$\tablenotemark{a}} &
\multicolumn{2}{c}{$M_{\rm BH}=M_{\rm primary}$\tablenotemark{b}} \\
\colhead{$V_{\rm kick}$\tablenotemark{c}} 
& \colhead{FBB} & \colhead{Maxw.} & \colhead{100} & \colhead{500}   
 & \colhead{100} & \colhead{500} 
 & \colhead{100} & \colhead{500}}
\startdata
{\bf DNS} II & 1.2 & 0.97 & 3.6 & 0.59 & 2.8 & 0.14 & 3.1 & 0.61 \nl
III & 0. & 0. & 0.0015 & 0. & 0.00083 & 0. & 0.0015 & 0. \nl
{\bf Total} & {\bf 1.2} & {\bf 0.7} & {\bf 3.6} & {\bf 0.59} & {\bf 2.8} & {\bf 0.14} & {\bf 3.2} & {\bf 0.61} \nl
{\bf BH/NS} IV & 0.55 & 0.057 & 1.0 & 3.6 & 0.068 & 0. & 4.9 & 0.0075 \nl
V & 2.6 & 0.20 & 6.6 & 1.9 & 8.0 & 0.001 & 14. & 0.091 \nl
VI & 0.00029 & 0. & 0. & 0. & 0. & 0.  & 0.0058 & 0. \nl
{\bf Total} & {\bf 3.1} & {\bf 0.26} & {\bf 7.6} & {\bf 5.5} & {\bf 8.1} & {\bf 0.001} & {\bf 19.} & {\bf 0.98} \nl
{\bf WD/BH} VII & 0.071 & 0.017 & 0.18 & 0.25  & 0. & 0. & 0.18 & 0. \nl
VIII & 0.00087 & 0.0012 & 0.0035 & 0. & 0. & 0. & 0. & 0. \nl
IX & 0.081 & 0.016 & 0.18 & 0.25 & 0. & 0. & 0.0099 & 0.0017 \nl
{\bf Total} & {\bf 0.15} & {\bf 0.034} & {\bf 0.36} & {\bf 0.5} & {\bf 0.} & {\bf 0.} & {\bf 0.19} & {\bf 0.0017} \nl
{\bf Collapsar} X & 4.5 & 4.6 & 4.5 & 4.8 & 0.44 & 0.012 & 4.6 & 5.0 \nl
XI & 620. & 610. & 600. & 590. & 140. & 30. & 600. & 620. \nl
XII & 5.8 & 0.9 & 1.5 & 0. & 0.46 & 0.06 & 0. & 0. \nl
{\bf Total} & {\bf 630.} & {\bf 620.} & {\bf 610.} & {\bf 590.} & {\bf 140.} 
& {\bf 30.} & {\bf 600.} & {\bf 630.} \nl
{\bf He-Merg.} XIII & {\bf 14.} & {\bf 3.0} & {\bf 35.} & {\bf 15.}& {\bf 23.} & {\bf 0.} & {\bf 46.} & {\bf 0.35} \nl

\tablenotetext{a}{This is the critical initial stellar mass above
which massive stars form black holes.  The higher number corresponds 
to the stellar mass required to collapse to a black hole without the 
launch of a supernova shock.  This higher number is required to 
form collapsars.  In our standard models, we assume that stars 
above $25 M_\odot$ collapse to form black holes and stars 
with mass above  $40 M_\odot$ do not launch supernova shocks.}
\tablenotetext{b}{A star may explode as a supernova but have 
sufficient fallback to drive its later collapse into a black hole.  
Thus, the resultant mass may not equal the star's mass just 
prior to its collapse.  Our standard model assumes that the 
star is 1/3 the collapsing star's mass.}
\tablenotetext{c}{We give several kick distributions.  
FBB stands for the Fryer, Burrows, Benz (1998) kick 
distribution: 40\% with $V_{\rm mean}=100 {\rm km \, s^{-1}}$ 
and 60\% with $V_{\rm mean}=600 {\rm km \, s^{-1}}$.  
Maxw. stands 1for a Maxwellian distribution with 
a mean of $450{\rm km \, s^{-1}}$  The other velocities 
refer to the mean velocity using our smoothed 
($\sigma=50 {\rm km \, s^{-1}}$) delta-function kick 
distributions.}

\enddata
\end{deluxetable}

\begin{deluxetable}{rlllllll}
\tablewidth{40pc}
\tablecaption{\bf Population Synthesis: Stellar Evolution}
\tablehead{  
& \multicolumn{7}{c}{\bf Formation Rate (Myr${^{-1}}$ per galaxy)} \\
\colhead{Scenario} & \multicolumn{2}{c}{Standard} & 
\multicolumn{2}{c}{$R_{\rm star} = 4 \times R_{\rm star}^{\rm standard}$} &
\colhead{$\alpha_{\rm wind}=0.$} & 
\colhead{$\alpha_{\rm wind}=0.5$} & 
\colhead{$\alpha_{\rm wind}=1.$} \\
\colhead{$V_{\rm kick}$} & \colhead{100}  
& \colhead{500}  & \colhead{100} & \colhead{500} 
 & \colhead{FBB} & \colhead{FBB} 
 & \colhead{FBB}}

\startdata

{\bf DNS} II& 3.2 & 0.61 & 0.60 & 0.013 & 1.2 & 0.91 & 2.2 \nl
III & 0.0012 & 0. &  0.00029 &  0. & 0. & 0. & 0. \nl
{\bf Total} & {\bf 3.2} & {\bf 0.61} & {\bf 0.60} & {\bf 0.013} & {\bf 1.2} & {\bf 0.91} & {\bf 0.22} \nl
{\bf BH/NS} IV & 1.7 & 0.00029 & 0.90 & 0. & 0.59 & 0.072 & 0.019 \nl
V & 10. & 0.009 & 3.4 & 0.0014 & 2.6 & 1.1 & 0.090 \nl
VI & 0.00029 & 0. & 0. & 0. & 0. & 0. & 0. \nl
{\bf Total} & {\bf 12.} & {\bf 0.0093} & {\bf 4.3} & {\bf 0.0014} & {\bf 3.2} & {\bf 1.2}& {\bf 0.11}  \nl
{\bf WD/BH} VII & 0.18& 0. & 0.85 & 0.00029 & 0.026 & 0.28 & 0.00020 \nl
VIII & 0.0032 & 0. & 0.0069 & 0. & 0.0035 & 0.00028 & 0. \nl
IX & 0.18 & 0.00058 & 0.36 & 0. & 0.15 & 0.066 & 0. \nl
{\bf Total} & {\bf 0.36} & {\bf 0.00058} & {\bf 1.2} & {\bf 0.00029} & {\bf 0.18} & {\bf 0.35} & {\bf 0.00020} \nl
{\bf Collapsar} X & 4.7 & 5.0 & 0.91 & 0.93 & 6.8 & 0.19 & 0.38 \nl
XI & 600. & 620. & 600. & 610. & 670. & 450. & 0. \nl
XII & 0. & 0. & 4.0 & 3.4 & 5.8 & 6.0 & 5.0 \nl
{\bf Total} & {\bf 600.} & {\bf 630.} & {\bf 600.} & {\bf 610.} & {\bf 680.} 
& {\bf 460.} & {\bf 5.4} \nl
{\bf He-Merg.} XIII & {\bf 39.} & {\bf 0.12} & {\bf 140.} & {\bf 0.046} & {\bf 14.} & {\bf 16.} 
& {\bf 6.2} \nl
\enddata
\end{deluxetable}

\begin{deluxetable}{rlllll}
\tablewidth{40pc}
\tablecaption{\bf Population Synthesis: Mass Transfer}
\tablehead{  
\multicolumn{6}{c}{\bf Formation Rate (Myr${^{-1}}$ per galaxy), $V_{\rm kick}=$FBB} \\
\colhead{Scenario} & 
\colhead{$\beta_{\rm MT}=1.0$}  &
\colhead{$\alpha_{\rm MT}=0.5$} &
\colhead{$\alpha_{\rm MT}=2.0$} &
\colhead{$\alpha_{\rm MT}=0.5$} &
\colhead{$\alpha_{\rm MT}=2.0$} \\
\colhead{} & \colhead{} & \colhead{$\beta_{\rm MT}=0.8$} &
\colhead{$\beta_{\rm MT}=0.8$} &
\colhead{$\beta_{\rm MT}=0.5$} &
\colhead{$\beta_{\rm MT}=0.5$}}

\startdata
{\bf DNS} II & 1.2 & 1.2 & 1.2 & 1.2 & 1.2 \nl
III & 0. & 0.00029 & 0.00058 & 0.00029 & 0. \nl
{\bf Total} & {\bf 1.2} & {\bf 1.2} & {\bf 1.2} & {\bf 1.2} & {\bf 1.2} \nl  
{\bf BH/NS} IV & 1.8 & 0.95 & 0.16 & 0.094 & 0.054 \nl
V & 8.1 & 3.9 & 1.2 & 0.76 & 0.62 \nl
VI & 0. & 0. & 0. & 0. & 0. \nl
{\bf Total} & {\bf 9.9} & {\bf 4.9} & {\bf 1.4} & {\bf 0.85} & {\bf 0.67} \nl 
{\bf WD/BH} VII & 0.069 & 0.071 & 0.073 & 0.073 & 0.074 \nl 
VIII & 0.0012 & 0.00087 & 0.0012 & 0.00088 & 0.0012 \nl
IX & 0.085 & 0.079 & 0.079 & 0.081 & 0.083 \nl
{\bf Total} & {\bf 0.16} & {\bf 0.15} & {\bf 0.15} & 
{\bf 0.16} & {\bf 0.16} \nl
{\bf Collapsar} X & 5.1 & 4.5 & 4.4 & 3.5 & 3.0 \nl
XI & 710. & 650. & 540. & 500. & 170. \nl
XII & 6.0 & 5.9 & 5.5 & 5.5 & 5.2 \nl
{\bf Total} & {\bf 720.} & {\bf 660.} & {\bf 550.} & {\bf 510.} 
& {\bf 180.} \nl  
{\bf He-Merg.} XIII & {\bf 36.} & {\bf 20.} & {\bf 9.5} & {\bf 8.0} & {\bf 7.2} \nl
\enddata
\end{deluxetable}

\begin{deluxetable}{rllllll}
\tablewidth{38pc}
\tablecaption{\bf Population Synthesis: Common Envelope Efficiency}
\tablehead{  
& \multicolumn{6}{c}{\bf Formation Rate (Myr${^{-1}}$ per galaxy)} \\
\colhead{Scenario} & \multicolumn{2}{c}{Standard:  $\alpha_{\rm CE}=0.5$} & 
\multicolumn{2}{c}{$\alpha_{\rm CE}=0.25$} &
\multicolumn{2}{c}{$\alpha_{\rm CE}=1.0$} \\
\colhead{$V_{\rm kick}$} & \colhead{100}  
& \colhead{500}  & \colhead{100} & \colhead{500} 
 & \colhead{100} & \colhead{500}}

\startdata

{\bf DNS} II& 3.2 & 0.61 & 3.3 & 0.78 & 3.8 & 0.51 \nl
III & 0.0012 & 0. &  0.0012 &  0. & 0.00058 & 0. \nl
{\bf Total} & {\bf 3.2} & {\bf 0.61} & {\bf 3.3} & {\bf 0.78} & {\bf 3.8} & {\bf 0.51} \nl
{\bf BH/NS} IV & 1.7 & 0.00029 & 2.0 & 0.00058 & 1.4 & 0.00058 \nl
V & 10. & 0.0090 & 4.9 & 0.0058 & 10.0 & 0.007 \nl
VI & 0.00029 & 0. & 0. & 0. & 0.00058 & 0. \nl 
{\bf Total} & {\bf 12.} & {\bf 0.0093} & {\bf 6.9} & {\bf 0.0064} & {\bf 11.} & {\bf 0.0076} \nl 
{\bf WD/BH} VII & 0.18 & 0. & 0.022 & 0.00029 & 0.54 & 0.00029 \nl  
VIII & 0.0032 & 0. & 0.0029 & 0. & 0.018 & 0. \nl
IX & 0.18 & 0.00058 & 0.078 & 0.0017 & 0.29 & 0. \nl  
{\bf Total} & {\bf 0.36} & {\bf 0.00058} & {\bf 0.10} & {\bf 0.0020} 
& {\bf 0.86} & {\bf 0.00029} \nl 
{\bf Collapsar} X & 4.7 & 5.0 & 4.5 & 4.6 & 4.5 & 4.7 \nl
XI & 600. & 620. & 580. & 600. & 630. & 650. \nl
XII & 0. & 0. & 1.3 & 1.2 & 1.6 & 1.4 \nl
{\bf Total} & {\bf 600.} & {\bf 630.} & {\bf 590.} & {\bf 610.} & {\bf 640.} 
& {\bf 660.} \nl  
{\bf He-Merg.} XIII & {\bf 39.} & {\bf 0.12} & {\bf 33.} & {\bf 0.091} & 
{\bf 49.} & {\bf 0.038} \nl 
\enddata
\end{deluxetable}

\begin{deluxetable}{rllllllll}
\tablewidth{42pc}
\tablecaption{\bf Population Synthesis: Hypercritical Accretion}
\tablehead{  
& \multicolumn{8}{c}{\bf Formation Rate (Myr${^{-1}}$ per galaxy)} \\
\colhead{Scenario} & \multicolumn{2}{c}{Standard} & 
\multicolumn{2}{c}{No HA} &
\multicolumn{2}{c}{No He-Mrg.} &
\multicolumn{2}{c}{No He-Mrg./HA } \\
\colhead{$V_{\rm kick}$} & \colhead{100}  
& \colhead{500}  & \colhead{100} & \colhead{500}  & \colhead{100}
& \colhead{500}  & \colhead{100} & \colhead{500}}

\startdata

{\bf DNS} I & 0. & 0. & 6.6 & 0.0081 & 0. & 0. & 20. & 0.018. \nl
II & 3.2 & 0.61 & 3.2 & 0.61 & 63. & 15. & 63. & 15. \nl
III & 0.0012 & 0. & 0.0015 &  0. & 0.00058 & 0. & 0.0015 & 0. \nl
{\bf Total} & {\bf 3.2} & {\bf 0.61} & {\bf 9.8} & {\bf 0.62} & {\bf 63.} & {\bf 15.} & {\bf 83.} & {\bf 15.} \nl
{\bf BH/NS} IV & 1.7 & 0.00029 & 1.0 & 0.00029 & 11. & 0.22 & 9.9 & 0.021 \nl
V & 10. & 0.0090 & 0. & 0. & 26. & 0.019 & 0. & 0. \nl
VI & 0.00029 & 0. & 0. & 0. & 0.00029 & 0. & 0.00029 & 0. \nl
{\bf Total} & {\bf 12.} & {\bf 0.0093} & {\bf 1.0} & {\bf 0.00029} & {\bf 37.} & {\bf 0.040} & {\bf 10.} & {\bf 0.021} \nl 
{\bf WD/BH} VII & 0.18 & 0. & 0.18 & 0. & 0.19 & 0.0012 & 0.19 & 0.0012 \nl
VIII & 0.0032 & 0. & 0.0032 & 0. & 0.0035 & 0. & 0.0035 & 0. \nl
IX & 0.18 & 0.00058 & 0.18 & 0.00058 & 0.18 & 0.00087 & 0.18 & 0.00087 \nl
{\bf Total} & {\bf 0.36} & {\bf 0.00058} & {\bf 0.36} & {\bf 0.00058} & {\bf 0.37} & {\bf 0.0020} & {\bf 0.37} & {\bf 0.0020}  \nl
{\bf Collapsar} X & 4.7 & 5.0 & 4.7 & 5.0 & 4.4 & 4.6 & 4.4 & 4.6 \nl
XI & 600. & 620. & 600. & 620. & 610. & 630. & 610. & 630. \nl
XII & 0. & 0. & 0. & 0. & 0. & 0. & 0. & 0. \nl
{\bf Total} & {\bf 600.} & {\bf 630.} & {\bf 600.} & {\bf 630.} & {\bf 610.} & {\bf 630.} & {\bf 610.} & {\bf 630. } \nl 
{\bf He-Merg.} XIII & {\bf 39.} & {\bf 0.12} & {\bf 39.} & {\bf 0.12} & {\bf 0.} & {\bf 0.} & {\bf 0.} & {\bf 0.} \nl

\enddata
\end{deluxetable}

\begin{deluxetable}{rcc}
\tablewidth{30pc}
\tablecaption{\bf Formation Rate Summary}
\tablehead{
\colhead{model} & \colhead{\bf Range} & 
\colhead{Standard Parameters} \\
\colhead{} & \colhead{(Myr$^{-1}$ per gal.)\tablenotemark{a}} 
& \colhead{(Myr$^{-1}$ per gal.)\tablenotemark{a}}}

\startdata

{\bf DNS} & 0.01-80 & 1.2 \nl
{\bf BH/NS} & 0.001-50 & 3.1 \nl
{\bf WD/BH} & 0.0001-1 & 0.15 \nl
{\bf Collapsar} & $\sim$10-1000 & 680 \nl
{\bf He-Merg.} & $\sim$0.1-50 & 14 \nl

\tablenotetext{a}{The Formation Rate is for a galactic supernova rate 
of 0.02\,yr$^{-1}$.}

\enddata
\end{deluxetable}

\begin{deluxetable}{rllllll}
\tablewidth{40pc}
\tablecaption{\bf BHAD GRB Models}
\tablehead{  
\colhead{model} & \colhead{\bf Form. Rate} & 
\colhead{Log E$_{\nu \bar{\nu}}$\tablenotemark{a}} & 
\colhead{Log E$_{\rm Mag.}$} & \colhead{Duration}\tablenotemark{a} 
& \colhead{Fraction} 
& \colhead{z$^{\rm median}_{GRB}$} \\
\colhead{} & \colhead{(GRB/day)\tablenotemark{b}} 
& \colhead{(Erg)} &\colhead{(Erg)} & \colhead{(s)} & 
\colhead{$>$\,1\,Mpc\tablenotemark{c}} & \colhead{(z$^{\rm median}_{SN}$)}}

\startdata

{\bf DNS} & 1-300 & 49.7-50.5 & $\lesssim 53$ & $\sim 0.1$s 
& 3-8,17-39\% & 0.5-0.8 \nl 
{\bf BH/NS} & 1-1200 & 50-51 & $\lesssim 53$ & $\sim 0.1$s 
& 3-8,17-39\% & 0.5-0.8 \nl 
{\bf WD/BH} & 0.1-50 & $\lesssim 50$ & $\lesssim 53$ & 15-150s 
& 0,0\% & $\lesssim 1$ \nl 
{\bf Collapsar} & $100-5.\times10^4$ & $\lesssim 52$ 
& $\lesssim 53$ & 10s & 0,0\% & $\gtrsim 1$ \nl 
{\bf He-Merg.} & $10-4\times10^3$ & $\lesssim 50$ 
& $\lesssim 53$ & 15-500s & 0,0\% & 1 \nl 

\tablenotetext{a}{The energies and durations are taken from Popham, Woosley, 
\& Fryer(1998), Ruffert \& Janka (1998), or Janka, Ruffert, \& Eberl (1998).
These models assumed high viscous forces.  The durations of the DNS 
and BH/NS GRBs could be much higher if the viscosity is lower 
(Meszaros \& Rees 1997a).}
\tablenotetext{b}{The formation rates are based on a flat star 
formation history with a z=5 cut-off and a limited set of parameter 
values to produce the ``most-likely'' range.}
\tablenotetext{c}{We list values for galaxies with respective masses 
equal to and 1/4th that of the Milky Way.}

\enddata
\end{deluxetable}

\begin{deluxetable}{rcccccc}
\tablewidth{42pc}
\tablecaption{\bf DNS+BH/NS Relation to Host Galaxy\tablenotemark{a}}
\tablehead{  
\colhead{Model} & \multicolumn{2}{c}{$M_{\rm host}=M_{\rm MW}$} & 
\multicolumn{2}{c}{$M_{\rm host}=0.25M_{\rm MW}$} & 
\multicolumn{2}{c}{$M_{\rm host}=0.01M_{\rm MW}$} \\
\colhead{} & \colhead{$>$100 kpc\tablenotemark{b}} & \colhead{$>$1 Mpc} 
& \colhead{$>$100 kpc} & \colhead{$>$1 Mpc}
& \colhead{$>$100 kpc} & \colhead{$>$1 Mpc}} 

\startdata

{\bf Stan., FBB Kick} & 13\% & 3.6\% & 30\% & 8.8\% & 56\% & 17\% \nl
{\bf High Wind, FBB Kick} & 9.7\% & 3.4\% & 24\% & 8.6\% & 71\% & 21\% \nl
{\bf Large R, FBB Kick} & 11\% & 4.1\% & 37\% & 13\% & 77\% & 39\% \nl
{\bf Stan., Maxw. Kick} & 25\% & 7.2\% & 39\% & 14\% & 44\% & 17\% \nl
{\bf Large R, Maxw. Kick} & 21\% & 8.2\% & 52\% & 23\% & 62\% & 30\% \nl

\tablenotetext{a}{WD/BH, helium-merger, and collapsar 
GRBs all occur within their host galaxy.}
\tablenotetext{b}{The columns list the percentage of 
DNS and BH/NS binaries which merge beyond 100,1000 kpc 
respectively.}

\enddata
\end{deluxetable}

\begin{deluxetable}{rcccc}
\tablewidth{40pc}
\tablecaption{\bf DNS+BH/NS Redshift Distribution and Rates\tablenotemark{a}}
\tablehead{  
\colhead{Model} & \multicolumn{2}{c}{Flat\tablenotemark{b}} & 
\multicolumn{2}{c}{Peaked\tablenotemark{c}} \\
\colhead{} & \colhead{median z} & \colhead{Rate} 
& \colhead{median z} & \colhead{Rate} \\
\colhead{} &  \colhead{} & \colhead{day$^{-1}$} &  \colhead{} &  
\colhead{day$^{-1}$}}

\startdata

{\bf Stan., FBB Kick} & 1.2 & 540 & 0.71 & 53 \\
{\bf High Wind, FBB Kick} & 0.97 & 380 & 0.57 & 36 \\
{\bf Large R, FBB Kick} & 0.85 & 128 & 0.54 & 12 \\
{\bf Stan., Maxw. Kick} & 1.5 & 156 & 0.81 & 16 \\
{\bf Large R, Maxw. Kick} & 1.2 & 24 & 0.71 & 2.3 \\

\tablenotetext{a}{The median redshift of helium-merger 
GRBs will trace the mean supernova redshift: 1.8, 0.96
respectively, for the flattened and peaked supernova histories.  
The median redshift of collapsar GRBs will be even larger due to the 
weaker winds at high redshifts (low metallicities).  
For the flat star formation history,
the supernova, he-merger, collapsar rates are, respectively, 
$1.3\times 10^6, 900, \sim 4\times10^4$\,per day.
The corresponding rates for a peaked star formation history
are $1.3\times10^5, \sim 90, \sim 4000$\,per day.}
\tablenotetext{b}{Flat:  A=3.0, B=0.0, z$_{\rm p}$=1.5}
\tablenotetext{c}{Peaked: A=1.0, B=0.5, z$_{\rm p}$=1.0}

\enddata
\end{deluxetable}

\begin{figure}
\plotfiddle{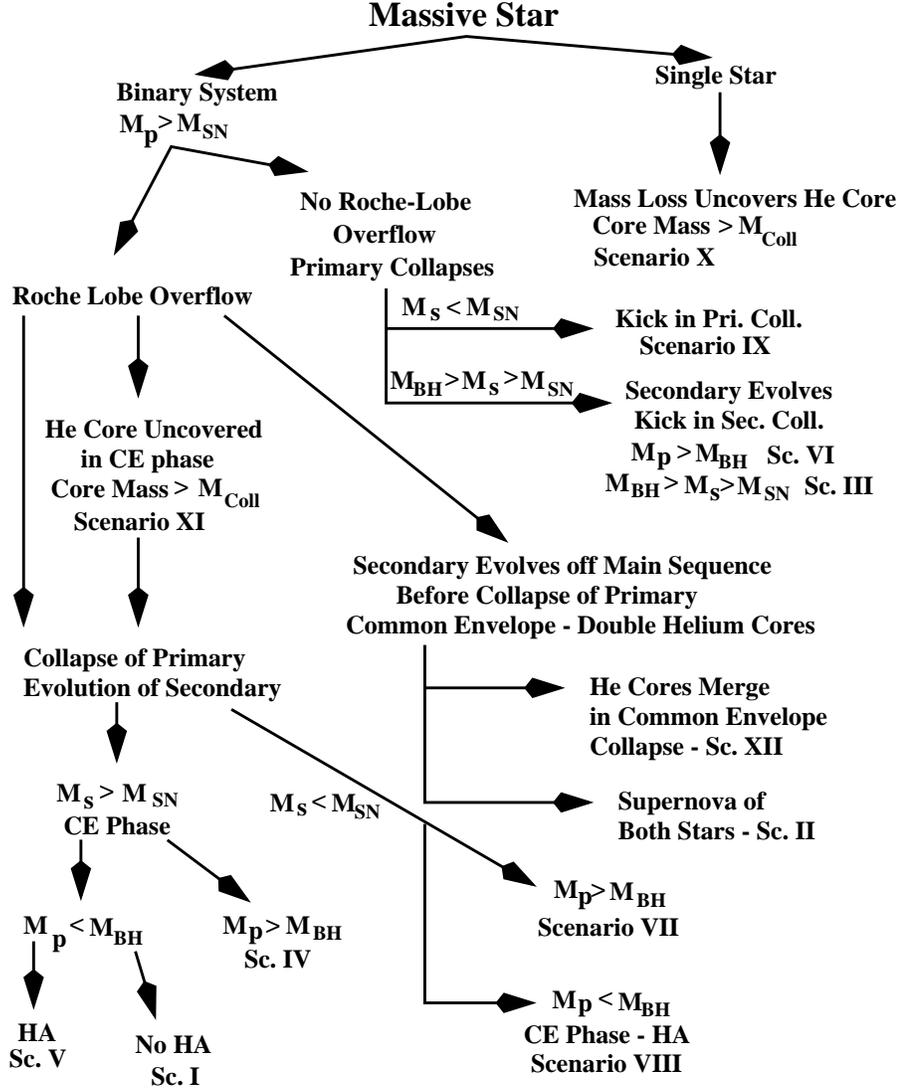}{6in}{0}{60}{60}{-200}{0}
\caption{Summary of the BHAD GRB formation scenarios studied 
in this paper.  Many of the scenarios are quite similar.  
For instance, scenarios III, VI, and IX are differentiated 
simply by the initial mass of the companion stars.  Scenarios 
I and IV differ only by the mass of their primary star.  
Scenarios I and V are identical except that, for the case of 
scenario I, we assume that photons limit the accretion onto the 
neutron star and, for scenario V, hypercritical accretion occurs.  
The helium merger models are not listed here but occur when a 
compact object (either black hole or neutron star) merges with 
its helium companion during a common envelope phase.  We use 
the following abbreviations:  $M_{\rm p}$, $M_{\rm s}$ are the 
primary and secondary masses respectively, $M_{\rm SN} \approx 
10 M_\odot$ are the critical masses above which a massive stars
collapse to form supernovae, $M_{\rm BH} \approx 25 M_\odot$ 
is the critical mass where massive stars form black holes, 
$M_{\rm Coll} \approx 40 M_\odot$ is the mass of helium 
cores above which no explosion occurs and the entire star 
collapses to form a black hole, CE indicates a common 
envelope phase, and HA stands for hypercritical accretion.}
\end{figure}
\clearpage

\begin{figure}
\plotfiddle{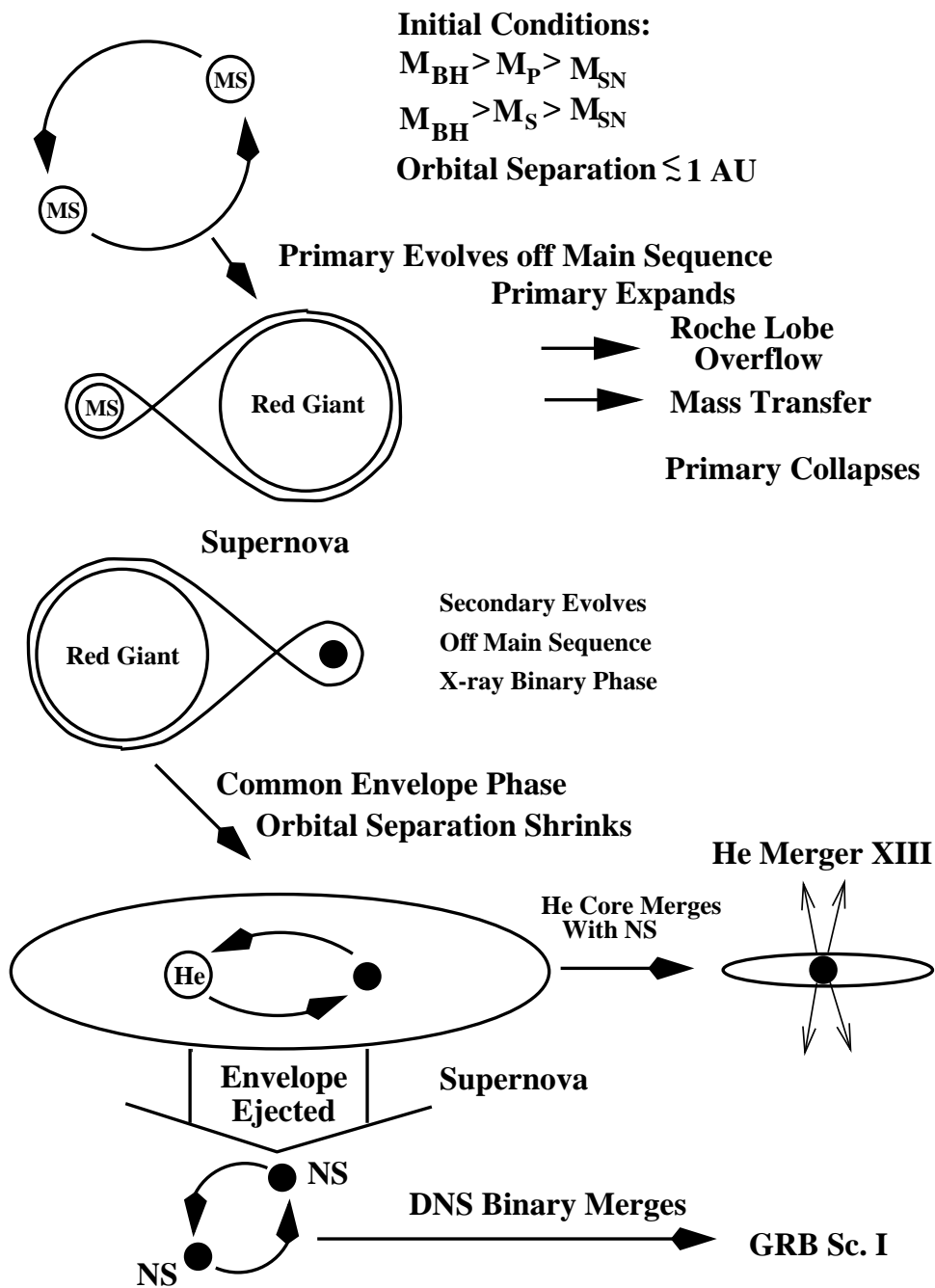}{7in}{0}{70}{70}{-200}{0}
\caption{Scenario I:  the ``standard'' double neutron 
star formation scenario.  All symbols are as described in 
Fig. 1, MS denotes a main-sequence star, NS and BH are 
neutron stars and black holes respectively.  Note that if 
the neutron star merges with its 
helium companion in the common envelope phase, a He-merger 
GRB is produced.  This scenario assumes the accretion onto 
the neutron star during this phase is limited to the 
photon Eddington rate.}
\end{figure}
\clearpage

\begin{figure}
\plotfiddle{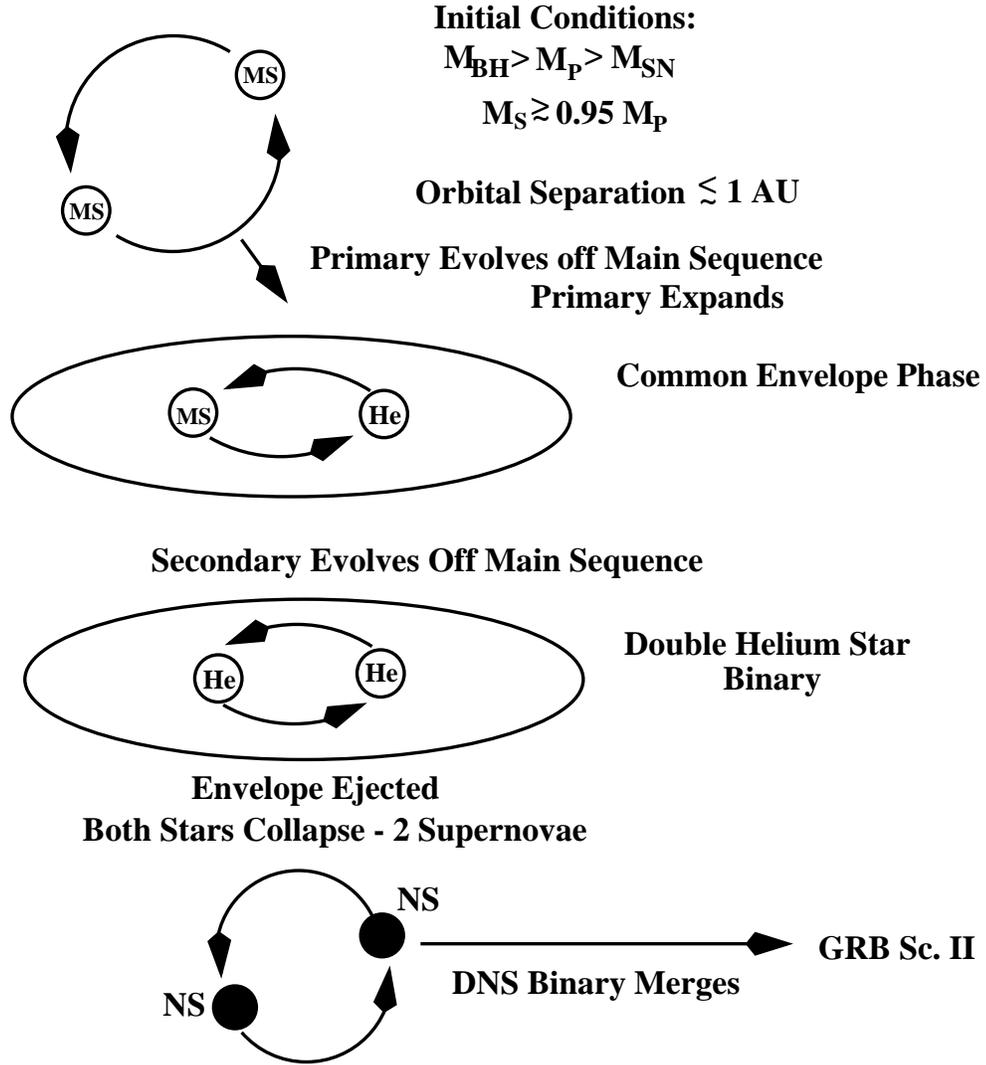}{7in}{0}{70}{70}{-200}{0}
\caption{Scenario II:  The ``Brown'' mechanism for 
forming DNS binaries (Brown 1995).  All symbols are as described in 
Figs. 1 and 2.  In this phase, the two stars have nearly the same 
mass and the secondary evolves off the main sequence before 
the primary collapses, forming a double helium star binary.  
The collapse of the helium cores form a DNS binary.}
\end{figure}
\clearpage

\begin{figure}
\plotfiddle{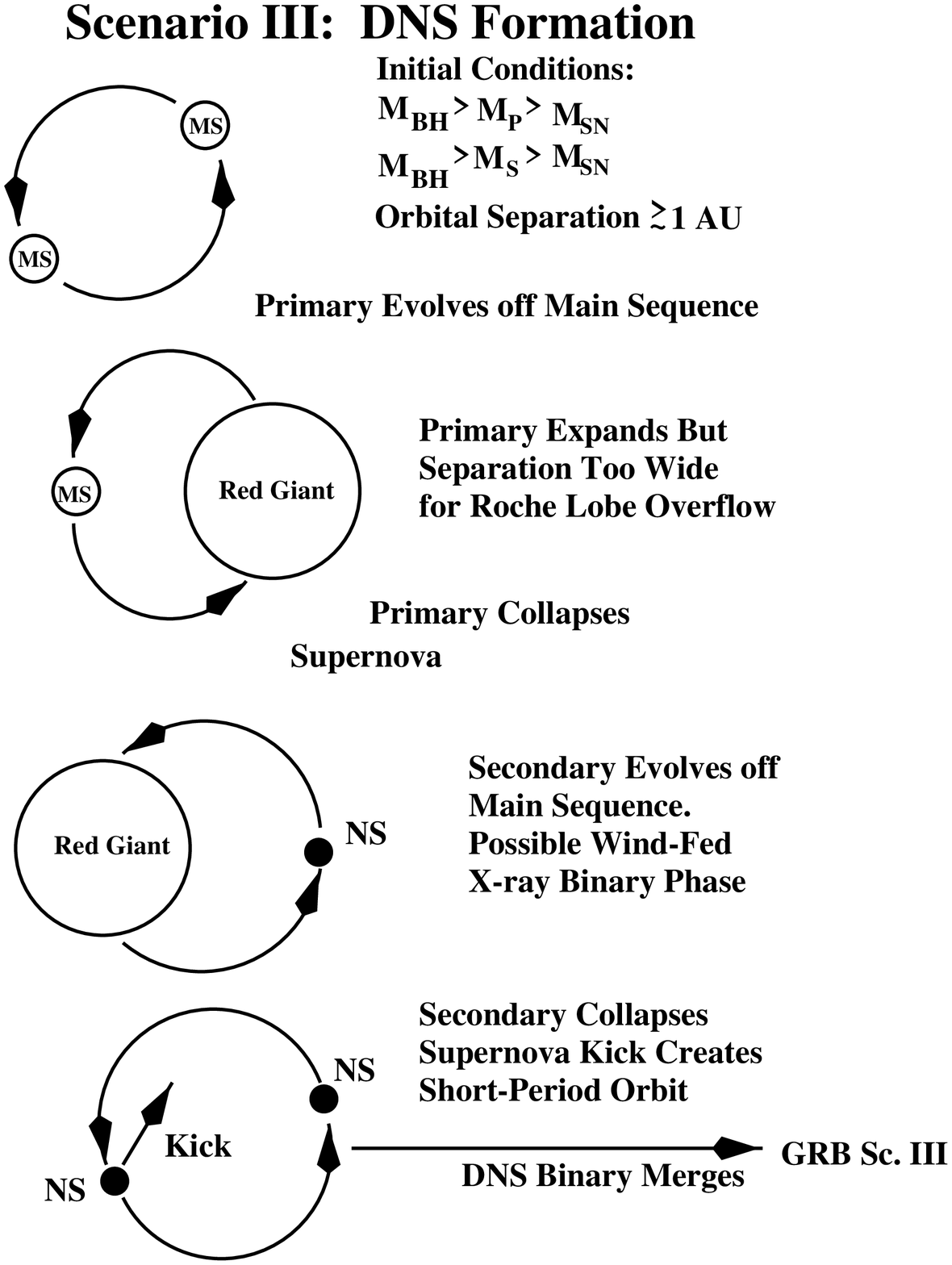}{7in}{0}{70}{70}{-200}{0}
\caption{Scenario III: The kick scenario for DNS binaries.  
In this scenario, the binary stars avoid any 
common envelope phase.  All symbols are as 
described in Figs. 1 and 2.  Only a small range 
of kicks can tighten the orbit and allow the 
DNS binary to merge within a Hubble time.}
\end{figure}
\clearpage

\begin{figure}
\plotfiddle{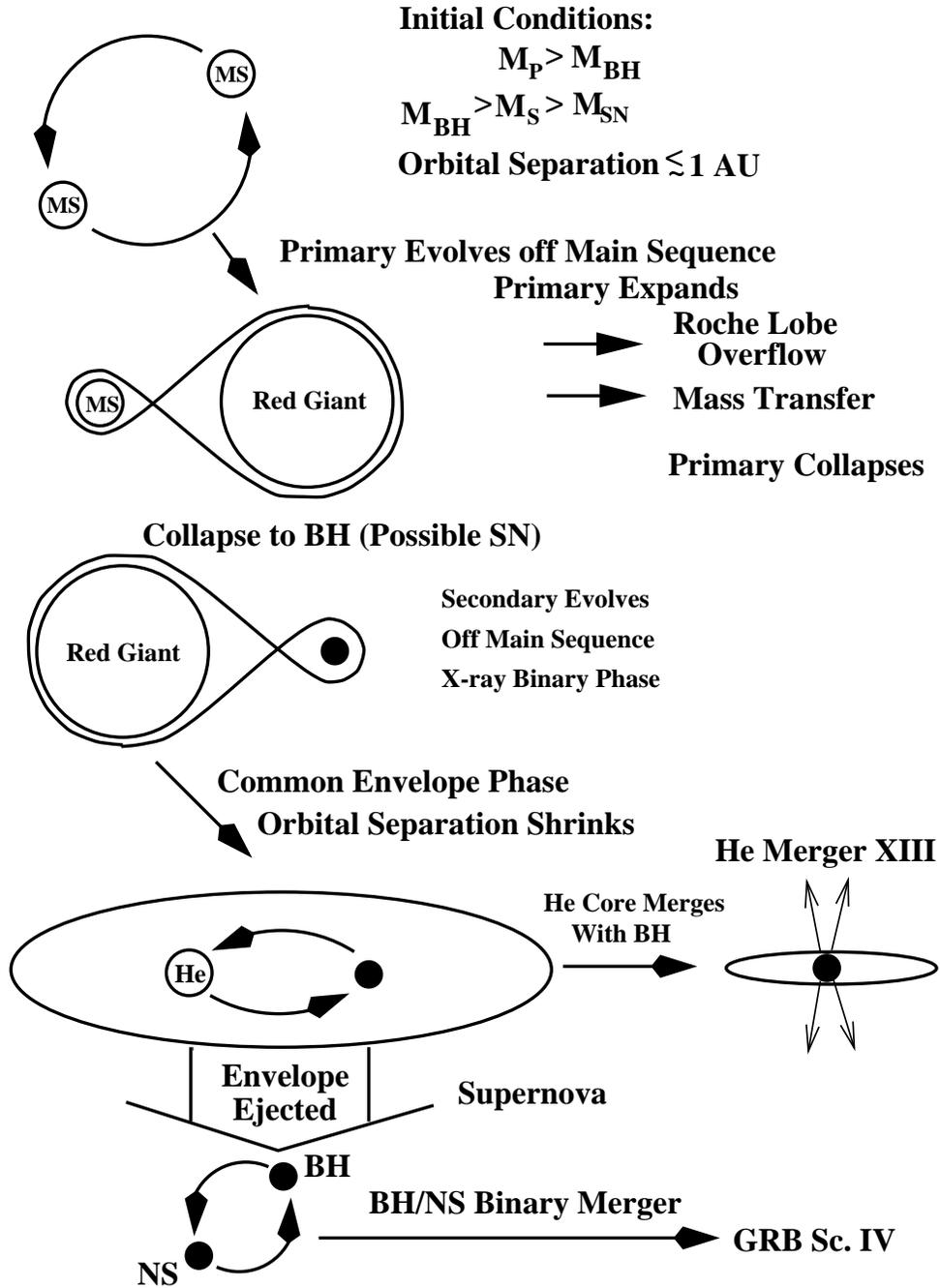}{7in}{0}{70}{70}{-200}{0}
\caption{Scenario IV: The ``standard'' BH/NS 
binary formation phase.  This scenario is identical 
to Scenario I (Fig. 2) except that the primary mass ($M_{\rm p}$) 
is greater than the critical mass for black hole formation.}
\end{figure}
\clearpage

\begin{figure}
\plotfiddle{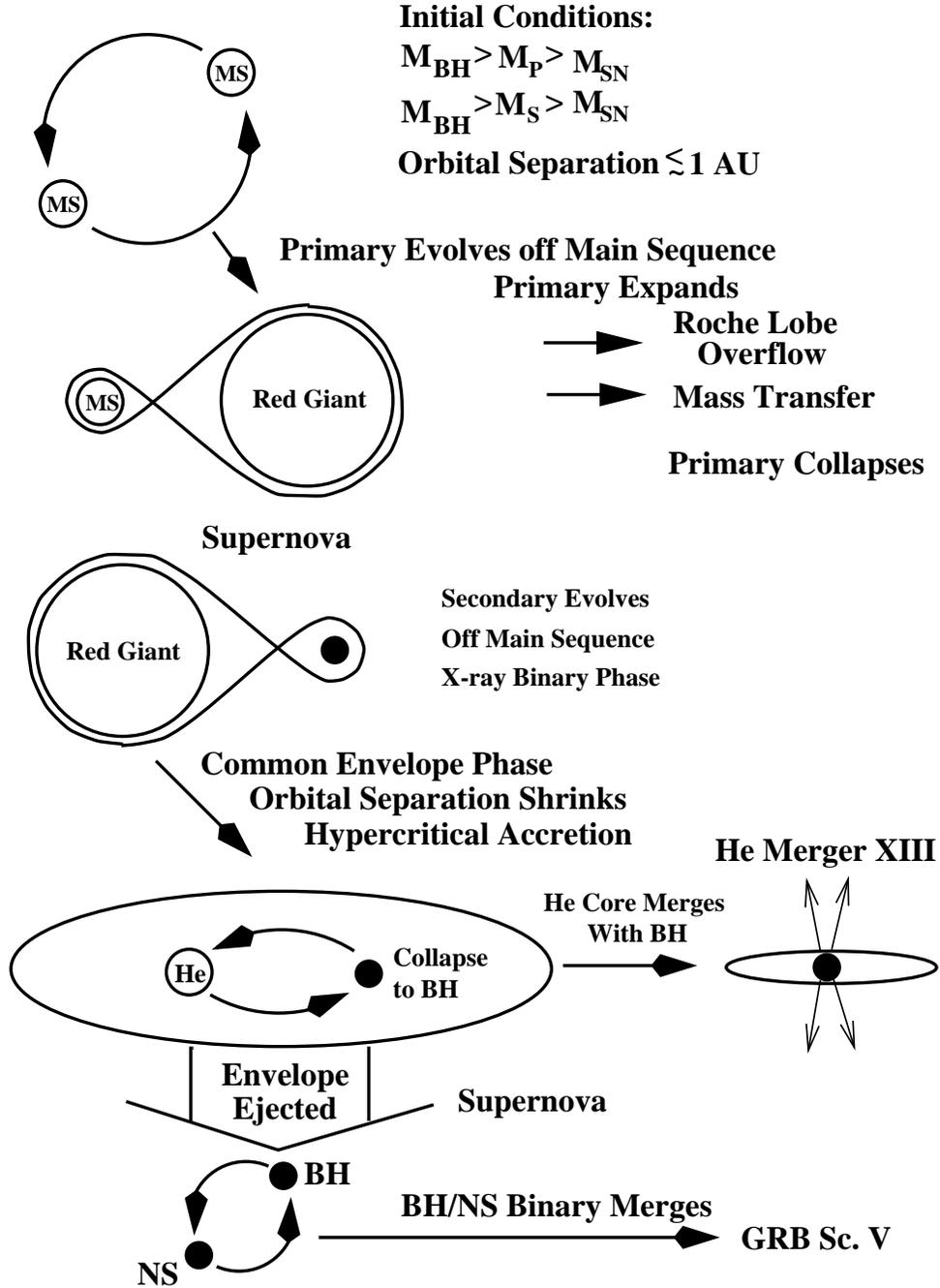}{7in}{0}{70}{70}{-200}{0}
\caption{Scenario V:  low-mass black hole + neutron 
star binary formation scenario.  This scenario is 
identical to Scenario I (Fig. 2) except that hypercritical 
occurs during the common envelope phase, causing the 
neutron star to collapse into a black hole.}
\end{figure}
\clearpage

\begin{figure}
\plotfiddle{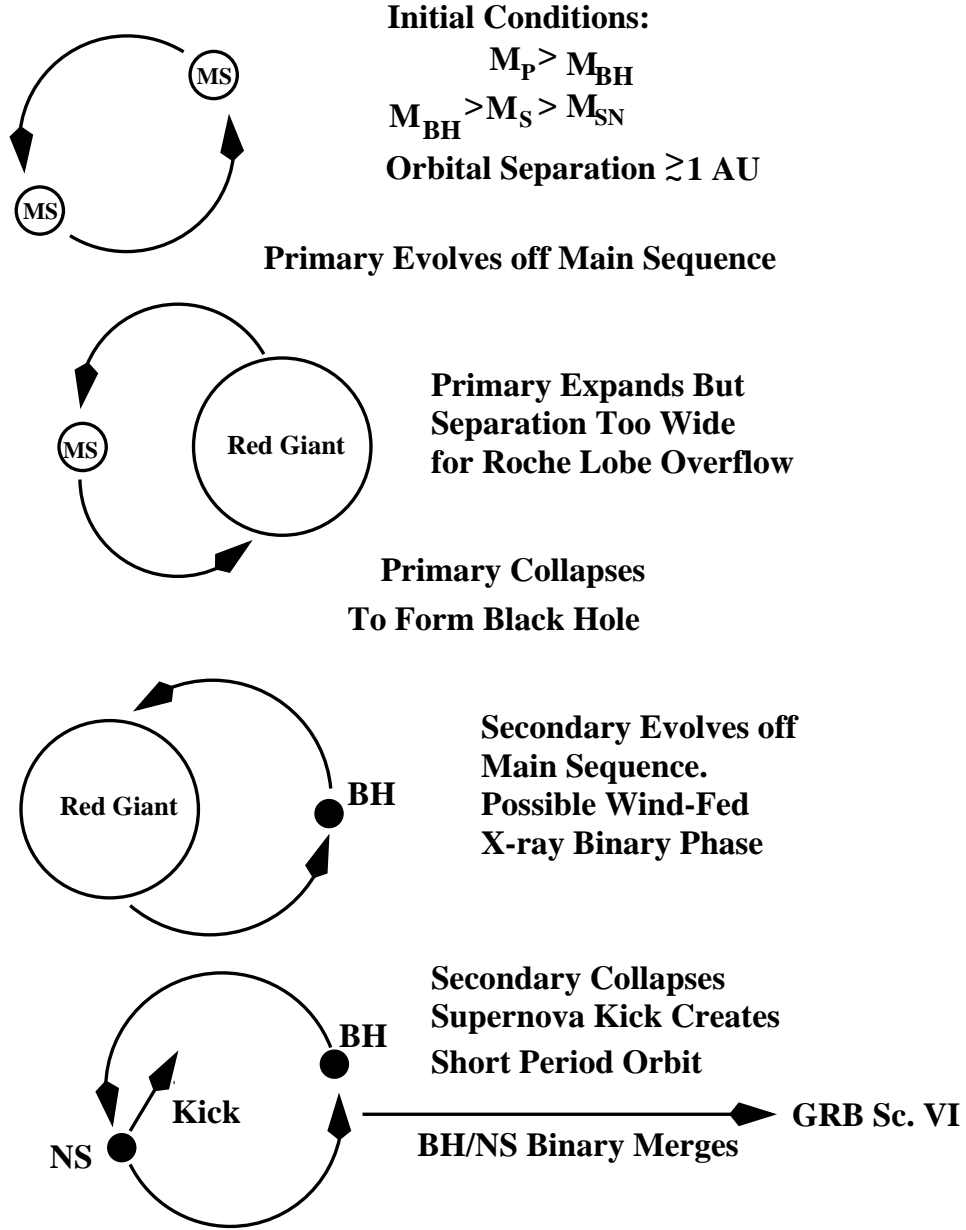}{7in}{0}{70}{70}{-200}{0}
\caption{Scenario VI:  The kick scenario for BH/NS binaries.  
Identical to scenario III (Fig. 4) except the primary mass 
is less than the critical black hole formation mass.}  
\end{figure}
\clearpage

\begin{figure}
\plotfiddle{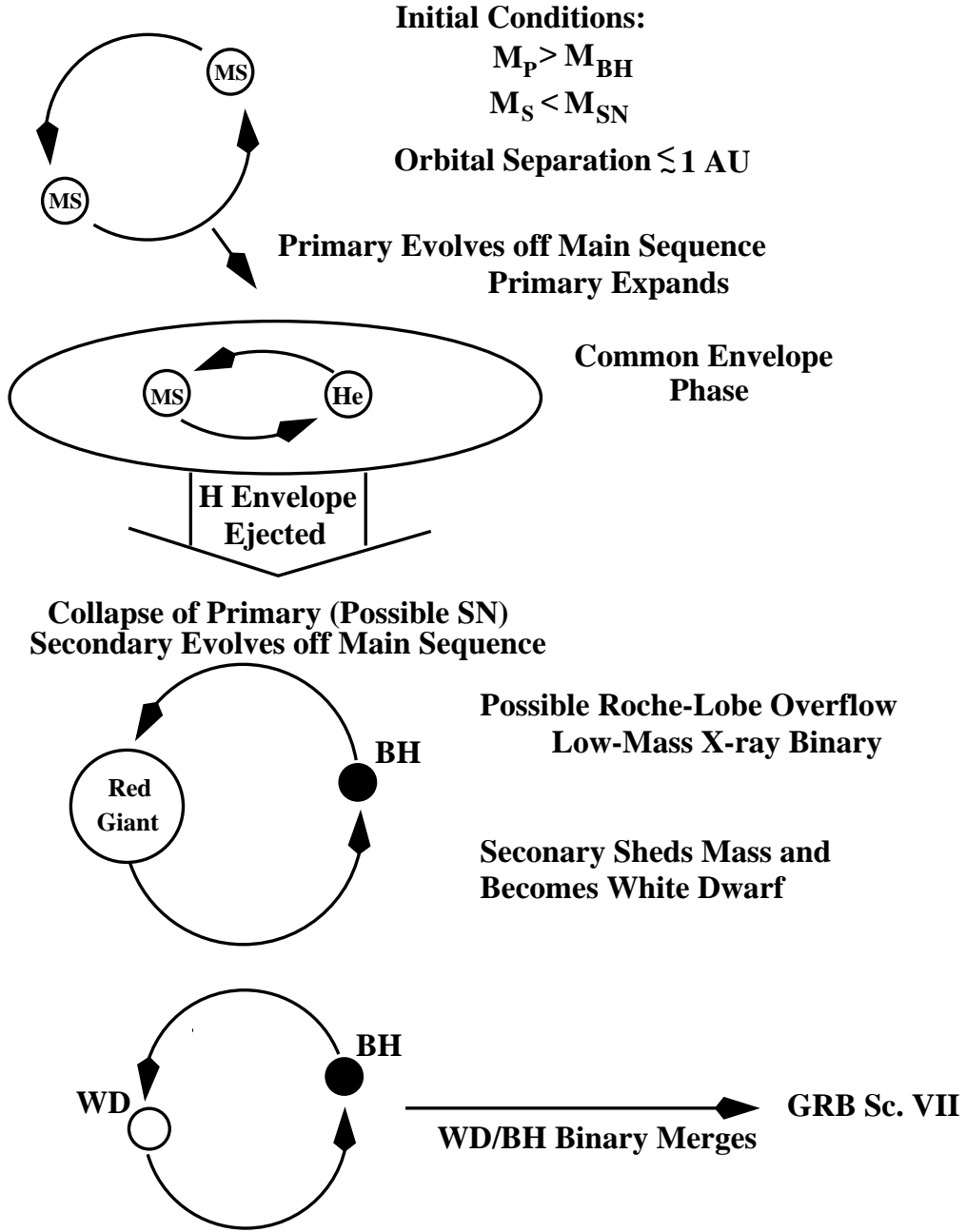}{7in}{0}{70}{70}{-200}{0}
\caption{Scenario VII:  ``standard'' WD/BH formation 
scenario.  This scenario probably produces the intermediate 
mass X-ray binaries such as LMC X-3.  All symbols are 
described in Figs. 1 and 2.}  
\end{figure}
\clearpage

\begin{figure}
\plotfiddle{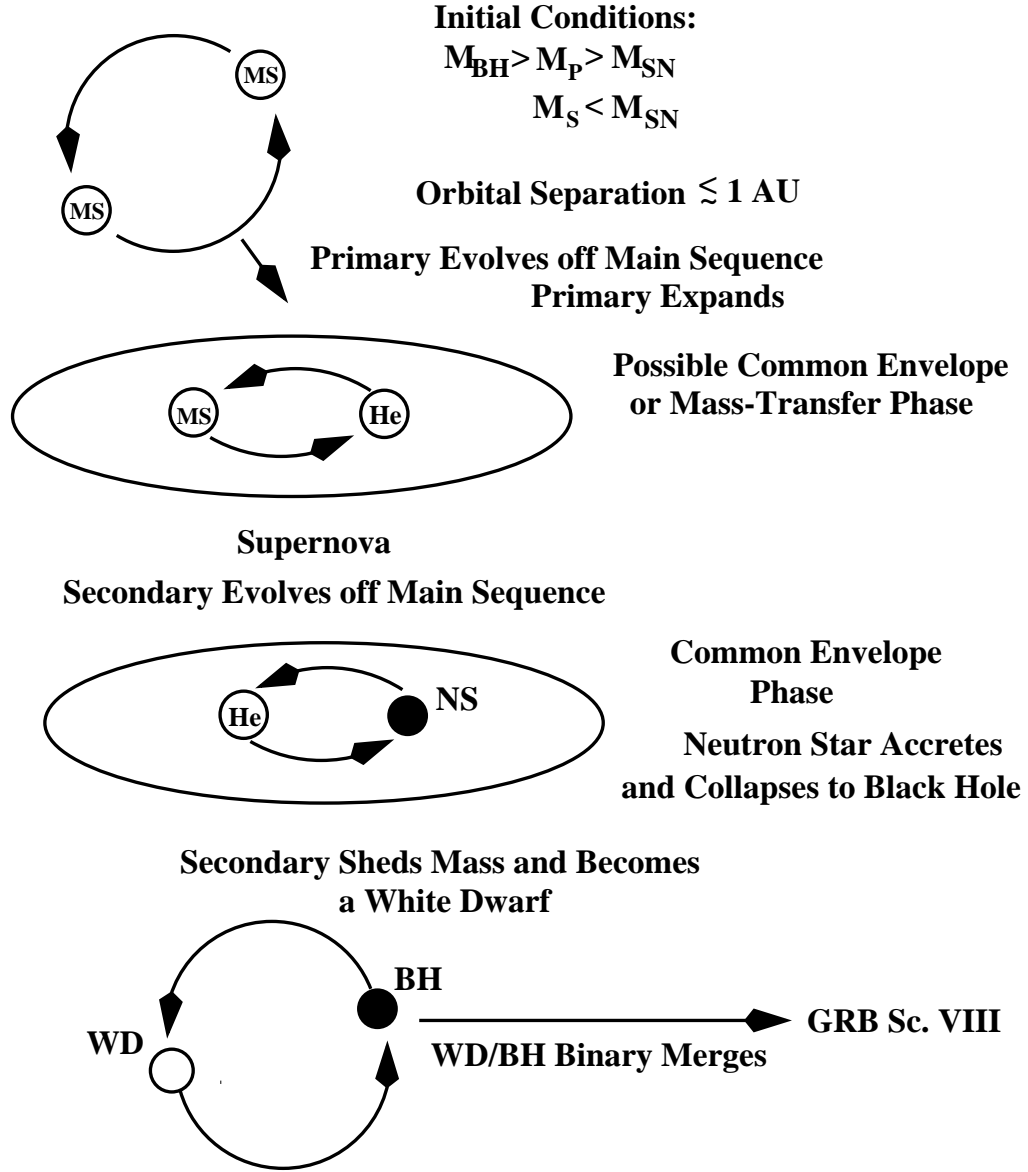}{7in}{0}{70}{70}{-200}{0}
\caption{Scenario VIII:  The hypercritical accretion 
scenario for WD/BH binaries.  This scenario is 
identical to scenario V (Fig. 6) except that the 
secondary forms a white dwarf instead of a neutron 
star.  Not that in the common envelope phase, the 
neutron star can merge and form a helium merger.}
\end{figure}
\clearpage

\begin{figure}
\plotfiddle{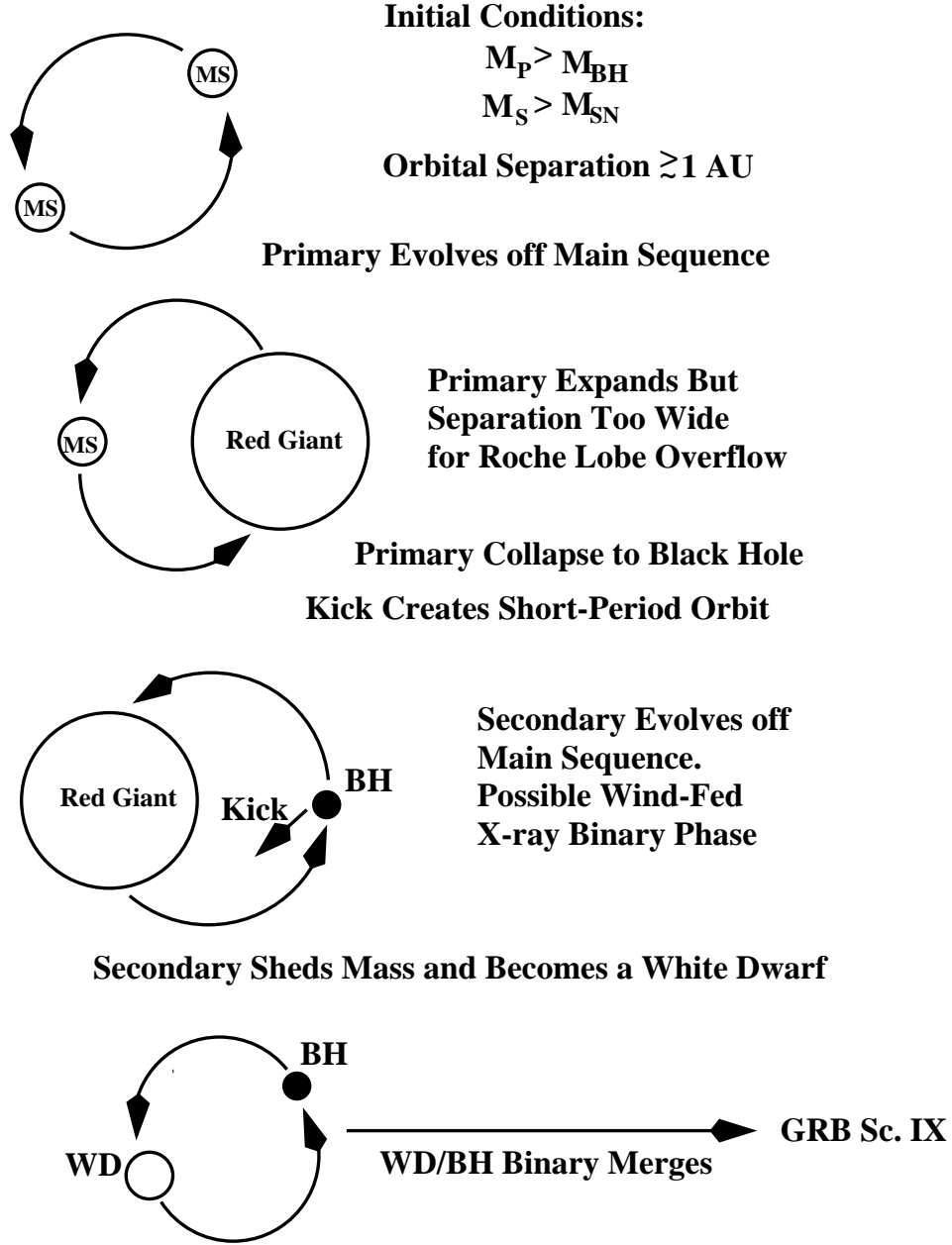}{7in}{0}{70}{70}{-200}{0}
\caption{Scenario IX:  The kick scenario for WD/BH binaries.  
Identical to scenario III (Fig. 4) except that the primary 
forms a black hole and the secondary forms a white dwarf.}  
\end{figure}
\clearpage

\begin{figure}
\plotfiddle{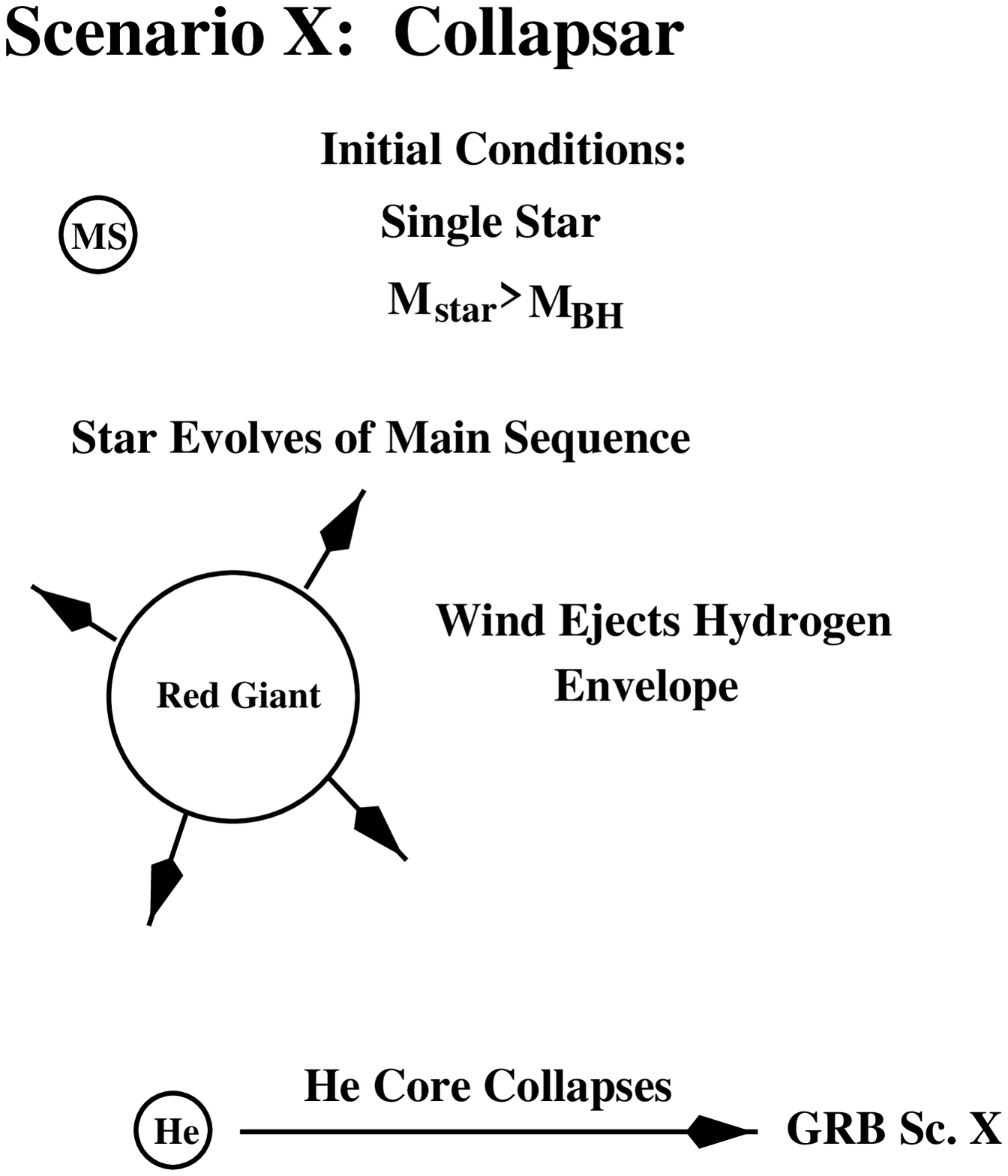}{7in}{0}{70}{70}{-200}{0}
\caption{Scenario X:  The single star Collapsar 
formation scenario.  Wolf-Rayet winds blow off the 
hydrogen mantle of a rotating massive star leaving behind a 
massive helium core which then collapses to a black 
hole.  The helium core must have a mass $\gtrsim 10 M_\odot$ to 
insure that the core immediately forms a black hole without 
any supernova explosion.}
\end{figure}
\clearpage

\begin{figure}
\plotfiddle{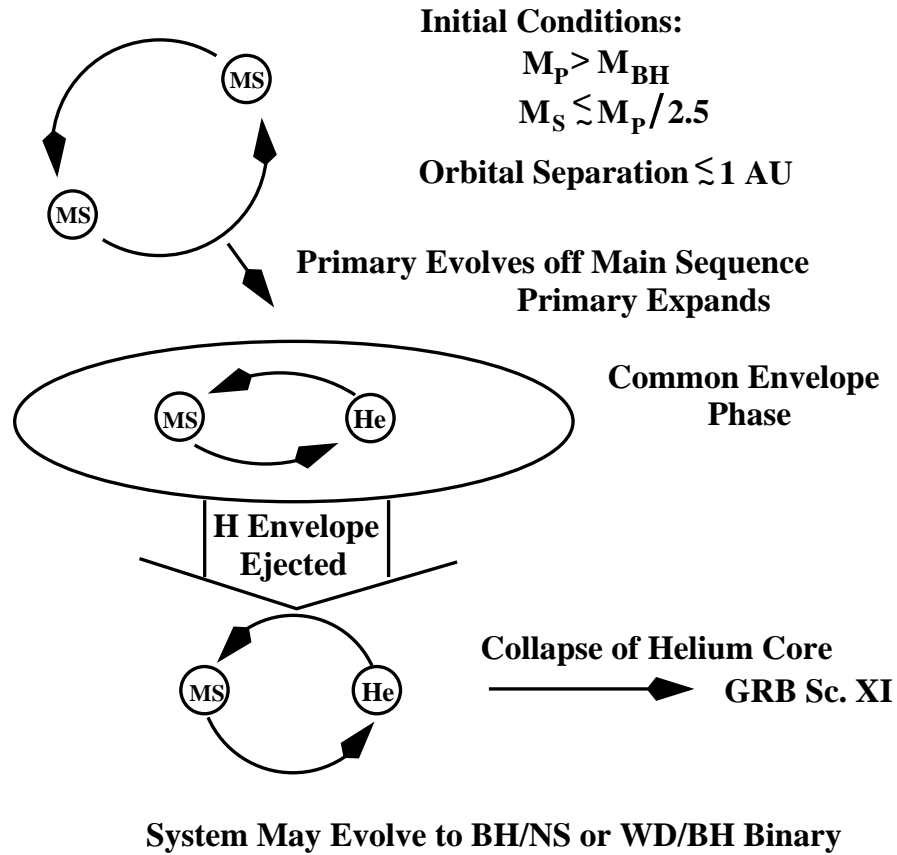}{7in}{0}{70}{70}{-200}{0}
\caption{Scenario XI:  The dominant collapsar formation 
scenario.  Common envelope evolution drives off the 
hydrogen mantle of a rotating massive star.  The 
helium core collapses to form a GRB.  The system 
may then go on to form a WD/BH or BH/NS binary.}
\end{figure}
\clearpage

\begin{figure}
\plotfiddle{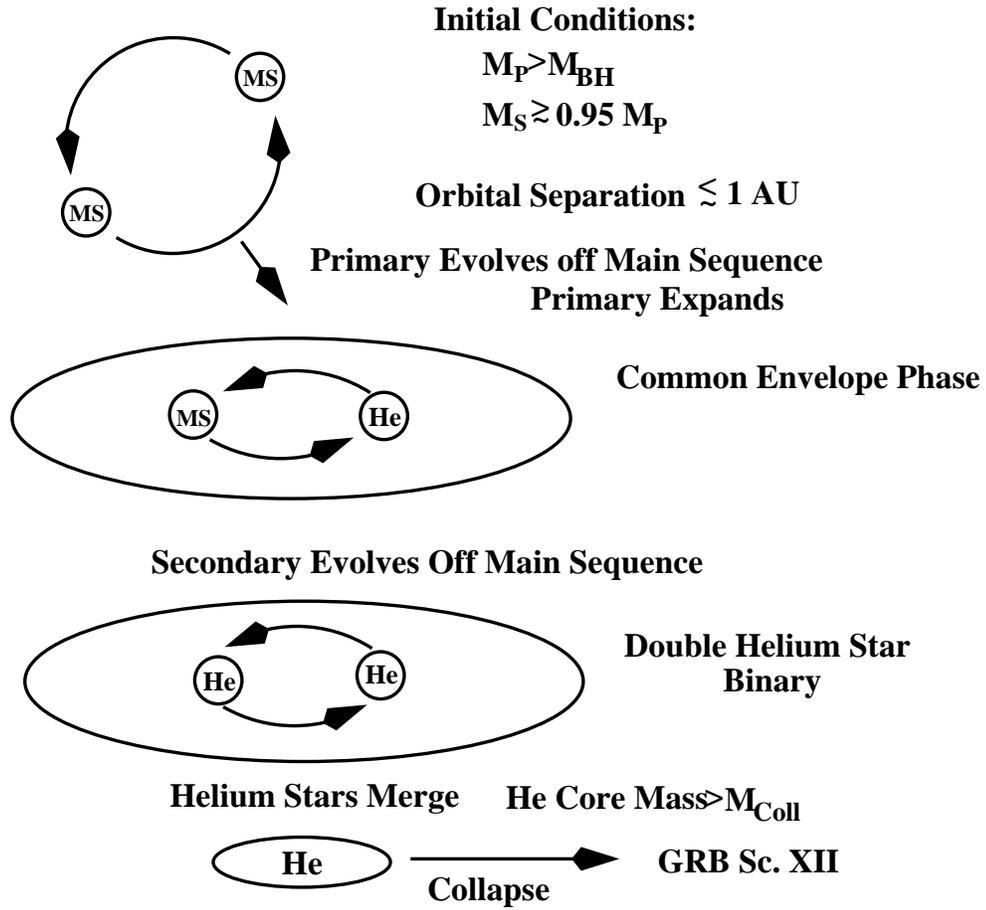}{7in}{0}{70}{70}{-200}{0}
\caption{Scenario XII:  collapsar scenario from the 
merger of a double helium binary.  This model follows 
the same path as scenario II (Fig. 3) except that the 
helium stars merge, producing a flattened helium 
core which then powers the collapsar.}
\end{figure}
\clearpage

\begin{figure}
\plotfiddle{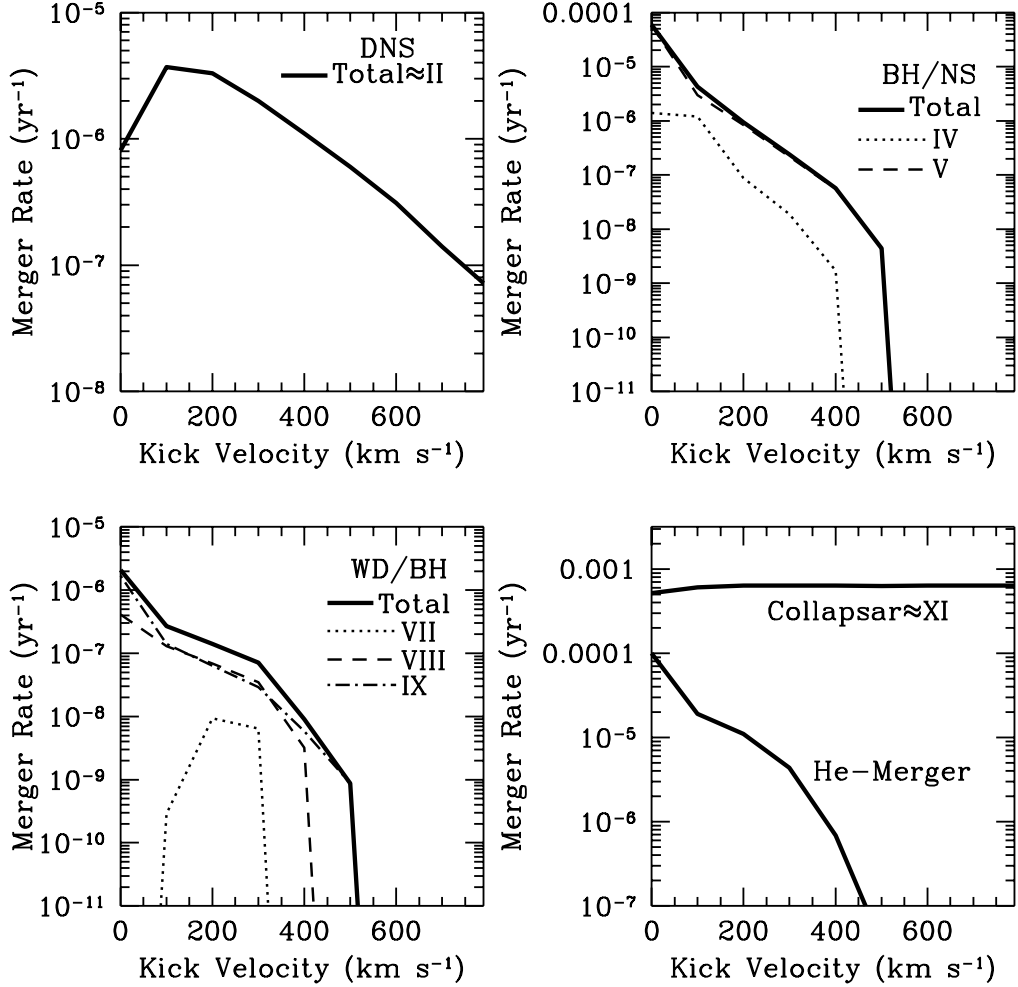}{7in}{0}{70}{70}{-200}{0}
\caption{Merger rates of the various GRB systems vs. 
kick magnitude for isotropic kicks with delta 
function velocities.  In the standard set 
of parameters, Scenario I is not allowed because hypercritical 
accretion converts such systems into BH/NS binaries via 
scenario V.  Kicks strongly effect the black hole binaries, 
but primaries which collapse to form black holes may receive 
lower kicks.  The drop at low velocities of the DNS 
mergers is caused because kicks are actually required to 
produce short-period (merging) binaries.  Collapsars are 
not affected by the kick as most collapsars are produced 
in the collapse of the primary or in single systems.}
\end{figure}
\clearpage

\begin{figure}
\plotfiddle{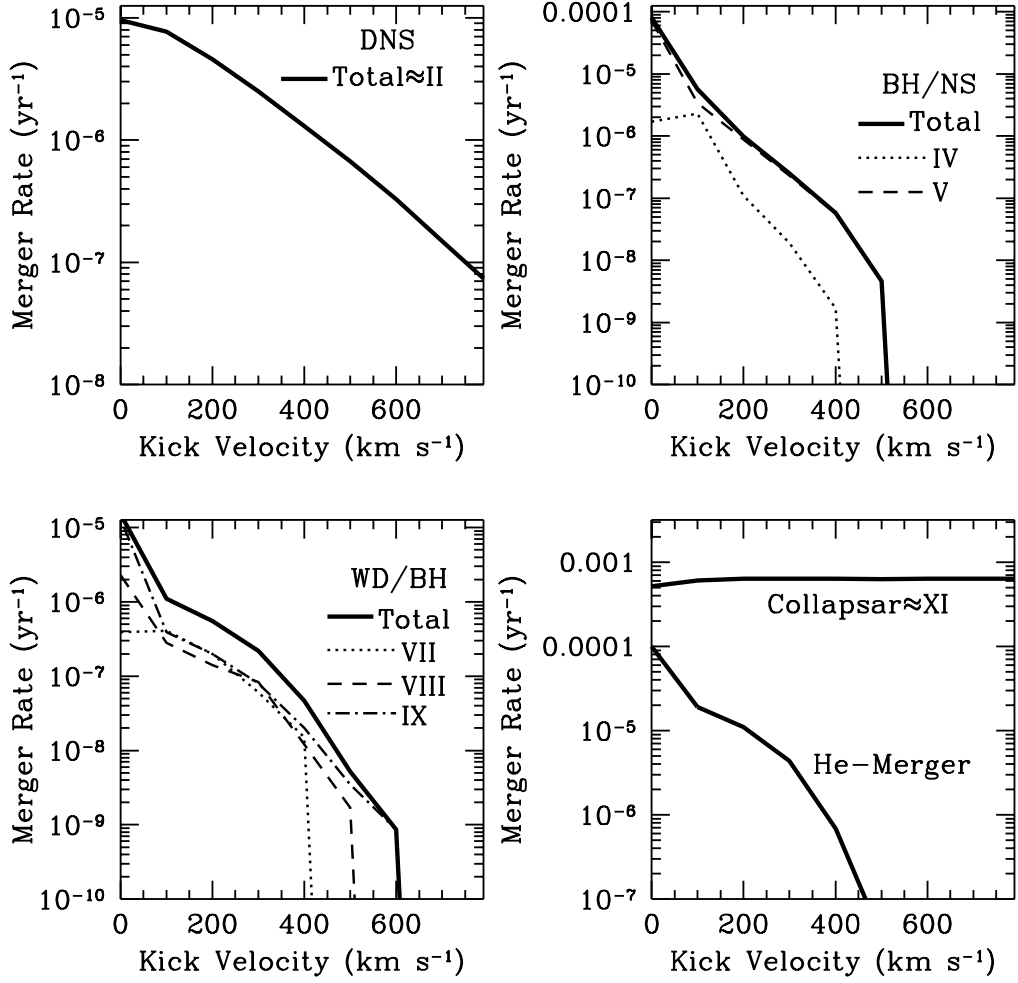}{7in}{0}{70}{70}{-200}{0}
\caption{Formation rates (rather than merger rates) vs. kick 
velocity from Fig. 14.  Note that the merger rate can be nearly 
an order of magnitude less than the formation rate for the binary 
mergers.}
\end{figure}
\clearpage

\begin{figure}
\plotfiddle{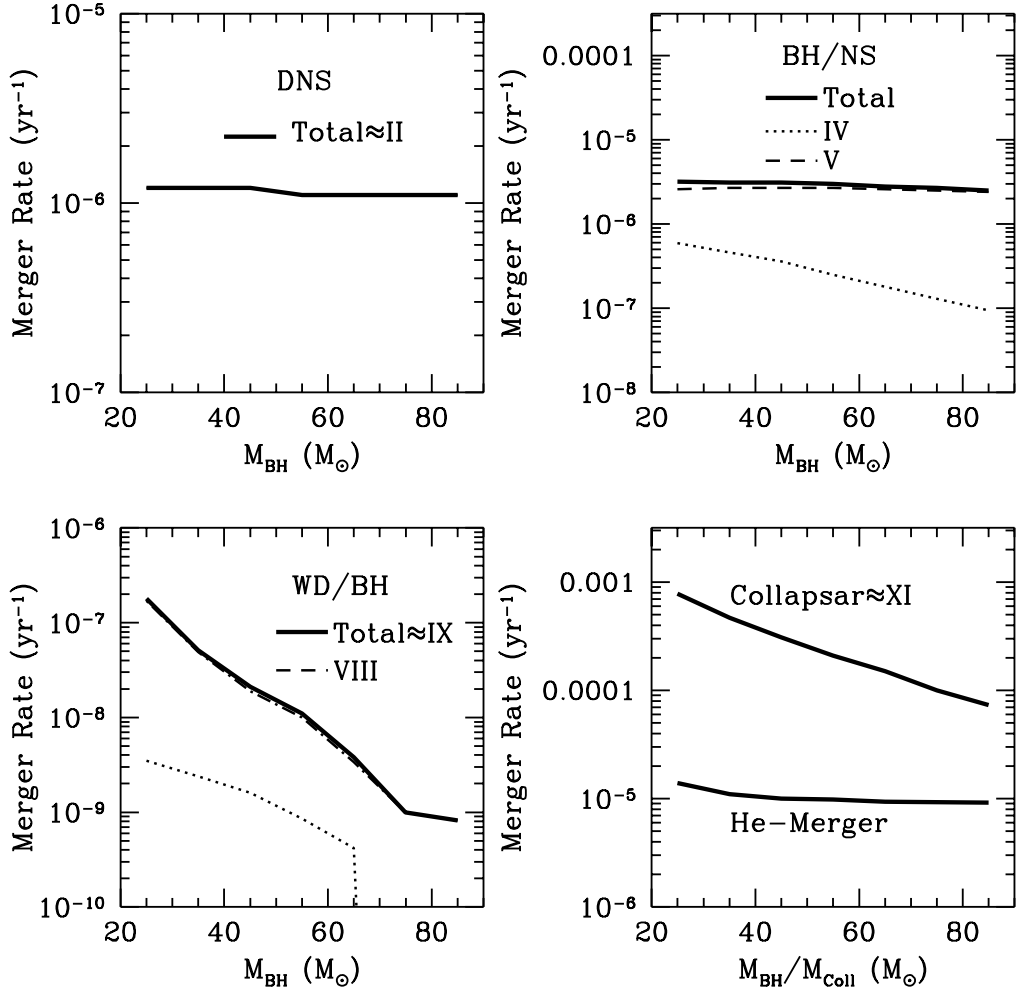}{7in}{0}{70}{70}{-200}{0}
\caption{Merger rates vs. critical black hole mass using 
the FBB kick distribution (40\% of neutron stars given kicks 
with $V_{\rm mean}=100 {\rm km \, s^{-1}}$ and 60\% with 
$V_{\rm mean}=600 {\rm km \, s^{-1}}$).  $M_{\rm Coll}=
M_{\rm BH}+15\,M$\sun.  Note that if 
hypercritical accretion does not occur, the BH/NS binary
merger rate decreases rapidly with increasing critical 
mass.}
\end{figure}
\clearpage

\begin{figure}
\plotfiddle{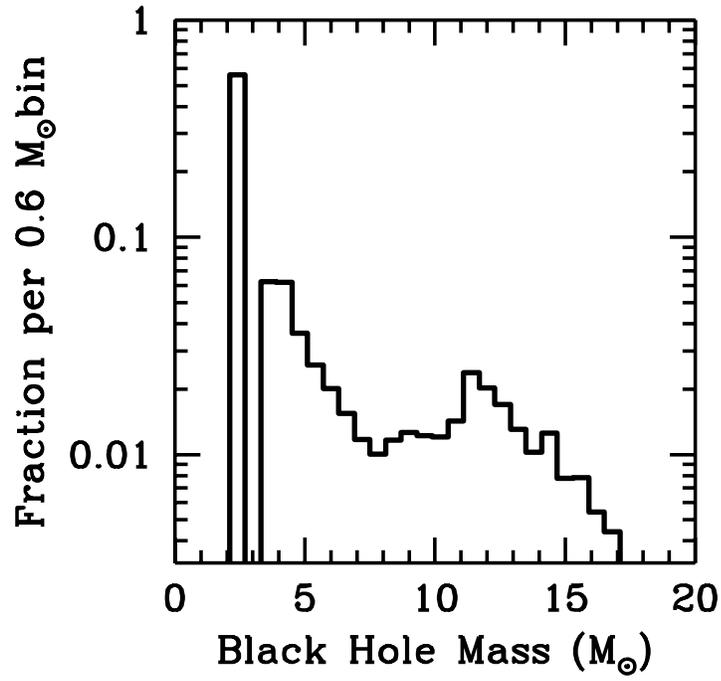}{7in}{0}{100}{100}{-175}{0}
\caption{Mass distribution of black holes in BH/NS binaries 
assuming the black hole mass is set to 1/3 the mass of 
its progenitor at the time of collapse.  The peak at 
2.4\,M\sun is due to those neutron stars which collapse 
to form black holes after accreting $\sim 1$M\sun\, 
in a common envelope phase.}
\end{figure}
\clearpage

\begin{figure}
\plotfiddle{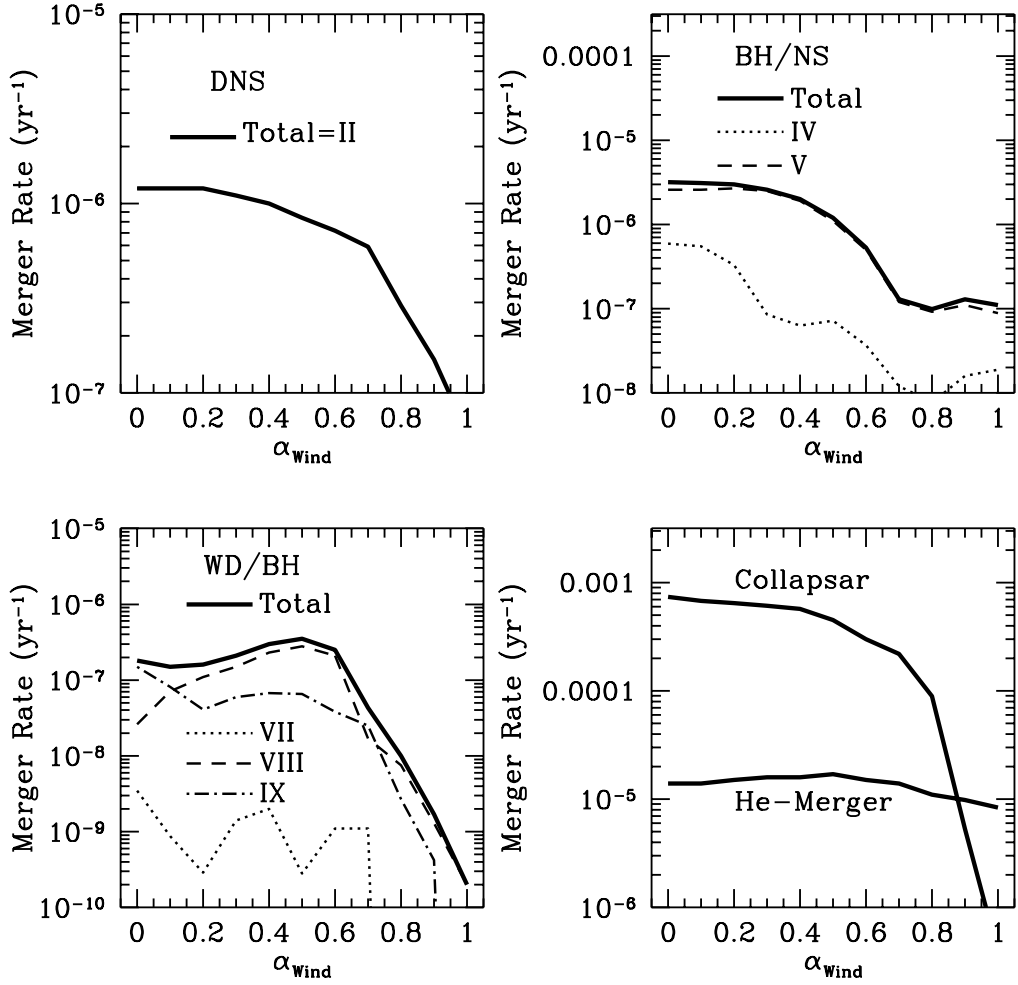}{7in}{0}{70}{70}{-200}{0}
\caption{Merger rates vs. mass loss fraction 
$\alpha_{\rm Wind}=$(Mass Loss)$/$(Mass Loss$_{\rm WLW}$).
The Collapsar rate depends most sensitively upon the 
mass loss rate, but for the range of reliable values 
for $\alpha_{\rm Wind} (0.-0.5$), the dependence is slight.  
However, if one uses the full Mass Loss of Woosley, Langer, 
\& Weaver (1993,1995), the collapsar and BH/NS GRB rates 
decrease dramatically (Table 4).}
\end{figure}
\clearpage

\begin{figure}
\plotfiddle{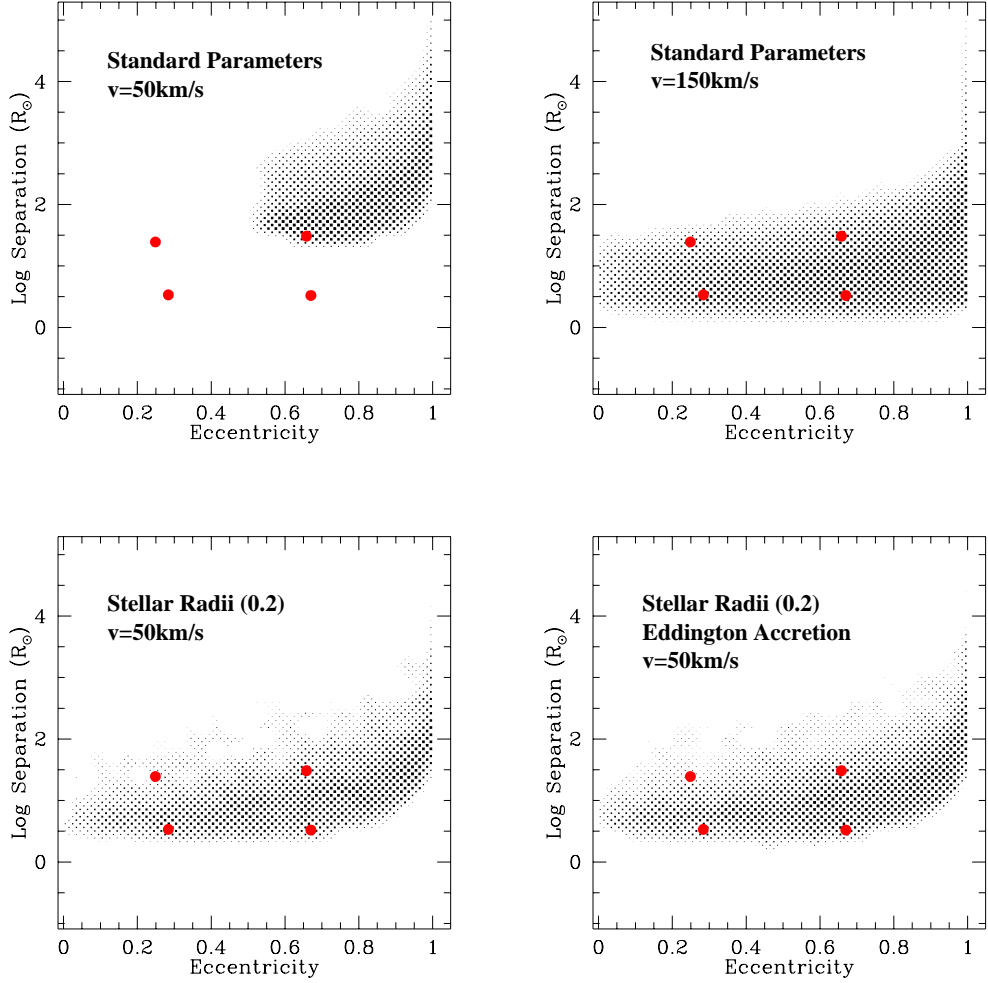}{7in}{0}{70}{70}{-200}{0}
\caption{Separation/eccentricity distribution of DNS binaries 
compared to the 4 observed galactic binaries.  Although 
we can constrain some sets of parameter choices, we 
do not constrain the allowed parameter space enough 
to severely limit the uncertainties.}
\end{figure}
\clearpage

\begin{figure}
\plotfiddle{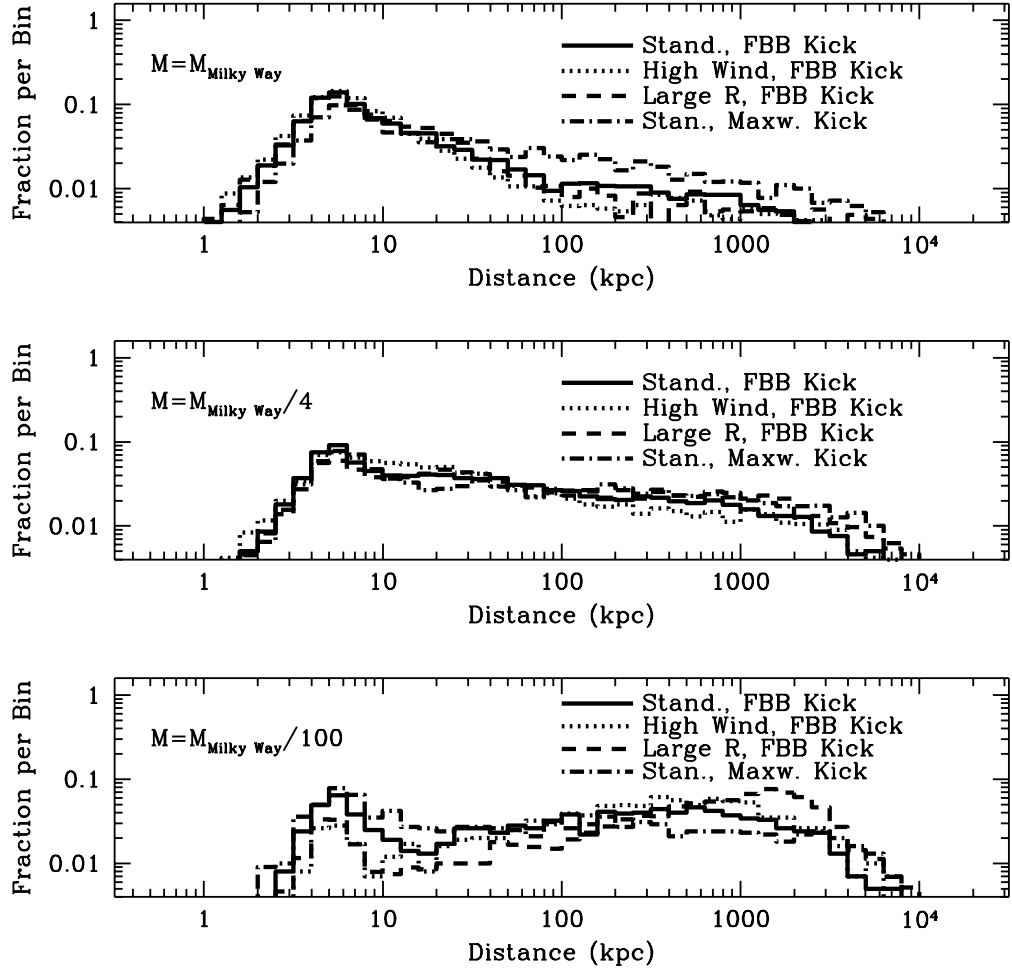}{7in}{0}{70}{70}{-200}{0}
\caption{Distribution of distance from the host galaxy 
for DNS and BH/NS binaries using a range of population 
synthesis parameters for (a) a galaxy mass equal to that 
of the Milky Way and (b) a galaxy mass set to 1/4th that 
of the Milky Way.  A distance of 100 kpc corresponds to 
roughly 20 arc seconds at a redshift of 1-3.} 
\end{figure}
\clearpage

\begin{figure}
\plotfiddle{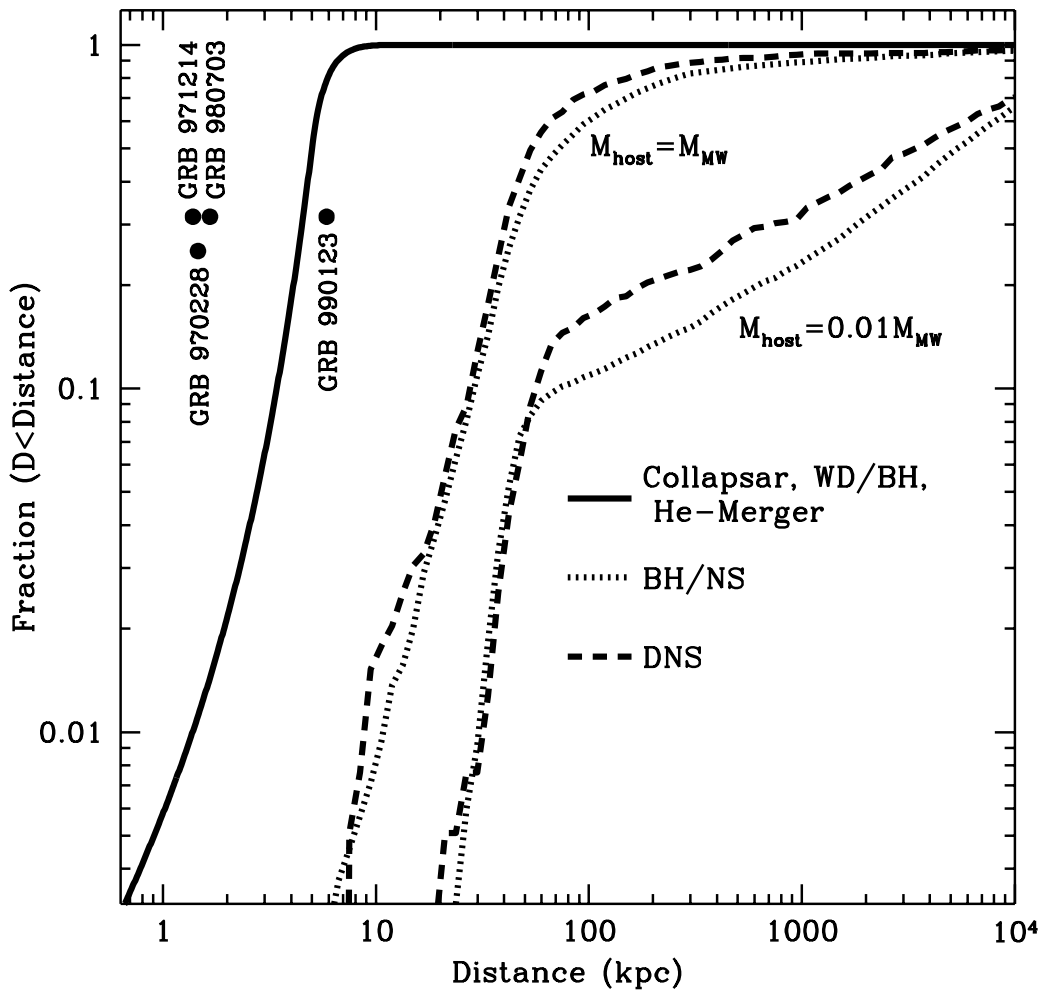}{7in}{0}{70}{70}{-200}{0}
\caption{The fraction of GRBs within a given distance of 
their host galaxy using the standard set of population 
synthesis parameters with large radii and the FBB kick 
distribution.  Note that the DNS and BH/NS mergers occur 
at distances well beyond the observed long-duration 
bursts.  The distribution of WD/BH mergers, collapsars 
and helium-mergers trace our distribution of star formation 
regions.   The separations of the localized GRBs from the 
center of the host galaxy are given for comparison assuming 
$h_0=65$\,km\,s$^{-1}$\,Mpc$^{-1}$ and $q_0=0.$.  We do 
not include GRB 970508 which was at the host center).}
\end{figure}
\clearpage

\begin{figure}
\plotfiddle{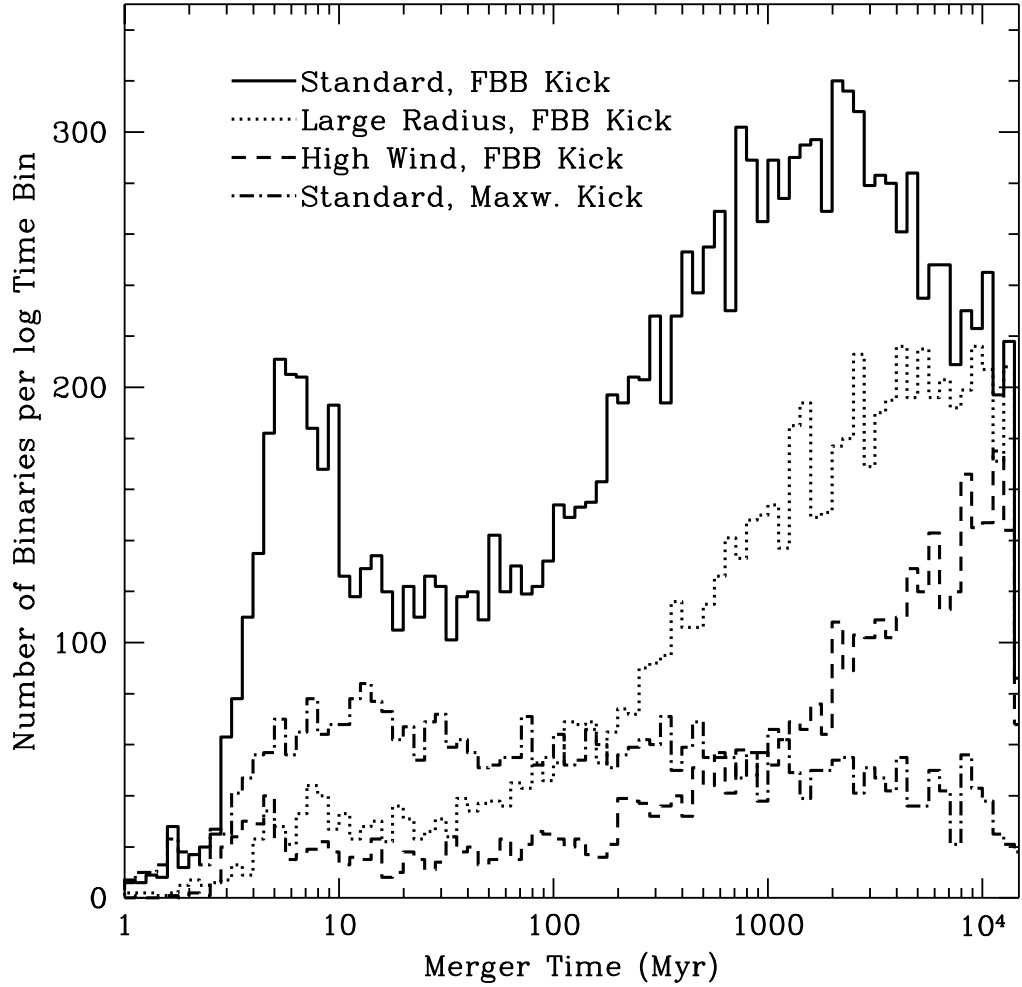}{7in}{0}{70}{70}{-200}{0}
\caption{Merger times for DNS and BH/NS binaries using a range 
of population synthesis parameters.  Higher stellar radii or 
winds produce wider binaries and hence, longer merger times.  
The double peaked merger distribution for the standard model 
(solid line) is a signature of the double peaked FBB kick 
distribution.  $>$90\% of the binaries with merger times 
less than 30 Myr received kicks of $300 {\rm km \, s^{-1}}$ 
or greater.}
\end{figure}
\clearpage

\begin{figure}
\plotfiddle{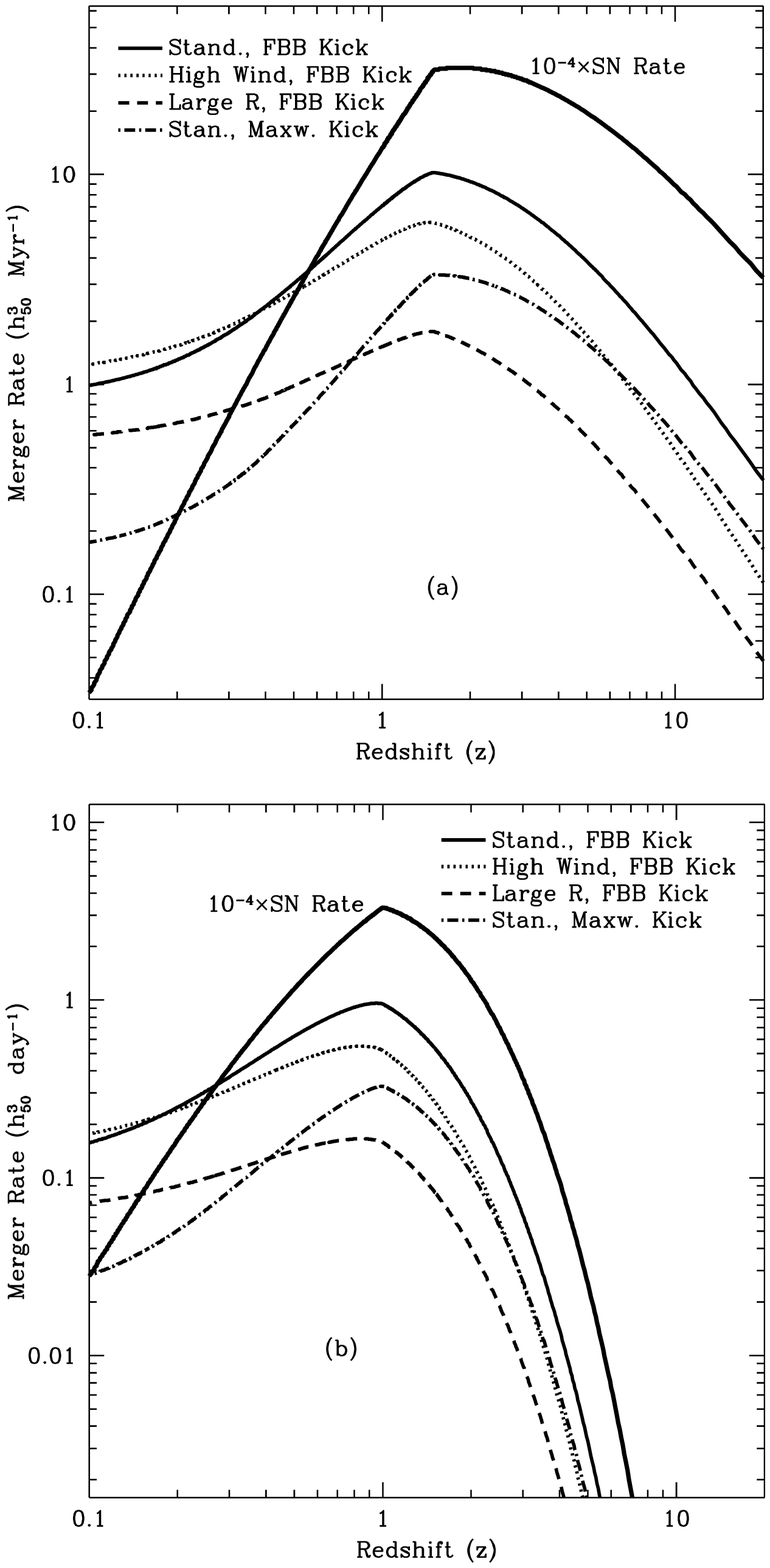}{7in}{0}{70}{70}{-225}{-10}
\caption{The redshift distribution of the merger rate 
of DNS and BH/NS binaries for an Einstein-deSitter universe 
and three different star formation histories:  a) flattened 
star formation rate - A=3.0, B=0.0, z$_{\rm p}$=1.5; b) peaked
star formation rate - A=1.0, B=0.5, z$_{\rm p}$=1.0.
The star formation history is proportional to the supernova history 
(thick solid line).  The merger times cause a delay in the bursts 
produced by these mergers.  The median redshift of DNS and 
BH/NS GRBs is much less than that of supernovae.  The 
median redshift of WD/BH mergers, helium-mergers, and collapsars 
are all roughly equal to the supernova median redshift.}
\end{figure}
\clearpage


\begin{thebibliography}{}

\bibitem[1998]{Arz98}
Arzoumanian, Z., Cordes, J.M., \& Wasserman, I., 1999, 
submitted to ApJ, astro-ph/9811323

\bibitem[1996]{Bai96}
Bailes, M. 1996, Pulsars: Problems and Progress, ASP Conf. Series, 
105, 3, ed. Johnston, Walker, \& Bailes

\bibitem[1998]{Bel98}
Belczy\'nski, K., \& Bulik, T., 1998, submitted to A\&A, astro-ph/9901193

\bibitem[1998]{Bet98} 
Bethe, H., \& Brown, G.E., 1998, ApJ, 506, 780

\bibitem[1977]{Bla77}
Blandford, R.D., \& Znajek, R.L., 1977, MNRAS, 179, 433

\bibitem[1998]{Blo98}
Bloom, J.S., Djorgovski, S.C., Kulkarni, S.R., \& Frail, D.A., 
1998, 507, L25

\bibitem[1998]{Blo99a}
Bloom, J.S., {\it et al.} 1999a, ApJ Letters, submitted

\bibitem[1998]{Blo99b}
Bloom, J.S., Sigurdsson, S., \& Pols, O. 1999b, MNRAS, in press 

\bibitem[1995]{Bra95}
Brandt, N., \& Podsiadlowski, P., 1995, MNRAS, 274, 461

\bibitem[1995]{Bro95} 
Brown, G.E. 1995, ApJ, 440, 270

\bibitem[1998]{Bul98}
Bulik, T., Belczy\'nski, K., \& Zbijewski, W.  1999, 
submitted to MNRAS, astro-ph/9903407

\bibitem[1986]{Bur86}
Burrows, A. \& Woosley, S.E., 1986, ApJ, 308, 680

\bibitem[1997]{Cap97}
Cappellaro, E., Turatto, M., Tsvetkov, D. Yu., Bartunov, O. S.,
Pollas, C., Evans, R., Hamuy, M., 1997, A\&A, 322, 431

\bibitem[1993]{Car93}
Caraveo, P., 1993, ApJ, 415, L111

\bibitem[1993]{Che93} 
Chevalier, R.A., 1993, ApJ, 411, L33

\bibitem[1996]{Che96} 
Chevalier, R.A., 1996, ApJ, 459, 322

\bibitem[1979]{Cla79}
Clark, J.P.A., van den Heuvel, E.P.J., \& Sutantyo, W., 
1979, A\&A, 72, 120

\bibitem[1993]{Cor93}
Cordes, J.M., Romani, R.W., Lundgren, S.C., 1993, Nature, 362, 133

\bibitem[1998]{Cor98}
Cordes, J.M., \& Chernoff, D.F., astro-ph/9707308

\bibitem[1995]{Cur95}
Curran, S. J., Lorimer, D. R. 1995, MNRAS, 276, 347

\bibitem[1995]{Dal95}
Dalton, W.W., \& Sarazin, C.L. 1995, ApJ, 448, 369

\bibitem[1994]{Dav94}
Davies, M.B., Benz, W., Piran, T., Thielemann, F.K.,
1994, ApJ, 431, 742

\bibitem[1987]{Dew87}
Dewey, R.J. \& Cordes, J., 1987, ApJ, 321, 780

\bibitem[1998]{Djo98}
Djorgovski, S. G., Kulkarni, S. R., Bloom, J. S., Goodrich, R.,
Frail, D. A., Piro, L., $\&$ Palazi, E. 1998, ApJ 508, L17

\bibitem[1999]{Djo99}
Djorgovski, S., Kulkarni, S. R., Bloom, J. S., Frail, D. A., Chaffee, F.,
$\&$ Goodrich, R. 1999, GCN Circ. No 189

\bibitem[1998]{Ebe98}
Eberl, T., Diploma Thesis, MPA-garching

\bibitem[1998]{Ebe98b}
Eberl, T., Janka, H.-T., Ruffert, M., \& Fryer, C.L.,
in preparation

\bibitem[1998]{Erg98}
Ergma, E., \& van den Heuvel, E.P.J., 1998, A\&A, 331, L29

\bibitem[1999]{Fin99}
Finn, L. S. 1999, Lectures given at XXVI SLAC Summer Institute on Particle 
Physics ``Gravity: From the Hubble Length to the Planck Length'', 
August 3-14, 1998 (gr-qc/9903107)

\bibitem[1995]{Fis95}
Fishman, G. J., \& Meegan, C. A. 1995, ARA\&A, 33, 415

\bibitem[1975]{Fla75}
Flannery, B.P., \& van den Heuvel, E.P.J., 1975, A\&A, 39, 61

\bibitem[1994]{Fra94}
Frail, D.A., Goss, W.M., \& Whiteoak, J.B.Z., 1994, ApJ, 342, 260

\bibitem[1997]{Fra97} 
Frail, D.A., Kulkarni, S.R., Nicastro, S.R., Feroci, M., \& 
Taylor, G.B 1997, Nature, 389, 261

\bibitem[1996]{Fry96} 
Fryer, C.L., Benz, W., \& Herant, M., 1996, ApJ, 460, 801

\bibitem[1998]{Fry98a}
Fryer, C.L., \& Kalogera, V., 1998, ApJ, 489, 244

\bibitem[1998]{Fry98b}
Fryer, C.L., Burrows, A., \& Benz, W., 1998, ApJ, 496, 333

\bibitem[1998]{Fry98c}
Fryer, C.L., \& Woosley, S.E., 1998, ApJ, 502, L9

\bibitem[1999]{Fry99a}
Fryer, C. L. 1999, accepted by ApJ

\bibitem[1999]{Fry99b}
Fryer, C. L., Woosley, S. E., Herant, M., \& Davies, M. B., 
1999, accepted by ApJ

\bibitem[1995]{Gae95}
Gaensler, B.M, \& Johnston, S., 1995, MNRAS, 273, L73

\bibitem[1980]{Gar80}
Garmany, C.D., Conti, P.S., Massey, P., 1980, ApJ, 242, 1063

\bibitem[1990]{HEW90}
Hartmann, D. H., Epstein, R. I., \& Woosley, S. E. 1990, ApJ 348, 625

\bibitem[1995]{HW95}
Hartmann, D. H., \& Woosley, S. E. 1995, Advances in Space Research, 
vol. 15, no. 5, 143

\bibitem[1998]{Heg98}
Heger, A. 1998, PhD Thesis, MPA-Garching

\bibitem[1991]{Hil91}
Hills D., Bender, P.L., \& Webbink, R.F., 1991, 
ApJ, 369, 271

\bibitem[1983]{Hil83}
Hills, J.G. 1983, ApJ, 267, 322

\bibitem[1990]{Hog90}
Hogeveen, S.J., 1990, Ap\&SS, 173,315

\bibitem[1999]{HF99}
Hogg, D. W., \& Fruchter, A. S. 1999, ApJ, in press

\bibitem[1998]{Hol98}
Holz, D.E., Miller, M.C., \& Quashnock, J.M., 1999, 
submitted to ApJ

\bibitem[1991]{Hou91} 
Houck, J.C., \& Chevalier, R.A., 1991, ApJ, 376, 234

\bibitem[1996]{Jan96}
Janka, H.-T., \& Ruffert, M., 1996, A\&A, 306, L33

\bibitem[1998]{Jan98}
Janka, H.-T., \& Ruffert, M., \& Eberl, 1998, astro-ph/9810057

\bibitem[1996]{Kal98a}
Kalogera, V., \& Webbink, R.F., 1998, ApJ, 493, 351

\bibitem[1998]{Kal98b}
Kalogera, V. 1998, ApJ, 493, 368

\bibitem[1999]{Kal99}
Kalogera, V., 1999, accepeted by ApJ

\bibitem[1996]{Kas96}
Kaspi, V. M., Manchester, R. N., Johnston, S., Lyne, A. G.,
D'Amico, N. 1996, AJ, 111, 2028

\bibitem[1997]{Kat97}
Katz, J.I., 1997, ApJ, 490, 633

\bibitem[1997]{Kol97} 
Kobulnicky, H.A., and Skillman, E.D., 1997, ApJ, 489, 636

\bibitem[1979]{Kra79}
Kraicheva, Z. T., Popova, E. I., Tutukov, A. V., \& Yungelson, 
L. R. 1979, Soviet Astron., 56, 520

\bibitem[1998]{Kul98}
Kulkarni, S. R., {\it et al.} 1998, Nature 393, 35

\bibitem[1995]{Lee95}
Lee, W. H., \& Kluzniak, W. 1995, AcA 45, L705

\bibitem[1987]{Lip87}
Lipunov, V.M., Postnov, K.A., \& Prokhorov, M.E., 1987,
A\&A, 176, L1

\bibitem[1995]{Lip95}
Lipunov, V.M., Nazin, S.N., Panchenko, I.E., Postnov, K.A.,
\& Prokhorov, M.E., 1995, A\&A, 298, 677

\bibitem[1999]{Liv99}
Livio, M., Ogilvie, G. I., Pringle, J. E. 1999, ApJ, 512, L100

\bibitem[1994]{Lyn94}
Lyne, A.G., Lorimer, D.R., 1994, Nature, 369, 127L

\bibitem[1992]{Mae92}
Maeder, A., 1992, A\&A, 264, 105

\bibitem[1986]{Mac86}
MacDonald, D.A., Thorne, K.S., Price, R.H., \& Zhang, X.-H., 
1986, in ``Black Holes, the Membrane Paradigm'', Eds. Thorne, K.S., 
Price, R.H., \& MacDonald, D.A., Yale Univ. Press

\bibitem[1986]{Mac98}
MacFadyen, A., \& Woosley, S.E., 1999, accepted by ApJ

\bibitem[1997]{Mat97}
Mathews, G.J., Wilson, J.R., 1997, ApJ, 482, 929

\bibitem[1992]{MR97b}
Meszaros, P., \& Rees, M. J. 1997, ApJ 476, 232 

\bibitem[1992]{MR97a}
Meszaros, P., \& Rees, M. J. 1997, ApJ 482, L29

\bibitem[1997]{Met97} 
Metzger, M.R., Djorgovski, S.G, Kulkarni, S.R., Steidel, C.C., 
Adelberger, K.L., Frail, D.A., Costa, E., \& Frontera, F., 1997, 
Nature, 387, 389

\bibitem[1991]{Nar91}
Narayan R., Piran T., Shemi A., 1991, ApJ, 379, L17

\bibitem[1991]{Nar92}
Narayan R., Mahadevan, R., \& Quataert, E. 1998, to appear in 
``The Theory of Black Hole Accretion Discs'', eds. M. A. Abramowicz,
G. Bjornsson, and J. E. Pringle

\bibitem[1991]{Pac91b}
Paczynski, B., 1991, AcA, 41, 257

\bibitem[1998]{P98}
Paczysnki, B. 1998, ApJ 494, L45

\bibitem[1998]{PLF98}
Pascarelle, S. M., Lanzetta, K. M., \& Fernandez-Soto, A. 1998, 
astro-ph/9810060

\bibitem[1993]{Pee93}
Peebles, P. J. E., 1993, Principles of Physical Cosmology,
Princeton Univ. Press (Princeton)

\bibitem[1991]{Phi91}
Phinney E.S., 1991, ApJ, 380, L17

\bibitem[1994]{Phi94}
Phinney E.S., \& Kulkarni, S.R., 1994, ARA\&A, 32, 591

\bibitem[1992]{Pod92}
Podsiadlowski, P., Joss, P.C., Hsu, J.J.L., 1992, ApJ, 391, 246

\bibitem[1999]{Pop99}
Popham, R., Woosley, S. E., \& Fryer, C. L., 1999, accepted by 
ApJ, astro-ph/9807028

\bibitem[1996]{Por96a}
Portegies-Zwart, S.F., \& Spreeuw, H.N., 1996, A\&A, 312, 670

\bibitem[1996]{Por96b}
Portegies-Zwart, S.F., Verbunt, F., \& Ergma, E., 1997, A\&A, 321, 207

\bibitem[1998]{Por98}
Portegies-Zwart, S.F., \& Yungelson, L.R., 1998, accepted 
in A\&A

\bibitem[1996]{Ras96}
Rasio, F. A., \& Livio, M. 1996, ApJ, 471, 366

\bibitem[1994]{Ras94}
Rasio, F. A., \& Shapiro, S. L., 1994, ApJ, 432, 242

\bibitem[1996]{Ruf96}
Ruffert, M., Janka, H.-T., Sch\"afer, G.,
1997, A\&A, 311, 532

\bibitem[1997]{Ruf97}
Ruffert, M., Janka, H.-T., Takahashi, K., Schaefer, G.,
1997, A\&A, 319, 122

\bibitem[1998]{Ruf98}
Ruffert, M., \& Janka, H.-T., 1998, submitted to A\&A

\bibitem[1998]{San98}
Sandquist, E.L., Taam, R.E., Chen, X., Bodenheimer, P., \& 
Burkert, A., 1998, accepted by ApJ

\bibitem[1986]{Sca86}
Scalo, J. M. 1986, Fund. of Cosmic Phys., 11, 1

\bibitem[1998]{Spr98}
Spruit, H. C., \& Phinney, E. S. 1998, Nature, 393, 193

\bibitem[1995]{Ter95}
Terman, J.L., Taam, R.E., \& Hernquist, L., 1995, ApJ, 445, 367

\bibitem[1993]{Tut93}
Tutukov, A.V., \& Yungelson, L.R., 1993, MNRAS, 260, 675

\bibitem[1986]{van86}
van den Heuvel, E.P.J., \& Rappaport, S., 1986, Physics
of Be stars, Proc. of the 92nd IAU Coll., 291, Cambridge
University Press

\bibitem[1999]{vKw99}
van Kerkwijk, M. H., Kulkarni, S. R. 1999, accepted by ApJ

\bibitem[1993]{Ver93}
Verbunt, F., 1993, ARA\&A, 31, 93

\bibitem[1996]{Was96}
Wasserman, I., Cordes, J., \& Chernoff, D., 1997, in preparation

\bibitem[1998]{Wat98}
Watanabe, K., Hartmann, D.H., Leising, M.D., \& The, L.-S. 1999, 
in press

\bibitem[1979]{Web79}
Webbink, R.F., 1979, ApJ, 227, 178

\bibitem[1984]{Web84}
Webbink, R.F., 1984, ApJ, 277, 355

\bibitem[1999]{Wel99}
Wellstein, S. \& Langer, N. 1999, submitted to A\&A

\bibitem[1996]{Whi96}
White, N.E., \& Paradijs, J.V., 1996, ApJ, 473, L25

\bibitem[1996]{Wil96}
Wilson, J.R., Mathews, G.J., \& Marronetti, P., 1996, 
Phys. Rev. D, 54, 1317

\bibitem[1993]{Woo93}
Woosley, S.E., 1993, ApJ, 405, 273

\bibitem[1995]{Woo93b}
Woosley, S.E., Langer, N., \& Weaver, T.A., 1993, ApJ, 411, 823

\bibitem[1995]{Woo95b}
Woosley, S.E., Langer, N., \& Weaver, T.A., 1995, ApJ, 448, 315

\bibitem[1995]{Woo95}
Woosley, S.E., \& Weaver, T.A., 1995, ApJS, 101, 181

\bibitem[1993]{Yam93}
Yamaoka, H., Shigeyama, T., Nomoto, K., 1993, ApJ, 267, 433




\end{thebibliography}
\end{document}